\documentclass[leqno, a4paper, 11pt]{article}
\usepackage[utf8]{inputenc}
\usepackage[margin=1in]{geometry}
\usepackage{amsmath,amssymb,amsfonts}
\usepackage{amsthm,comment}
\usepackage{float}              
\floatstyle{plaintop}           
\restylefloat{table}
\usepackage[flushleft]{threeparttable}
\usepackage{booktabs}
\usepackage{dcolumn}
\newcolumntype{d}[1]{D..{#1}}

\usepackage{xcolor}
\newtheorem{theorem}{Theorem}
\newtheorem{assumption}{Assumption}

\newtheorem{corollary}{Corollary}
\newtheorem{lemma}{Lemma}
\newtheorem{remark}{Remark}
\newtheorem{example}{Example}
\usepackage{bbm}
\usepackage{natbib}

\newcommand\independent{\protect\mathpalette{\protect\independenT}{\perp}}
\def\independenT#1#2{\mathrel{\rlap{$#1#2$}\mkern2mu{#1#2}}}

\usepackage{setspace}
\usepackage{graphicx,subfigure}
\usepackage[colorlinks=true,linkcolor=blue,citecolor=blue,urlcolor=blue]{hyperref} %citecolor=black

\usepackage[margin=1in]{geometry}

\title{  Testing for Underpowered Literatures }
\author{ Stefan Faridani\footnote{Georgia Institute of Technology. sfaridani6@gatech.edu. I thank Graham Elliott, Nikolay Kudrin,  Xinwei Ma, Paul Niehaus,   Sabareesh Ramachandran,  Jordan van Rijn, Andr\'es Shahidinejad, Davide Viviano, Kaspar W\"{u}thrich, Karen Yan,  several anonymous referees, and the participants in the Berkeley Initiative for Transparency in the Social Sciences Annual Meeting for very helpful comments at various stages of this project. I thank Xianfang Xiong for excellent research assistance. An earlier arXiv version of this paper can be found at: \url{https://arxiv.org/abs/2406.13122}}}

\begin{document}

\maketitle \begin{center}
\end{center}
\begin{abstract}
    How many experimental studies would have come to different conclusions had they been run on larger samples? I show how to estimate the expected number of statistically significant results that a set of experiments would have reported had their sample sizes all been counterfactually increased. The proposed deconvolution estimator is asymptotically normal and adjusts for publication bias. Unlike related methods, this approach requires no assumptions about the distribution of true intervention treatment effects and allows for point masses. Simulations find good coverage even when the $t$-score is only approximately normal. An application to randomized trials (RCTs) published in economics journals finds that doubling every sample would increase the power of $t$-tests by 7.2 percentage points on average. This effect is smaller than for non-RCTs and comparable to systematic replications in laboratory psychology where previous studies enabled more accurate power calculations. This suggests that RCTs are on average relatively insensitive to sample size increases. Research funders who wish to raise power should generally consider sponsoring better-measured and higher-quality experiments---rather than only larger ones.
\end{abstract} 

\vspace{0.1in}

\vspace{0.1in}

\pagenumbering{arabic}
\onehalfspacing

\clearpage
\section{Introduction}

A key tradeoff in experimental design is balancing the risk of drawing an erroneous conclusion with project cost. Nearly all experiments published in top economics journals use $t$-tests to interpret their results. Sampling variance means that the $t$-test can fail to reject the null hypothesis of zero treatment effect even when there is in fact an effect of meaningful magnitude.  Every experiment therefore runs the risk of a false negative. Collecting more data raises statistical power and reduces the risk of a false negative, but by an amount that depends on the true effect which is ex ante unknown. A central question faced by funders and researchers is when to focus resources on larger samples and when to direct funds toward other drivers of power, e.g. measurement quality \citep{McKenzie}.

 This paper provides research funders and meta-analysts with a statistical procedure to estimate how much larger the expected fraction of statistically significant $t$-scores would have been had every experiment in some population been counterfactually run with $c^2$ times the sample size where $c > 1$. This power gain is a number between zero and one that I call $\Delta_c$. If this quantity is large within a given scientific literature or funding initiative, then the rejection decisions of $t$-tests reported by that population of experiments are sensitive on average to sample size choices. In this case, the returns to collecting more data points could be relatively high. 

 To estimate $\Delta_c$ with the proposed method, the meta-analyst needs a dataset of $n$ $t$-scores reported by a sample of experiments. The method works by first finding the distribution of true intervention treatment effects that best fits a smoothed version of the empirical distribution of $t$-scores. The fitted distribution of true effects is then integrated to calculate an estimate of $\Delta_c$. I show that this procedure is consistent, asymptotically normal, and converges in a power of $n$. 
 
 The main contribution of this paper is that it allows the set of experiments being studied to be highly heterogeneous. The outcomes, designs, interventions, scales, and settings can vary across studies in any way. This is accomplished by removing all assumptions about the distribution of true intervention treatment effects that related meta-analyses have so far relied upon. The methods of \cite{PowerOfBias}, \cite{Brunner}, and others are only consistent when the population distribution of true intervention treatment effects has a specific shape. Restrictions of this kind need not hold when the literature being studied is, for example, a set of experiments united by a publisher or funder rather than by a type of intervention. This paper makes no assumptions of any kind about the distribution of true effects and so any kind of effect heterogeneity is allowed. 
 
A second contribution of this paper is to make its method robust to simple forms of publication bias. Recent empirical evidence shows that statistically insignificant $t$-scores are less likely to be published in academic journals \citep{Franco1502,Andrews, Brodeur,bb, Elliott}. Selective reporting can distort the distribution of reported $t$-scores and confound estimation. This paper addresses selective reporting by parameterizing a simple model of publication bias, estimating the model, and then reweighting the observed $t$-scores to remove publication bias. Accommodating this first stage requires that I use a particular smoothing method which results in a different estimator than those used to solve mathematically related {\it deconvolution} problems like \cite{CarrascoPaper}.

 An empirical estimate of $\Delta_c$ is useful to meta-analysts and funders because they may otherwise lack a way to evaluate how efficiently sample sizes are being chosen in practice. Sample size decisions are typically made under high uncertainty. If funders knew ex ante how quickly the statistical power of a proposed experiment responded to its sample size, they could optimally balance power and cost. But power also depends on how effective the treatment actually is, a quantity which is ex ante unknown. Grant-giving organizations try to address this challenge by requiring experimenters to collect samples large enough to guarantee at least 80\% power to detect a true effect larger than some chosen threshold \citep{jpal_power_calcs}. One of the main aims of these {\it power calculations} is to direct resources to projects that will only fail to detect an effect when that effect is truly small.

Power calculations are not guaranteed to achieve this goal in practice because they involve a great deal of guesswork. Statistical power depends on many unknown parameters besides the true effectiveness of the intervention, e.g. how heterogeneous treatment effects are across individuals.  Even power calculations based on data from previous experiments are likely to systematically under-target sample sizes \citep{VU2024105868}. If the power calculations are noisy or biased, then the choice of sample size made on their basis may trade off power and cost sub-optimally.

Even ex post it is difficult to determine what the consequences of a larger sample size would have been for a single experiment. This problem arises because it is not in general possible to infer how powerful an individual experiment was. Plugging the estimated treatment effect into the power function is known to be highly misleading \citep{posthocbad}. In fact, precisely estimating the full distribution of true power over a literature is infeasible without very strong assumptions \citep{Carroll,Fan91}. Experimenters and funders therefore lack a rigorous way to assess how efficiently sample sizes are being chosen in practice. This paper addresses this need by showing how to estimate {\it average} power over a heterogeneous collection of studies under counterfactual sample sizes.

This paper applies its method to an empirical question of broad interest: how sensitive are randomized controlled trials (RCTs) published in top economics journals to counterfactual sample size increases? I use the data from \cite{bb} which contains  $t$-tests of main hypotheses reported by articles published in 25 top economics journals during 2015-2018. This is a very diverse set of experiments and was chosen to highlight the econometric contribution of this paper: a method robust to arbitrary heterogeneity in interventions, scales, settings, and designs. I estimate that counterfactually doubling the sample sizes of every RCT in that set would only increase the expected number of $t$-scores clearing the critical value of 1.96 by 7.2 percentage points with a standard error of 2.5 percentage points.

This power gain is small compared to several benchmarks. First, I estimate that doubling the sample size of all non-RCTs in the same dataset would increase average power by 17.3 percentage points which is significantly larger. As a second benchmark, consider a hypothetical literature where every experiment is adequately powered against its true effect. By conventional standards, a literature where every experiment has exactly 80\% power would meet this criterion \citep{jpal_power_calcs}. For such a hypothetical literature, we can calculate that doubling every sample size would increase power by 17.8 percentage points, which is also much larger than the estimate for economics RCTs. 

This paper constructs a third benchmark using data from the Many Labs systematic replication project \citep{ManyLabs}.  In this project, 36 laboratories each independently attempted to replicate 11 published effects from laboratory psychology. We would expect these replication experiments to be very well powered because they were designed using previous experimental data and the consequences of failure to detect an already published effect could be great. Yet I cannot reject the null hypothesis that $\Delta_c$ is the same for both RCTs in economics and the Many Labs replications. This means that RCTs are on average about as sensitive to sample size as replication experiments that were designed using a great deal of prior knowledge. The non-rejection is not simply due to high uncertainty because we can indeed reject the null that $\Delta_c$ is the same for non-RCTs in economics vs replications from Many Labs. I also leverage the fact that the Many Labs experiments are replications of one another to estimate power gain conditional on the in-sample true effects. The conditional Many Labs power gain is precisely estimated at $7.8$ percentage points with a standard error of $0.18$ percentage points and is also statistically indistinguishable from the $\Delta_c$ for economics RCTs.

The takeaway from the application is that randomized trials in economics are on average relatively insensitive to doubling their sample sizes.\footnote{This is not to say that individual reported results do not need to be confirmed in replication. Rather, this application says that had the original sample sizes been larger, replicability would not have improved much.} Power calculations appear to be surprisingly effective in practice---despite requiring guesswork {ex ante} and being impossible to directly verify {ex post}. Funders looking to improve power could consider alternative reforms instead. \cite{McKenzie} argues that many field experiments could raise power substantially without collecting more data points by improving measurement quality and compliance. This paper's evidence suggests that the power gain from growing the sample size of economics RCTs is modest on average and therefore quality improvements may deserve more attention. 

This paper is organized as follows. Section \ref{sec:literature} discusses the contributions of this paper to existing literatures. Section \ref{sec:setup} sets up the  problem without publication bias and explains why the deconvolution approach is needed. Section \ref{sec:id} shows identification and Section \ref{sec:estimation} proposes an estimator and derives its rate of consistency. Section \ref{sec:pb} introduces publication bias. Section \ref{sec:inference} shows asymptotic normality and discusses inference. Section \ref{sec:simulations} uses simulations to recommend tuning parameters that yield high coverage of the confidence intervals for a variety of DGPs. This section also shows that coverage rates usually remain high when the $t$-score is only approximately normal. Section \ref{sec:empirical_applications} presents two empirical applications and Section \ref{sec:conclusion} concludes. Proofs of the main theorems, followed by the technical lemmas, appear after the conclusion. The Online Appendix provides a key robustness check.

\section{Related Literature}\label{sec:literature}

This paper contributes to two literatures. The first is an applied literature that empirically studies average statistical power over a population of experiments under very strict assumptions.  \cite{PowerOfBias}  estimates median power in empirical economics under the assumption that papers can be sorted into groups within which every study is estimating the same true effect.  A similar assumption is used by \cite{DellaVigna}, \cite{bundock}, and \cite{Ferraro}. Other methods instead assume that the distribution of true effects has a specific shape, e.g. a Gamma distribution or a finite mixture model with fixed means \citep{Brunner,Sotola,zcurve2,NBERw31666}. Assumptions like these need not hold in practice and are virtually impossible to check. 

This paper's main methodological contribution  is to remove all assumptions about the distribution of true effects. This relaxation is important in theory because it produces robust results, allows true nulls to have positive probability, and accommodates outlier treatment effects that are of interest to the theory of experimental design \citep{ABfattails}. These assumptions matter in practice because my nonparametric method yields substantively different results than the mixture model of \cite{zcurve2} in this paper's empirical application.

This paper also contributes on a conceptual level by suggesting that meta-analysts think of power ``on the margin" and not just ``in levels." This means that we should think of a collection of studies as ``too small" when average power against true effects is easy to increase by growing the experiments, i.e. $\Delta_c$ is large. While many meta-studies have identified literatures where the level of power is low, this paper suggests refining this search and looking for places where power is easy to raise \citep{neuroPowerIoannidis,PowerOfBias}. This has direct implications for research funding and design. When the returns to increasing sample size are low, other improvements in measurement and outcome choice might offer more fruitful solutions to power problems \citep{McKenzie}.

This paper studies populations of experiments with arbitrary heterogeneity and therefore avoids specifying an experimenter utility function. It is not clear how to choose a function that maps effect magnitudes into welfare when studies can vary in their treatments, outcomes, scales, and settings---especially when not all studies are program evaluations. For this reason many analyses use the error rate of the hypothesis rejection decision as their welfare concept, e.g.  \cite{neuroPowerIoannidis,Franco1502,Head,PowerOfBias,doi:10.1126/science.aac4716,young_ri,Brodeur,bb,Elliott}. This paper estimates $\Delta_c$ because it is interpretable even when the meta-sample encompasses a diverse set of interventions and can be used to direct research funding towards where it can improve power most. This estimand is relevant to, e.g. a grant-giver evaluating whether a particular funding initiative would have found more results if it had been given more resources.

The second related literature is technical. The estimation problem studied in this paper belongs to a class of problems called {\it deconvolutions} which aim to ``de-blur" a smooth probability density. Specifically, we wish to learn features of the distribution $\Pi_0$ of true intervention treatment effects $H$, but we can only observe $T=H+Z$ where $Z\sim N(0,1)$.  Deconvolutions are a classic and active problem.\footnote{See for example \cite{Carroll,Stefanski01011990,Fan91,Fan_global,Johannes,Meister,CarrascoPaper,handbookinverseproblems,Koenker03042014,Hohage}} 

At first glance, the deconvolution problem posed by this paper appears difficult because the error distribution is Gaussian and the usual polynomial rates of convergence do not apply \citep{Fan_global}.  \cite{Carroll} showed that when the error is normally distributed and $\Pi_0$ has a density with at most $k$ bounded derivatives, then $\Pi_0$ cannot be recovered faster than $(\log n)^{-k/2}$. Since we do not necessarily know anything about $\Pi_0$---perhaps it does not even have a density at all---the possibly logarithmically slow rate of convergence makes nonparametric estimation of $\Pi_0$  appear impractical. 

Fortunately, this paper's estimand of interest is $\Delta_c$, which depends on only some features of $\Pi_0$. In fact, $\Delta_c$ can be rewritten as a functional of the distribution of the $t$-score under a counterfactual sample size. This distribution has already been mollified by the Gaussian and therefore has an infinitely differentiable PDF---so the impossibility result of \cite{Carroll} does not apply. Recovering mollified distributions is known to be a better-posed problem \citep{Hohage}.

Another important contribution of this paper is to account for the fact that in the presence of publication bias all of the above-cited deconvolution methods will be inconsistent. There is growing evidence that statistically insignificant $t$-scores are less likely to be reported in economics publications \citep{Chrestensen, Andrews,Havranek24}. Such omissions can create a discontinuity in the density of $t$-scores which is problematic for all methods of deconvolution \citep{caliperOriginal,Kudrinjmp}.

 Addressing publication bias is not as simple as concatenating deconvolution with publication bias removal because the two steps interact. This paper proposes a new deconvolution step in order to manage such interactions. The method begins with the same singular value decomposition as \cite{CarrascoPaper} but uses a different method of regularization---called spectral cutoff---that prevents uncertainty about the extent of selective reporting from magnifying the regularization bias. This choice of smoothing method yields a new estimator that converges faster and interacts minimally with publication bias removal. 

 Spectral cutoff is also attractive because our true goal is to estimate the distribution of the counterfactual $t$-score which must be smooth. \cite{Stefanski01011990} note that this method of regularization ``takes full advantage of smoothness properties of [the target distribution] by allowing bias to decrease at rates dictated by the tail behavior of [its characteristic function]."  Similar to \cite{Johannes}, I find that the regularization bias is a well-behaved polynomial in the point at which the spectrum is truncated. 

A known disadvantage of spectral cutoff is the {\it Gibbs phenomenon} where the reconstructed density converges in the $L^2$ norm but not in the sup-norm \citep{on_Gibbs,Hohage}. This makes any estimate of the deconvolved distribution unlikely to be a valid probability distribution. Fortunately, the target scalar $\Delta_c$ depends on the estimated PDF only via an integral over a compact interval and therefore  $L^2$ convergence is all we need. The Gibbs fluctuations could be high in amplitude but are nevertheless integrated out of the final estimate.

Some recent meta-analyses  identify publication bias via the joint distribution of standard errors and point estimates \citep{Andrews,Duval2000,Havranek24,vu2024pb}. This strategy is not appropriate for my setting because it requires that the true standard errors and true effects are not correlated. In large meta-samples that encompass many interventions this independence breaks down because each researcher's choice of experimental design may be driven by knowledge about their true effect. This paper identifies selective reporting via the $t$-ratio like \cite{bb} and \cite{Elliott}.

\section{Setup and Estimand}\label{sec:setup}

First define a key piece of notation. Let $\varphi(z)$ denote the probability density function of the standard normal distribution. Adding a subscript $\varphi_{\sigma^2}(z)$ denotes the density of the normal distribution with variance $\sigma^2$. No subscript means that the variance is unity.

Consider a population of experiments. Each experiment studies a unique intervention with its own treatment effect $b \in \mathbb{R}$. The treatment effect $b$ is unobserved, but the experimenter estimates it with an estimator $\widehat{b}$ which is assumed to be asymptotically normal. The experimenter knows the standard error $\sigma$ and summarizes the evidence against the null hypothesis of zero treatment effect by reporting the $t$-score: $T = \frac{\widehat{b}}{\sigma}$. Let the random variable $H$ be called the ``true effect" and define it as: $H \equiv \frac{b}{\sigma}$. The distribution of $T$ can now be described using $H$ only (making further references to $b,\sigma$ unnecessary). Assumption \ref{assum:normality_of_T} below states that $T$ is conditionally normally distributed centered on $H$ with unit variance. This assumption is justified by the same appeal to the central limit theorem that justified the use of the $t$-score in the first place.
\begin{assumption}\label{assum:normality_of_T}
    $$ T = H+Z \qquad \text{ where } Z\sim N(0,1), \text{ and } H\independent Z $$
\end{assumption}

\noindent Consequently, the  conditional probability density function $ f_{T \mid H}\left(t\mid h\right)$ is the following:
\begin{equation}
   f_{T \mid H}\left(t\mid h\right) = \varphi(t-h)
\end{equation}

\begin{remark}\normalfont

    In practice, $Z$ might only be approximately normal. Since we do not observe a sample from $Z$ itself and only one draw of $T$ is observed per $H$, we cannot estimate the exact error distribution like, e.g. \cite{Johannes}  or Section 3.4 of \cite{CarrascoPaper}. In theoretical meta-analysis it is common to abstract from these concerns and use the normal distribution because it is justified by the same central limit theorem that motivates the $t$-test in the first place \citep{Elliott}. In this paper, realistic deviations from normality are unlikely to matter much. Section \ref{sec:simulations} presents a set of simulations where the numerator of the $t$-score is a sample mean of log-normals and another set where the $t$-score is distributed $t(30)$. In both cases the decline in coverage rates is small.
\end{remark}

 $T$ can be interpreted as the test statistic of a two-sided $t$-test of the null hypothesis that $H= 0$. The power of a size-$\alpha$ test run by an individual experiment is the  probability that $|T|$ exceeds the critical value $CV(\alpha)$ conditional on the true effect $H$. I suppress the dependence of the critical value on $\alpha$ for ease of notation. I call this conditional probability the {\it conditional power}. Conditional power can be written in terms of an integral over the normal density.
\begin{equation}
 \text{Pr}\left(|T| > CV\:|\:H=h\right)  =  1-\int_{-CV}^{CV}  \varphi(t-h) dt 
\end{equation}

Since  $H$ is unobserved, the meta-analyst cannot condition on it. So conditional power is always unknown.  In practice this means that it is not possible to recover the power of any {\it individual} experiment by plugging its reported $t$-score into the power function \citep{posthocbad}. To see why, notice that Jensen's Inequality guarantees that even though $\mathbb{E}[T\:|\: H=h]=h$, nevertheless $\mathbb{E}\left[\varphi(t-T)  \:|\: H= h\right] \neq \varphi(t-h)$.\footnote{Plugging the $t$-score into the power function also cannot be justified by invoking consistency of the point estimate because the $t$-score never converges in probability to a number.}  Fortunately, the meta-analyst does not need to know the true power of any individual experiment. Instead the meta-analyst wishes to know the expected statistical power of an experiment {\it randomly drawn} from the population. I call this expectation the ``unconditional power." 

Unconditional power is defined as the expectation of conditional power over  $H$. Unconditional power therefore depends crucially on the true probability distribution $\Pi_0$ of $H$. This paper will not require any restrictions on $\Pi_0$ of any kind---not even regularity conditions. This means that $\Pi_0$ could in principle be any mixture of discrete and continuous distributions and the moments of $H$ need not necessarily exist. This level of generality is necessary because we must accommodate the possibility of true nulls, i.e. probability mass at $H=0$.

 Fubini's Theorem allows unconditional power to be expressed as an integral in Equation (\ref{eq:expectedpower}).
\begin{equation}\label{eq:expectedpower}
    \text{Pr}\left(|T| > CV\right)   =  1- \int_{-CV}^{CV}\int_{-\infty}^\infty \varphi(t-h)  d\Pi_0(h)  \: dt   
\end{equation}

Experimenters can influence the statistical power of their experiments by choosing the sample size. Increasing the sample size of an experiment will increase its conditional power whenever $H\neq 0$. The meta-analyst wishes to learn how much larger unconditional power would have been had the sample sizes of every experiment in the population been counterfactually increased while holding the distribution of true effects constant. 

 Let $c > 1$. Define the random variable $T_c$ as a $t$-score randomly drawn from a counterfactual population where every experiment has been run at $c^2$ times the actual sample size while holding $\Pi_0$ constant.\footnote{In other words, this paper considers situations where there is an opportunity to collect more data without changing the treatment effect--i.e. the experimenter can sample more individuals without altering the sampling frame. This is often possible in practice given appropriate funding.} Since it is counterfactual, no draw of $T_c$ is ever actually observed. Multiplying the sample size by $c^2$ shrinks the standard error and grows $H$ by a factor of $c$. Conditional on the true effect, $T_c$ is normally distributed but with a larger mean, i.e. $T_c=cH+Z$. The unconditional counterfactual power is defined as the probability that $T_c$ exceeds the critical value. This can be expressed as the following double integral.
\begin{equation}\label{eq:ucp}
  { \text{Pr}\left(|T_c| > CV\right)} =  1-\int_{-CV}^{CV} \int_{-\infty}^\infty\varphi(t-ch) d\Pi_0(h)\:dt 
\end{equation}

This paper proposes a method to determine whether the unconditional power of a given population of experiments is sensitive to sample size increases. High sensitivity represents an ``opportunity missed" because there are in expectation many false negatives that could have been avoided had the experiments been larger. To make this idea precise, define the estimand $\Delta_{c}$ as the power gain resulting from increasing every sample size in the population of experiments by a factor of $c^2$. 
\begin{equation}
    \Delta_c \equiv \text{Pr}\left(|T_c| > CV\right) - \text{Pr}\left(|T| > CV\right)
\end{equation}
$\Delta_c$ will be the estimand throughout this paper. Example \ref{example:powerexample} and Figure \ref{fig:powerexample} illustrate why I interpret a large value of $\Delta_c$ as an indicator of an underpowered literature.

\begin{example}\label{example:powerexample}
\normalfont How unconditional power responds to sample size choices depends on the population distribution of true effects. The example in Figure \ref{fig:powerexample} illustrates. Consider two different literatures each with their own distribution of true effects. For literature (1), there are many ``medium-sized" effects and for literature (2) effects are either extremely large or close to zero.  Figure \ref{fig:powerexample} contrasts how unconditional power (y-axis) responds to counterfactual sample size increases (x-axis) for the two literatures. Increasing the sample size leads to rapid gains for the ``underpowered" literature (1) but not the ``well-powered" literature (2). Literature (2) is less responsive because it contains either studies where there is nothing to find or studies that are already fully powered.  The aim of this paper is to determine whether a sample of $t$-scores was drawn from a population more like literature (1) or more like literature (2).\begin{figure}[h!] 
    \centering
    \includegraphics[width=4.5in]{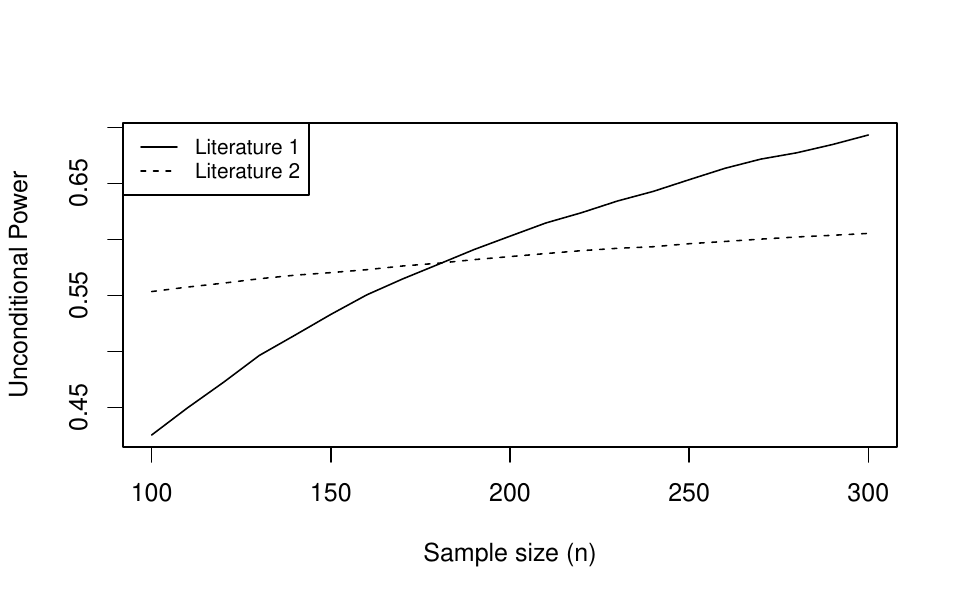}
    \caption{\footnotesize Underpowered literature (1) vs well-powered literature (2).}
    \label{fig:powerexample}
\end{figure}

\end{example}

\subsection{Why $\Delta_c$ is Difficult to Estimate}

To identify and estimate $\Delta_c$, we will need to draw on the classic theory of deconvolution. To motivate the use of this mathematical framework, I pause here to briefly explain why counterfactual power cannot just be estimated using the simple and intuitive approaches that often come to mind. The fundamental difficulty is that power depends on how large the true effects tend to be, which is challenging to learn. 

   $T$ can be interpreted as the sum of $H$ plus an independent ``noise" random variable $Z$ that has the normal distribution: $T=H+Z$. While it is not possible to estimate the power of an individual study conditional on $H$, it is straightforward to estimate the unconditional power at the status quo sample sizes in a simple way: just compute the fraction of $t$-scores that exceed the critical value. \begin{equation}
       \frac{1}{n}\sum_{i=1}^n \mathbf{1}\{|t_i|>1.96 \}\to_p \underbrace{\text{Pr}\left(|Z+H|>1.96\right)}_{\text{``Status Quo" Power}}
   \end{equation}
    
    Unfortunately, this method cannot be extended to estimate {\it counterfactual} power (or $\Delta_c$). That is, multiplying each $t$-score by $c$ and computing the fraction of those over $1.96$ will not be informative about what power would have been under larger sample sizes. The equation below shows that the probability limit of this procedure does not equal counterfactual power.
\begin{equation}\label{eq:basic_method_fail}
    \frac{1}{n}\sum_{i=1}^n \mathbf{1}\{c|t_i|>1.96 \} \to_p \underbrace{\text{Pr}\left(|cZ+cH|>1.96\right)}_{\text{Incorrect Target}} \neq \underbrace{\text{Pr}\left(|Z+cH|>1.96\right)}_{\text{Counterfactual Power}}
\end{equation}
    
    To see how consequential this gap is, consider an example literature where the true effect $H$ is equal to zero for all studies. Here, power will be equal to size no matter how much the sample size grows. Nevertheless, the simple estimator in Equation (\ref{eq:basic_method_fail}) will say that doubling sample sizes raises power to 16.6\%! Why does this simple procedure fail? It is based on the idea that increasing the sample size by $c^2$ would decrease the denominator of the $t$-score (the standard error) by a factor of $c$. While this is true, collecting more data will also make the point estimate less volatile, pulling the numerator of the $t$-score towards the true effect of zero. Thus, while increasing the sample size of the experiment will decrease the denominator of $T$, it also changes the numerator as well.

The problem of learning $\Delta_c$ cannot be solved so easily because $\Delta_c$ depends on the unknown distribution $\Pi_0$ of $H$. In order to estimate $\Delta_c$ without any assumptions about $\Pi_0$, we need the convolutional framework developed in the following pages. 

\subsection{Convolutions}

Assumption \ref{assum:normality_of_T} says that $T$ is the sum of $H$ plus independent normal ``noise" $Z$. The sum of two independent random variables is called a {\it convolution} and an operator that maps the distribution of $H$ into the distribution of $H$ plus noise is called a {\it convolution operator}. It will be useful to express the distributions of $T$ and $T_c$ in terms of convolution operators. The first step is to write down the densities of $T$ and $T_c$ as expectations over the distribution of true effects $\Pi_0$.
\begin{align*}
    f_T(t)  = \int_{-\infty}^\infty \varphi(t-h)  d\Pi_0(h)  \qquad\quad  f_{T_c}(t)  &= \int_{-\infty}^\infty \varphi(t-ch)  d\Pi_0(h)
\end{align*}

Both densities are the outcomes of closely related mappings of $\Pi_0$. Consider the operator $K_{\sigma^2}$ below that maps the distribution of $H$ to the density of $H+\sigma Z$ where $Z$ is a standard normal random variable independent of $H$ and $\sigma > 0$. This maps any probability distribution to a probability density. 
\begin{align*}
   (K_{\sigma^2}\Pi)[t] &\equiv  \int_{-\infty}^\infty \varphi_{\sigma^2}(t-h)  d\Pi(h)
\end{align*}

 A helpful fact about normal convolutions is that they can be decomposed.  Lemma \ref{lem:sumnoise}  says that adding normal noise of unit variance is equivalent to adding normal noise of variance $c^{-2}$ and then adding further independent normal noise of variance $1-c^{-2}$.
\begin{lemma}\label{lem:sumnoise}

For any probability distribution $\Pi$ and any $c> 1$:
   $$K_{1}\Pi = K_{1-c^{-2}}K_{c^{-2}}\Pi $$
  
\end{lemma}

 The decomposition in Lemma \ref{lem:sumnoise} is not new and comes almost immediately from the basic properties of sums of independent normal random variables. Its proof is therefore omitted for brevity. It is immediate to see that $f_T = K_{1-c^{-2}}K_{c^{-2}}\Pi_0$.  Lemma \ref{lem:conv_op} expresses the estimand $\Delta_c$ in terms of $K_{c^{-2}}\Pi_0$ as well. The proof is in Appendix \ref{proof:conv_op}. The motivation for this decomposition is that $f_{T_c}$ can be written in terms of the Gaussian mollification $K_{c^{-2}}\Pi_0$ of $\Pi_0$; i.e. $K_{c^{-2}}\Pi_0$ is $\Pi_0$ convolved with the $N(0,c^{-2})$ kernel. This mollification smooths away high-frequency features of $\Pi_0$, which is why $K_{c^{-2}}\Pi_0$ and $f_{T_c}$ are much easier to estimate than $\Pi_0$ itself. The next section shows why this is the case. 

\begin{lemma}\label{lem:conv_op} 
If Assumption \ref{assum:normality_of_T} holds, then:

    \begin{align*}
    \Delta_c &=\int_{-CV}^{CV}(K_{1-c^{-2}}K_{c^{-2}}\Pi_0)[t]dt- \int_{-CV/c}^{CV/c}(K_{c^{-2}}\Pi_0)[t]dt
\end{align*}
\end{lemma}

\section{Identification}\label{sec:id}

The meta-analyst observes $n$ draws of $T$ and wishes to estimate $\Delta_c$. This section will start by showing identification when $\Pi_0$ is continuous and has a probability density function $\pi_0$ with bounded height and there is no publication bias. Then we will see that the identification result in fact holds for all probability distributions $\Pi_0$---even those that are not continuous. Publication bias will be introduced in Section \ref{sec:pb}.

For now, assume that $H$ is continuous with PDF $\pi_0$ and that $\pi_0$ has finite height. The problem of recovering the PDF $\pi_0$ from $f_T$ is known to be severely ill-posed because very large changes in $\pi_0$ can result in very small changes in $f_T$. The intuition is that  $f_T$ is a smoothed-out version of $\pi_0$. Since $\pi_0$ is not necessarily smooth itself, many of its ``high-frequency" features are destroyed by convolution and are therefore difficult to recover from $f_T$. However, the problem of recovering $K_{c^{-2}}\Pi_0$ from $f_T$ is much better-posed because $K_{c^{-2}}\Pi_0$ has already lost its rapid oscillations.

This intuition can be formalized using the {\it singular value decomposition} (SVD). Intuitively, the singular value decomposition expresses the densities $K_{c^{-2}}\Pi_0$ and $\pi_0$ in terms of orthonormal basis polynomials with a one-to-one correspondence. The decompositions presented in this section are modified versions of the decompositions in Example 1 of \cite{CarrascoPaper} and \cite{Wand1995KernelS}. 

The first task is to precisely define the domain and range of the two convolution operators $K_{c^{-2}}$ and $K_{1-c^{-2}}$. We will do this using Example 1 from \cite{CarrascoPaper}. The meta-analyst must start by choosing the scalar $\sigma_T^2>0$. In the simulations and applications of this paper I always choose $\sigma_T^2=1$ and never deviate from this choice. Define the following three Hilbert spaces of functions,
$\mathcal{L}_{T},\mathcal{L}_{T_c},\mathcal{L}_{H}$, which \cite{CarrascoPaper} constructed to guarantee that the operators $K_{c^{-2}}$ and $K_{1-c^{-2}}$ are Hilbert-Schmidt and will therefore have discrete spectrum: 
\begin{align*}
\mathcal{L}_{T} &\equiv \left\{g(x)\text{ such that } \int_{-\infty}^\infty g(x)^2\varphi_{\sigma_T^2}(x)dx < \infty \right\}\\
\mathcal{L}_{T_c} &\equiv \left\{g(x)\text{ such that } \int_{-\infty}^\infty g(x)^2\varphi_{\sigma_T^2+1-c^{-2}}(x)dx < \infty \right\}\\
    \mathcal{L}_{H} &\equiv \left\{g(x)\text{ such that } \int_{-\infty}^\infty g(x)^2\varphi_{\sigma_T^2+1}(x)dx < \infty \right\}
\end{align*}
These three spaces are all very large. Each contains every bounded probability density function of a real-valued random variable. Following Example 1 of \cite{CarrascoPaper}, equip each space with the following inner products:
\begin{align*}
     \langle g_1, g_2 \rangle_{T} &\equiv  \int_{-\infty}^\infty g_1(x)g_2(x)\varphi_{\sigma_T^2}(x)dx \\
     \langle g_1, g_2 \rangle_{T_c} &\equiv  \int_{-\infty}^\infty g_1(x)g_2(x)\varphi_{\sigma_T^2+1-c^{-2}}(x)dx\\
     \langle g_1,g_2 \rangle_{H} &\equiv  \int_{-\infty}^\infty g_1(x)g_2(x)\varphi_{\sigma_T^2+1}(x)dx 
\end{align*}

These inner products induce norms:  $\left|\left|g\right|\right|_T^2 \equiv \langle g,g\rangle_{T} $. The convolution operators can now be fully defined by specifying their domains: $ K_{c^{-2}} \::\: \mathcal{L}_H \to \mathcal{L}_{T_c}$ and $   K_{1-c^{-2}} \::\: \mathcal{L}_{T_c}\to \mathcal{L}_T$. Both  are compact linear operators and therefore must have { singular value decompositions}. In this case the singular value decomposition will express the outcome of a convolution as a weighted sum of known orthonormal polynomials where the weights are known to decay at a geometric rate. The  decompositions, also adapted from Example 1 of \cite{CarrascoPaper}, are  expressed below.  
\begin{align}
    K_{c^{-2}}\Pi &= \sum_{j=0}^\infty \underbrace{\eta_j}_{\text{scalar}} \langle \chi_j,\pi \rangle_{H} \underbrace{\phi_j}_{\text{polynomial}}  \label{eq:svdc2}\\
        K_{{1-c^{-2}}}G  &= \sum_{j=0}^\infty \underbrace{\lambda_j}_{\text{scalar}}  \langle \phi_j,g \rangle_{T_c} \underbrace{\psi_j}_{\text{polynomial}} \label{eq:svd1m2}
\end{align}

The singular values $\eta_j,\lambda_j$ are the sequences of scalars defined below. These decay geometrically fast to zero. The fast rate of decay implies that the problem of recovering $\pi_0$ from $K_1\Pi_0$ is ill-posed because components of $\pi_0$ with large $j$ play a small role in  $K_1\Pi_0$.
\begin{align*}
    \eta_j &\equiv \left(\frac{1+\sigma_T^2-c^{-2}}{1+\sigma_T^2}\right)^{j/2}\qquad \lambda_j \equiv \left(\frac{\sigma_T^2}{\sigma_T^2+1-c^{-2}}\right)^{j/2}
\end{align*}

The singular functions $\chi_j,\psi_j,\phi_j$ are the generalized Hermite Polynomials.\footnote{
The generalized Hermite Polynomials are $
He_j(t)=\frac{1}{\sqrt{j!}}\sum_{l=0}^{[j/2]}(-1)^l \frac{(2l)!}{2^ll!}\binom{j}{2l}t^{j-2l}$.
The singular functions are scalings of the Hermite polynomials: $\chi_j(t) = He_j\left(\frac{t}{\sqrt{1+\sigma_T^2}}\right)$, $\phi_j(t) = He_j\left(\frac{t}{\sqrt{1+\sigma_T^2 -c^{-2}}}\right)$, and $\psi_j(t)=He_j\left(\frac{t}{\sigma_T}\right)$.}  I will use four properties of the Hermite polynomials. First, the polynomials are normalized so that, for example $\langle \chi_j,\chi_j\rangle_{H} = 1$.  Second, each set forms a complete basis for its corresponding Hilbert space \citep{HermiteComplete}. This means that for any two probability densities $\pi_1,\pi_2$, if $\langle \pi_1-\pi_2, \chi_j\rangle_{H} = 0$ for all $j$, then $\pi_1=\pi_2$ almost everywhere. Third, the polynomials are orthogonal, so $\langle \chi_j,\chi_k\rangle_{H}=0$ when $j\neq k$. Fourth, while these polynomials are themselves unbounded, Lemma \ref{lem:bound_coeffs} in Appendix \ref{proof:lem:bound_coeffs} uses \cite{HermiteBound} to show that they are uniformly bounded over all $t,j$ when multiplied by the kernels from their respective inner products. For instance:\begin{equation}\label{eq:HermitBound}
   \sup_{t\in\mathbb{R}}\left|\psi_j(t)\varphi_{\sigma_T^2}(t)\right| \leq  \left(2\sigma_T^2\pi\right)^{-1/2}
\end{equation}

\begin{remark}\label{rem:translate_carrascoflorens}\normalfont
        To see how Equation (\ref{eq:svdc2}) is a special case of Example 1 from \cite{CarrascoPaper}, map their operator $T$ to my $K_{c^{-2}}$, their normal error variance $\sigma^2$ to my $c^{-2}$, their variables $(Y,X)$ to my $(T_c,H)$, and their reference variance $\sigma_Y^2$ to my $\sigma_T^2+1-c^{-2}$. Then their space $L^2_{\pi_Y}$ corresponds to my $\mathcal L_{T_c}$ and their space $L^2_{\pi_X}$ corresponds to my $\mathcal L_H$. Their eigenvalues $\lambda_j^2$ correspond to my $\eta_j^2$, their domain singular functions $\varphi_j$ correspond to my $\chi_j$, and their range singular functions $\psi_j$ correspond to my $\phi_j$. A similar correspondence gives Equation (\ref{eq:svd1m2}): there, their
operator $T$ instead maps to my $K_{1-c^{-2}}$, their error variance is $1-c^{-2}$, their variables $(Y,X)$ correspond to my $(T,T_c)$, and their reference variance $\sigma_Y^2$ equals my $\sigma_T^2$.
\end{remark}

 Notice that the singular functions $\phi_j$ appear in both the SVD of $K_{c^{-2}}$ in  (\ref{eq:svdc2}) and in the SVD of $K_{1-c^{-2}}$ in (\ref{eq:svd1m2}). This fact is exploited in Equation (\ref{eq:fT}) below to express the density of $T$ as a linear combination of the Hermite polynomials. This relationship is a direct consequence of Lemma \ref{lem:sumnoise}, the SVDs, and the orthonormality of $\{\phi_j\}$. 
\begin{align}
 f_T &= K_{1-c^{-2}} K_{c^{-2}}\Pi_0 
 = \sum_{j=0}^\infty  \lambda_j \left\langle \phi_j ,    \sum_{\ell=0}^\infty {\eta_\ell} \langle \chi_\ell,\pi_0 \rangle_{H} {\phi_\ell} \right\rangle_{T_c}\psi_j  %\\
   =\sum_{j=0}^\infty {\lambda_j\eta_j} {\langle \chi_j,\pi_0 \rangle_{H}} {\psi_j}  \label{eq:fT}
\end{align}
\noindent Notice that the singular values $\lambda_j$ and $\eta_j$ are forcing higher order polynomials to play a small role in $f_T$. The implication for the meta-analyst is that high-frequency information about $\pi_0$ is not easy to recover from $f_T$.

Lemma \ref{lem:betac_singularvalues} expresses the estimand $\Delta_c$ in terms of the singular values and Hermite polynomials. 

\begin{lemma}\label{lem:betac_singularvalues}

If $\Pi_0$ has a probability density $\pi_0$ with bounded height and Assumption \ref{assum:normality_of_T} holds, then:

    $$\Delta_c = \sum_{j=0}^\infty \eta_j \langle \chi_j,\pi_0 \rangle_{H} \left(\lambda_j\int_{-CV}^{CV}\psi_j(t)dt-\int_{-CV/c}^{CV/c}\phi_j(t)dt\right) $$
     
\end{lemma}

\noindent The proof is in Appendix { \ref{proof:betac_singularvalues}}. Lemma \ref{lem:betac_singularvalues} illustrates why $\Delta_c$ is a fundamentally easier quantity to learn than $\pi_0$. The sequence of coefficients $\langle \chi_j,\pi_0 \rangle_{H}$ is sufficient for both the distribution of the data and for the estimand $\Delta_c$. Equation (\ref{eq:fT}) shows that the larger $j$ is, the more difficult it is to recover $\langle \chi_j,\pi_0 \rangle_{H}$ from $f_T$ since it is damped away by the rapidly decaying coefficients $\eta_j\lambda_j$. But, Lemma \ref{lem:betac_singularvalues} shows that when $\langle \chi_j,\pi_0 \rangle_{H}$ is given small weight in $f_T$, it is also given small weight in $\Delta_c$ because $\eta_j$ is decaying and playing a mollifying role. So the pieces of $\pi_0$ that are hardest to recover thankfully also play the smallest role in $\Delta_c$.

The result in Lemma \ref{lem:betac_singularvalues} can be generalized to Theorem \ref{thm:id} which places no restrictions on $\Pi_0$ at all. Theorem \ref{thm:id} is an identification result because it expresses our estimand $\Delta_c$ in terms of the population distribution of the $t$-score under the factual sample sizes. 

\begin{theorem}{\label{thm:id}} Define the deterministic constants $a_j\equiv \int_{-CV}^{CV}\psi_j(t)dt-\frac{1}{\lambda_j}\int_{-CV/c}^{CV/c}\phi_j(t)dt$ for $j\in \{0,1,2,\cdots\}$. If  Assumption \ref{assum:normality_of_T} holds, then for all probability distributions $\Pi_0$ of $H$,
    $$\Delta_c = \sum_{j=0}^\infty \mathbb{E}\left[\psi_j(T)\varphi_{\sigma_T^2}(T)\right] a_j $$
\end{theorem}

The proof is in Appendix \ref{proof:id}. The intuition for the argument is that for any $\Pi_0$ we can construct a sequence of densities $\pi_n$ (each with finite height) that converge weakly to $\Pi_0$. By the Portmanteau Theorem, the sequence of $\Delta_c$ for $\pi_n$ must converge to the $\Delta_c$ for $\Pi_0$.  By dominated convergence, the infinite sum over the expectations over $\pi_n$ in Lemma \ref{lem:betac_singularvalues} converge to the infinite sum on the right hand side of Theorem \ref{thm:id}. Using Portmanteau a second time verifies that the limit of each sequence of expectations over $\pi_n$ equals the expectation over $\Pi_0$.

\section{Estimation}\label{sec:estimation}

 For now there is no selective reporting. The meta-analyst observes a sample of $n$ $t$-scores denoted  $t_{i}$  indexed by $i\in \{1,\cdots, n\}$ reported by $m$ studies indexed by $k \in \{1,\cdots , m\}$. Assumption \ref{assum:cross_study_independence} below says that the $t$-scores are identically distributed and independent across studies (but can be dependent within  a study) and  that the number of $t$-scores per study is uniformly upper bounded by $C_B>0$. 

\begin{assumption}
\label{assum:cross_study_independence}

Let $\Lambda$ be the $n\times n$ block-diagonal matrix  where the element $\Lambda_{ij}$ indicates whether $t_i$ and $t_j$ were reported in the same study. All of the following are satisfied:
\begin{enumerate}
    \item Each $t_i$ is identically distributed 
    \item  The study-level vectors of reported $t$-scores are independent across studies.
    \item There is a universal constant $C_B>0$ such that $\max_{i\in \{1,\cdots, n\}}\sum_{j=1}^n \Lambda_{ij}\leq C_B$ for all $n$
\end{enumerate}
\end{assumption}

Even though $\Pi_0$ is itself identified, estimating it is a severely ill-posed problem because the singular values $\lambda_j\eta_j$ decay rapidly to zero. This means that information about the high-frequency components of $\Pi_0$ is severely attenuated in the distribution of $T$. The rate at which it is possible to estimate $\Pi_0$ depends on how we measure the difference between the estimated density and the true density. If we take this difference to be the  $\mathcal{L}_\infty$ norm between CDFs, then the minimax rate is known to be logarithmic in $n$ \citep{Carroll,Fan91}. This is extremely slow. Fortunately, the meta-analyst only needs to estimate the features of $\Pi_0$ that matter for counterfactual power.

This paper proposes the estimator $\widehat{\Delta}_{c,n}$ defined below which is the sample analogue of Theorem \ref{thm:id}. 
\begin{equation}\label{eq:def_Delta_hat}
    \widehat{\Delta}_{c,n}= \sum_{j=0}^{J_n} \left(\frac{1}{n}\sum_{i=1}^n \psi_j(t_{i})\varphi_{\sigma_T^2}(t_i) \right)a_j
\end{equation}

This sample analogue replaces expectations with sample means. Instead of summing over all $j$, the estimator is regularized by summing only up to an integer $J_n$ that grows with $n$. This regularization method is called {\it spectral cutoff} and has been studied in the context of deconvolutions on the continuous Fourier domain by, e.g. \cite{Johannes}. 

\begin{remark}\label{rem:tikhonov}
\normalfont
    There are other methods of regularizing deconvolutions. For example \cite{CarrascoPaper} use Tikhonov regularization. Spectral cutoff is appropriate for this particular problem for three reasons. First, when publication bias is added later on, the regularization bias of spectral cutoff is not affected by uncertainty about the extent of publication bias. This separability does not necessarily arise for other regularization methods---e.g. Tikhonov will penalize publication bias estimates that imply a $\Pi_0$ with rapid oscillations. Second, choosing spectral cutoff exploits the fact that in this setting the singular values are guaranteed to decay exponentially fast, speeding up the rate of convergence as pointed out by \cite{Stefanski01011990}. Third, in this problem the rate-optimizing choice of smoothing parameter for spectral cutoff does not depend on $c$. This surprising result reduces the meta-analyst's researcher degrees of freedom.
\end{remark}

The estimation error of $\widehat{\Delta}_{c,n}$ can be upper bounded by the two sums below. The first sum is random sampling error. The key to controlling its variance is Lemma \ref{lem:bound_a_jpsi_j} in the technical appendix, which shows that $\sup_{t\in \mathbb{R}} \sum_{j=0}^{J_n} \left|a_j\psi_j(t)\varphi_{\sigma_T^2}(t)\right| =\mathcal{O}\left( \lambda_{J_n}^{-1}\right)$. The second sum is the deterministic ``regularization bias" that is the consequence of halting the sum at $J_n$. Here $\lesssim$ means that the left side is upper bounded by a universal constant times the right side.
\begin{align*}
     \left|\widehat{\Delta}_{c,n}-\Delta_c\right|  &\lesssim \underbrace{\sum_{j=0}^{J_n} \frac{1}{\lambda_j}\left|\frac{1}{n}\sum_{i=1}^n \psi_j(t_{i})\varphi_{\sigma_T^2}(t_i) -\mathbb{E}\left[\psi_j(T)\varphi_{\sigma_T^2}(T)\right]\right|}_{\text{Sampling Error}} \\
     &+\underbrace{\sum_{j=J_n+1}^{\infty} \eta_j\left|\mathbb{E}\left[\chi_j(H)\varphi_{\sigma_T^2+1}(H)\right]\right|}_{\text{Reg. Bias}} 
\end{align*}

To see why regularization is necessary, consider the ``sampling error" term. Since $\lambda_j$ is decaying to zero, if $J_n$ grows too quickly (or is infinite)  then the variance of $ \widehat{\Delta}_{c,n}$ can diverge. The expectation of the square of the sampling error can be bounded by studying the rate of decay of the $\lambda_j$ and by uniformly bounding the functions $\psi_j(t)\varphi_{\sigma_T^2}(t)$ over all $j,t$ \citep{HermiteBound}. This argument yields a bound on the sampling error.
\begin{equation}
    {\sum_{j=0}^{J_n} \frac{1}{\lambda_j}\left|\frac{1}{n}\sum_{i=1}^n \psi_j(t_{i})\varphi_{\sigma_T^2}(t_i) -\mathbb{E}\left[\psi_j(T)\varphi_{\sigma_T^2}(T)\right]\right|} = \mathcal{O}_p\left(n^{-\frac{1}{2}}\lambda_{J_n}^{-1}\right)
\end{equation}

The drawback of cutting the sum off at $J_n$ is regularization bias, i.e. approximation error.  The bias can be bounded using the fact that $\chi_j(h)\varphi_{\sigma_T^2+1}(h)$ is uniformly bounded over $j,h$ \citep{HermiteBound}. This means that the regularization bias  is bounded by a geometric series that is of the same order as its first summand. 
\begin{equation}
    {\sum_{j=J_n+1}^{\infty} \eta_j\left|\mathbb{E}\left[\chi_j(H)\varphi_{\sigma_T^2+1}(H)\right]\right|}  = \mathcal{O}\left(\eta_{J_n}\right)
\end{equation}

The meta-analyst wishes to increase $J_n$ with the meta-sample size $n$ at a rate that makes  $\widehat{\Delta}_{c,n}-\Delta_c$ converge in probability to zero as fast as possible. This means balancing regularization bias and sampling variance. Because we regularize with spectral cutoff, the rate-optimal choice of smoothing parameter does not depend on $c$. Theorem \ref{thm:consistency_nopb} gives the optimized rate that equalizes the order of the bias and sampling error.\begin{theorem}\label{thm:consistency_nopb}
   Assume there is no publication bias and that Assumptions \ref{assum:normality_of_T}-\ref{assum:cross_study_independence} hold. Then, for  any sequence $\{J_n\}_{n\geq 1}\subseteq \mathbb{N}$ chosen by the meta-analyst,
     \begin{align*}
    \mathbb{V}\left[\widehat{\Delta}_{c,n}\right] &= \mathcal{O}\left(n^{-1}\lambda_{J_n}^{-2}\right)  \\
    \mathbb{E}\left[\widehat{\Delta}_{c,n}\right]-\Delta_c&= \mathcal{O}\left(\eta_{J_n}\right)
\end{align*}
    If the meta-analyst chooses $J_n,\sigma_T^2$ such that $n^{1/2}\left(\frac{\sigma_T^2}{\sigma_T^2+1}\right)^{J_n/2}$ converges to a positive number, then:
    \begin{align*}
    \widehat{\Delta}_{c,n}-\Delta_c &= \mathcal{O}_p\left(n^{-\frac{q}{2}}\right),\qquad \text{ where } q \equiv \left. \log\left(\frac{1+\sigma_T^2}{1+\sigma_T^2-c^{-2}}\right) \middle/\log \left(\frac{1+\sigma_T^2}{\sigma_T^2}\right) \right.
\end{align*}
\end{theorem}

The proof is in  Appendix {\ref{proof:consistency_nopb}}. The rate of convergence is involved but it contains several insights. If the meta-analyst wishes to estimate $\Delta_c$ for a marginal increase in sample size, then they set $c=1+\epsilon$ with $\epsilon$ small. This makes $q\approx 1$ and the estimator approximately achieves the parametric rate $n^{-1/2}$. As the meta-analyst increases $c$, the rate of convergence slows down. This illustrates that $\Delta_c$ is harder to estimate when $c$ is large. The intuition is the following. The larger $c$ is, the more it matters whether $H$ is small and positive or exactly zero. Therefore for large $c$, the high-frequency components of $\Pi_0$ matter more for $\Delta_c$. Since the non-smooth components are harder to estimate, the rate of convergence slows down.

\begin{remark}\label{rem:decon_nopb}\normalfont
    It is possible to rewrite $\widehat{\Delta}_{c,n}$ as a plug-in estimator where we integrate a slowly-converging (or even inconsistent) deconvolution estimator for $\Pi_0$ in order to calculate a fast-converging estimator for $\Delta_c$. This interpretation is revealing. If $\Pi_0$ is continuous with density $\pi_0 \in \mathcal{L}_H$, this true density could be expressed in terms of the singular value decomposition:
    \begin{equation}
        \pi_0(h)  = \sum_{j=0}^{\infty} \frac{1}{\lambda_j\eta_j}\mathbb{E}\left[\psi_j(T)\varphi_{\sigma_T^2}(T)\right]\chi_j(h)
    \end{equation}
    
     Consider the sample analogue $ \widehat{\pi}_n(h)$, which always exists regardless of $\Pi_0$:
    \begin{equation}\label{eq:pihat_nopb}
    \widehat{\pi}_n(h) = \sum_{j=0}^{J_n}  \frac{1}{\lambda_j\eta_j} \left(\frac{1}{n}\sum_{i=1}^n \psi_j(t_{i})\varphi_{\sigma_T^2}(t_i)\right)  \chi_j(h)
\end{equation}

\noindent $\widehat{\Delta}_{c,n}$ can be rewritten as a deconvolution where we plug $\widehat{\pi}_n$ into Lemma \ref{lem:betac_singularvalues}. This is numerically identical to the estimator defined in Equation (\ref{eq:def_Delta_hat}).
\begin{equation}
    \widehat{\Delta}_{c,n} = \sum_{j=0}^{\infty}  \langle \chi_j,\widehat{\pi}_n \rangle_{H}\eta_j\left(\lambda_j\int_{-CV}^{CV}\psi_j(t)dt-\int_{-CV/c}^{CV/c}\phi_j(t)dt\right)
\end{equation}

\noindent Expressing $\widehat{\Delta}_{c,n}$ as a deconvolution yields two insights. First, $\widehat{\pi}_n$ is not necessarily a good estimator of the distribution of true effects. If $\Pi_0$ is not smooth enough for $\pi_0$ to exist, then $\widehat{\pi}_n$ diverges with $n$. Even if $\Pi_0$ is smooth enough for $\pi_0$ to exist and be in $\mathcal{L}_H$, the estimator $\widehat{\pi}_n$ is not  guaranteed to converge to $\pi_0$ in the sup-norm at a power-$n$ rate \citep{Carroll,Fan91}. Yet we have used $\widehat{\pi}_n$ to compute a $\widehat{\Delta}_{c,n}$ that does converge in a power of $n$ for all $\Pi_0$. Intuitively, this happens because the  distribution of $T_c$ only depends on the smooth parts of $\Pi_0$. The second  insight from the deconvolution interpretation of $\widehat{\Delta}_{c,n}$ is that since the rate-optimizing $J_n$ in Theorem \ref{thm:consistency_nopb} does not depend on $c$, $\widehat{\pi}_n$ does not depend on $c$ either. So a researcher interested in many $c$ only needs to estimate one  $\widehat{\pi}_n$ and then calculate each $\widehat{\Delta}_{c,n}$ from $\widehat{\pi}_n$. This reduces the meta-analyst's researcher degrees of freedom. Figures \ref{fig:Two_plots_MM} and \ref{fig:RCts_Alt} discussed later in this paper use this approach.
\end{remark}

\begin{remark}\label{rem:signs}
\normalfont
   One virtue of $ \widehat{\Delta}_{c,n}$ is that the signs of the observations $t_i$ do not matter. This is useful because in many meta-datasets the two-tailed $t$-scores have all been reported as positive. To see why the estimator is unaffected by signs, notice that the Hermite polynomials are all either odd or even. Since the integrals in $\widehat{\Delta}_{c,n}$ over $\phi_j(t)$ are symmetrical about zero, then the summands where $\phi_j$ is odd are zero. For the rest of the summands, the functions $\psi_j(t)$ are even so the signs of the data points $t_i$ do not matter.
\end{remark}

\begin{remark} \normalfont \label{rem:gibbs}
    A known disadvantage of spectral cutoff regularization is the {\it Gibbs phenomenon} where estimates of the density of $T_c$ that converge in the $L^2$ norm do not always converge in the sup-norm \citep{on_Gibbs}. Since the amplitude of these errors does not vanish, any estimate of $f_{T_c}$ is likely to be an invalid PDF \citep{Hohage}. Fortunately, we can still consistently estimate $\Delta_c$ despite the Gibbs phenomenon because it is a scalar that depends only on an integral of $f_{T_c}$ over a fixed interval: 
    \begin{align}
        \Delta_c =  \int_{-CV}^{CV} f_T(t)dt -  \int_{-CV}^{CV}f_{T_c}(t)dt
    \end{align}
    
    \noindent We can rewrite the estimation error in terms of estimates of $f_T$ and $f_{T_c}$:
    \begin{align}
        \widehat{\Delta}_{c,n} -\Delta_c &=     \int_{-CV}^{CV} (\widehat{f}_{T,n}(t)-f_T(t))-(\widehat{f}_{T_c,n}(t)-f_{T_c}(t))dt    \\
       \widehat{f}_{T,n}(t) &\equiv \sum_{j=0}^{J_n}\psi_j(t)\left(\frac{1}{n}\sum_{i=1}^n \psi_j(t_{i})\varphi_{\sigma_T^2}(t_i)\right) \\
  \widehat{f}_{T_c,n}(t) &\equiv \sum_{j=0}^{J_n}\frac{1}{c\lambda_j}\phi_j(t/c)\left(\frac{1}{n}\sum_{i=1}^n \psi_j(t_{i})\varphi_{\sigma_T^2}(t_i)\right)
    \end{align}

\noindent For each $n$, the estimates $\widehat{f}_{T_c,n}(t),\widehat{f}_{T,n}(t)$ are always very far from their corresponding true PDFs in the sup-norm over all $t\in \mathbb{R}$ because they are polynomials. Nevertheless, Theorem \ref{thm:consistency_nopb} shows that integrals over them on $[-CV,CV]$ have bias of order $\mathcal{O}\left(\eta_{J_n}\right)$ and $\widehat{\Delta}_{c,n}$ converges.
\end{remark}

\begin{remark}\label{rem:sigmay}
\normalfont
    The researcher could in principle optimize the rate of convergence by choosing the tuning parameter $\sigma_T^2$ that makes $q$ as large as possible. By L'Hopital's Rule, as $\sigma_T^2$ goes to infinity, $q$ converges to $c^{-2}$, making the rate approach $n^{-\frac{1}{2c^{2}}}$. But, taking $\sigma_T^2$ to infinity also causes the chosen $J_n$ to be large, which imposes computational burdens. The choice of $\sigma_T^2$ is therefore a balance between rate of convergence and computability. In my simulations and empirical application I always choose $\sigma_T^2=1$ for transparency and reproducibility. 
\end{remark}

\begin{remark}\label{rem:rate_Johannes}
\normalfont
    The rate of $n^{-\frac{1}{2c^{2}}}$ from Remark \ref{rem:sigmay} is consistent with the root-MISE rate of the spectral cutoff procedure in Theorem 3.2 of \cite{Johannes} up to log factors. To see the connection, consider the problem of recovering the counterfactual density $f_{T_c}=K_{c^{-2}}\Pi_0$ from a sample drawn from the factual density $f_T = K_{1-c^{-2}}K_{c^{-2}}\Pi_0$. In Johannes' notation, the target density $f_{T_c}$ has a characteristic
function that decays with a Gaussian of variance $c^{-2}$, while the convolution error is
Gaussian with variance $1-c^{-2}$. Therefore the polynomial source condition
(3.2) in \cite{Johannes} holds for every $s\geq 0$ and every
$\beta < \frac{1}{c^2-1}$ (see their Example 3.1). Theorem 3.2 of \cite{Johannes} then gives squared $L^2$ risk of order $
   \mathbb E\|\widehat f_{T_c}-f_{T_c}\|^2 \lesssim  \left(\mathbb E\|\widehat f_T-f_T\|^2\right)^{\beta/(\beta+1)}$.
Since $f_T$ is itself Gaussian-mollified, it can be estimated with MISE
$n^{-1}$ up to logarithmic factors.\footnote{See for example Theorem 1 of \cite{Butucea} under the no-noise case of their $s=0$.} Taking
$\beta\uparrow 1/(c^2-1)$ gives squared-risk rate $n^{-1/c^2}$ up to logs, or root-MISE rate $n^{-1/(2c^2)}$ up to logs. So the rate in this paper corresponds to the literature on spectral cutoff.
\end{remark}

\section{Publication Bias}\label{sec:pb}
It is not realistic in practice to assume that every $t$-score computed by an experimenter is reported with equal probability. After $t$-scores are computed, only a subset may be published. There is growing evidence that $t$-scores in the social sciences are selected for publication partially on the  basis of whether they cross certain significance thresholds \citep{Franco1502,Brodeur,bb,Andrews,Elliott}. In this section I add publication bias to the problem. I show that $\Delta_c$ is still identified for a broad class of models of publication bias. Then I show how to consistently estimate $\Delta_c$ under the simplest canonical model.

%\subsection{Identification Under Publication Bias}

This paper approaches publication bias by specifying a parametric model of selective reporting similar to \cite{Hedges1992}. Here I provide a relatively weak sufficient condition for the model to be identified. Let $R$ be the binary random variable that equals one if the $t$-score $T$ was reported and zero otherwise. Let the conditional probability of reporting be equal to:
\begin{equation}\label{eq:w_definition}
    \text{Pr}\left(R=1\:|\: T=t\right) = w_{\theta_0}(t)
\end{equation}
where the form of the weighting function $w_{\theta}(t)$ is known to the meta-analyst and the parameter $\theta_0$ is an unknown member of the compact set $\theta_0 \in \Theta\subseteq \mathbb{R}^K$. I make several assumptions about the publication bias model. Assumption \ref{assum:pblowerbound} below requires that $w_{\theta}(t)$ be bounded away from zero. Without an assumption like this, $\Pi_0$ is not necessarily identified.

\begin{assumption}\label{assum:pblowerbound}
There is a constant $\overline{M}>1$ such that $\overline{M}^{-1}\leq w_{\theta}(t)$ for all $t \in \mathbb{R}$ and all $\theta \in \Theta$.
\end{assumption}

Under publication bias the meta-analyst observes only $T$ conditional on $R=1$. This is problematic because the expectations $\mathbb{E}[\psi_j(T)\varphi_{\sigma_T^2}(T)]$ from Theorem \ref{thm:id} are no longer directly available in terms of the distribution of observed $t$-scores. Instead the population distribution available to the meta-analyst is the conditional distribution $T\mid R=1$. If the meta-analyst knew ${\theta_0}$, then to recover an expectation over $T$, they could take an expectation of $T\mid R=1$ times an appropriate weight. Lemma \ref{lem:pb} computes this weighting using Bayes' Theorem.   %For ease of notation, I denote ``$\mid R=1$" as just ``$\mid R$." 

\begin{lemma}\label{lem:pb}
If Assumptions \ref{assum:normality_of_T} and \ref{assum:pblowerbound} hold, then:
    \begin{align*}
    \mathbb{E}[\psi_j(T)\varphi_{\sigma_T^2}(T)] =\left.\mathbb{E}\left[\frac{\psi_j\left({T}\right)\varphi_{\sigma_T^2}\left({T}\right)}{w_{\theta_0}({T})}\:|\: R=1\right]\middle/\mathbb{E}\left[\frac{1}{w_{\theta_0}({T})}\:\mid\: R=1\right]\right. 
\end{align*}
\end{lemma}

The proof is in  Appendix \ref{proof:pb}. Combining this result with Theorem \ref{thm:id} immediately yields Corollary \ref{cor:id_pb} below. This shows that if $\theta_0$ is identified, then $\Delta_c$ must also be identified. The remaining task is to identify $\theta_0$ itself.

\begin{corollary}\label{cor:id_pb}
If Assumptions \ref{assum:normality_of_T} and \ref{assum:pblowerbound} hold, then:

      $$\Delta_c = \sum_{j=0}^\infty\frac{\mathbb{E}\left[\frac{\psi_j\left({T}\right)\varphi_{\sigma_T^2}\left({T}\right)}{w_{\theta_0}({T})}\:\mid\: R=1\right]}{\mathbb{E}\left[\frac{1}{w_{\theta_0}({T})}\:\mid\: R=1\right]} a_j $$
\end{corollary}

This paper will identify $\theta_0$ using only the distribution of published $t$-ratios. I cannot identify $\theta_0$ via the joint distribution of published point estimates and standard errors as \cite{Andrews} or \cite{Duval2000} do because this would require assumptions far too strong for this setting. Their {\it funnel plot strategy} requires true effects and true standard errors to be uncorrelated. If this holds, then observing correlation in published point estimates and their standard errors reveals publication bias. This symmetry assumption would be plausible in settings where every study is investigating a similar effect. 

This paper allows each experiment to investigate a completely different effect. This flexibility allows the meta-analyst to study a set of experiments united by their funders or publishers and not necessarily by their topics. Full study heterogeneity can introduce dependence between true effects and their standard errors because researchers who know {\it ex ante} that they are probably investigating a small effect may choose a larger sample size or pre-specify a less robust but more precise estimator. Dependence can arise for other reasons as well. For instance, if some studies measure effects in dollars and others in euros, heterogeneity in the units will induce correlation between standard errors and true effects. Therefore this paper identifies $\theta_0$ via the $t$-curve which is not confounded by such dependence.

For $\theta_0$ to be identified from the distribution of $T\mid R=1$, publication bias must always distort the distribution of $t$-scores in a way that could never happen ``naturally." I now add an assumption on the shape of $w_\theta(t)$ to this effect. In the absence of publication bias,  $T$ must always have an infinitely differentiable probability density function because it is the outcome of a convolution with the normal distribution. Most models of publication bias specify that reporting decisions are made on the basis of statistical significance, i.e. on whether $T$ crosses certain thresholds. Define $f_{T\mid R=1}(t)$ as the PDF of $T$ conditional on $R=1$. If publication bias introduces kinks, jumps, or non-smoothness into the density $f_{T\mid R=1}(t)$ (or its derivatives), then it is possible to recover the original smooth density. Assumption \ref{assum:wnonsmooth} stipulates that publication bias ``reveals itself" through breaks at a countable set of points. If these breaks take the form of jump discontinuities at traditional critical values they are sometimes called ``Caliper Gaps" and they have been well studied by others \citep{caliperOriginal,Elliott,Kudrinjmp}. However Assumption \ref{assum:wnonsmooth} is more general because it envisions publication bias that reveals itself through any non-smoothness in the $t$-curve.

\begin{assumption}{\label{assum:wnonsmooth}}
   (i) For all $\theta_1,\theta_2\in \Theta$ the function $g(t)\equiv \frac{w_{\theta_1}(t)}{w_{\theta_2}(t)}$ is continuous and infinitely differentiable in $t$ except possibly at finitely many points. (ii) If $\theta_1\neq \theta_2$, then there exists an integer $k\geq 0$ and a point $s\in\mathbb{R}$ such that: $\lim_{t\to s^+}g^{(k)}(t)\neq \lim_{t\to s^-}g^{(k)}(t)$.
\end{assumption}

Theorem \ref{thm:id_pb_general} states that Assumption \ref{assum:wnonsmooth} is sufficient for identification. 

\begin{theorem}\label{thm:id_pb_general}
    If Assumptions \ref{assum:normality_of_T}-\ref{assum:wnonsmooth} hold, then $\theta_0$ and $\Delta_c$ are identified.
    
\end{theorem}

The proof is in Appendix \ref{proof:thm:id_pb_general}. The intuition for Theorem \ref{thm:id_pb_general} is the following. Suppose that the meta-analyst has a guess $\theta$ for $\theta_0$ and attempts to remove the publication bias by reweighting the density $f_{T\mid R=1}$ by $1/w_{\theta}$ via Lemma \ref{lem:pb}. Only by reweighting with the true $\theta_0$ can the meta-analyst remove all of the jump discontinuities in all the derivatives. So  $\theta_0$ is identified from $f_{T\mid R=1}$. Since Corollary \ref{cor:id_pb} expresses $\Delta_c$ as a function of $f_{T\mid R=1}$ and $\theta_0$, then $f_{T\mid R=1}$ must identify $\Delta_c$  as well. 

\subsection{Estimation Under Simple Publication Bias}

To estimate $\Delta_c$ under publication bias the meta-analyst must specify a parametric function $w_\theta(t)\: :\: \mathbb{R}\to \mathbb{R}^+$ that satisfies Assumptions \ref{assum:pblowerbound} and \ref{assum:wnonsmooth}. Many models are possible. For ease of exposition in this section I specify the most straightforward canonical model of publication bias from \cite{Andrews}. Suppose that the probability of publication depends only on whether $|T|$ exceeds the critical value $1.96$. The conditional reporting probability ratio is equal to the scalar $\theta_0=\frac{\text{Pr}\left(R=1\:|\:|T|< 1.96\right)}{\text{Pr}\left(R=1\:|\:|T|\geq 1.96\right)}\in \left[\overline{M}^{-1},1\right]$. This means that the weighting function is:
\begin{equation}\label{eq:pb_caliper_form}
    w_{\theta_0}(t) = \theta_0\mathbf{1}\left\{|t| < 1.96\right\}+\mathbf{1}\left\{|t| \geq 1.96\right\} 
\end{equation}  %the absolute values are actually here, even if you cannot see them
\noindent This function satisfies Assumption \ref{assum:wnonsmooth}. Here publication bias reveals itself via a discontinuity in the t-curve at the traditional critical value of 1.96. This kind of ``Caliper" discontinuity is well-studied \citep{caliperOriginal,Kudrinjmp}. Equation (\ref{eq:id_theta_simple}) expresses $\theta_0$ in terms of the jump discontinuity  at $|T|=1.96$ in the distribution of published $t$-scores. 
\begin{equation}\label{eq:id_theta_simple}
    \theta_0 =\lim_{\epsilon\to 0}\frac{\text{Pr}\left(|T|\in (1.96-\epsilon,1.96] \: \mid \: R=1\right)}{\text{Pr}\left(|T|\in (1.96,1.96+\epsilon] \: \mid \: R=1\right)} 
\end{equation}The meta-analyst can construct a plug-in estimator $\widehat{\theta}_n$ by estimating the caliper jump in the histogram of published $t$-scores. To do this the meta-analyst chooses a bin width $\epsilon_n>0$ and takes the ratio between the number of $t$-scores within $\epsilon_n$ below vs above the critical threshold of  $1.96$ in absolute value. 
\begin{equation}
    \widehat{\theta}_n \equiv \frac{\frac{1}{n}\sum_{i=1}^n \mathbf{1}\left\{|t_i| \in (1.96-\epsilon_n,1.96 ]\right\}}{\frac{1}{n}\sum_{i=1}^n \mathbf{1}\left\{|t_i| \in (1.96,1.96+\epsilon_n]\right\}} 
\end{equation}

\begin{remark}\normalfont
    There is now another tuning parameter $\epsilon_n$  that trades off bias and variance. The optimal rate at which to scale $\epsilon_n$ as $n$ increases turns out to be $n^{-1/3}$ which is the same as the optimal rate for pointwise convergence of histograms. In Section \ref{sec:simulations}, I provide specific recommendations for $\epsilon_n$ in finite samples.
\end{remark}

The meta-analyst can plug $\widehat{\theta}_n$  and the sample means into Corollary \ref{cor:id_pb} to obtain an estimator $ \widehat{\Delta}_{c,n}^{pb}$ for $\Delta_c$ under simple publication bias.
\begin{equation}
  \widehat{\Delta}_{c,n}^{pb} \equiv  \frac{\sum_{j=0}^{J_n} a_j \frac{1}{n}\sum_{i=1}^n \frac{\psi_j(t_i)\varphi_{\sigma_T^2}(t_i)}{w_{\widehat{\theta}_n}(t_i)} }{ \frac{1}{n}\sum_{i=1}^n \frac{1}{w_{\widehat{\theta}_n}(t_i)}}
\end{equation}

\begin{remark}\normalfont
   Spectral cutoff is a particularly convenient method of regularization when publication bias is involved. The regularization bias of $\widehat{\Delta}_{c,n}^{pb}$ incurred by cutting the spectrum off at $J_n$ is still of order $\sum_{j=J_n+1}^{\infty} \eta_j\mathbb{E}\left[\chi_j(H)\varphi_{\sigma_T^2+1}(H)\right]\lambda_ja_j$. Notice that this was completely unchanged by the addition of publication bias to the problem! Uncertainty about $\theta_0$ does not play any part in the approximation error incurred by spectral cutoff. If instead we had used Tikhonov regularization like \cite{CarrascoPaper}, then the sum over $j$ would be infinite, each term in the sum would depend on $\widehat{\theta}_n$. So non-smoothness incurred by estimation error in $\widehat{\theta}_n$ is subject to the Tikhonov penalty. The approximation error incurred by Tikhonov therefore depends on $\widehat{\theta}_n$, which significantly complicates the analysis. Moreover, Tikhonov regularization would slow down the rate even in the absence of publication bias, so spectral cutoff is appropriate. 
\end{remark}

Theorem \ref{thm:consistency_withpb} shows the consistency of $\widehat{\Delta}_{c,n}^{pb} $. Adding publication bias has slowed the rate of convergence down to $n^{-q/3}$. This happens because a small change in the histogram near $1.96$ can change the weight that every $t_i$ gets in each of the sample averages in the estimator. The proof is in Section \ref{proof:consistency_withpb}. 

\begin{theorem}\label{thm:consistency_withpb}
   Assume that publication bias follows Equations (\ref{eq:w_definition}) and (\ref{eq:pb_caliper_form}) and that Assumptions \ref{assum:normality_of_T}-\ref{assum:wnonsmooth} hold. If the meta-analyst chooses sequences  $\epsilon_n \propto n^{-\frac{1}{3}}$ and  $\{J_n\}_{n\geq 1}\subseteq \mathbb{N}$, then:
    $$  \widehat{\Delta}_{c,n}^{pb}-\Delta_c =\mathcal{O}_p\left(\lambda_{J_n}^{-1}n^{-\frac{1}{3}}+\eta_{J_n}\right) $$

If the meta-analyst also chooses $J_n$ such that $n^{1/3}\left(\frac{\sigma_T^2}{\sigma_T^2+1}\right)^{J_n/2}$ converges to a positive number, then:\begin{align*}
   \widehat{\Delta}_{c,n}^{pb}-\Delta_c &= \mathcal{O}_p\left(n^{-\frac{q}{3}}\right),\qquad \text{ where } q \equiv \log\left(\frac{1+\sigma_T^2}{1+\sigma_T^2-c^{-2}}\right)  /\log \left(\frac{1+\sigma_T^2}{\sigma_T^2}\right)
\end{align*}
\end{theorem}

\begin{remark}
    \normalfont We can interpret $\widehat{\Delta}_{c,n}^{pb}$ as a deconvolution using the same reasoning as Remark \ref{rem:decon_nopb}. The polynomial $\widehat{\pi}_n^{pb}$ below converges logarithmically to a density that itself approximates $\Pi_0$.
    \begin{equation}\label{eq:pihat}
    \widehat{\pi}_n^{pb}(h) =\sum_{j=0}^{J_n} \frac{\sum_{i=1}^n \frac{\psi_j(t_i)\varphi_{\sigma_T^2}(t_i)}{w_{\widehat{\theta}_n}(t_i)}}{\eta_j\lambda_j\sum_{i=1}^n \frac{1}{w_{\widehat{\theta}_n}(t_i)}} \chi_j(h)
\end{equation}
    
    \noindent We can express $\widehat{\Delta}_{c,n}^{pb}$ as an integral over $\widehat{\pi}_n^{pb}$.\begin{equation}\label{eq:delta_pihat_pb}
        \widehat{\Delta}_{c,n}^{pb} = \sum_{j=0}^\infty \eta_j \langle \chi_j,\widehat{\pi}_n^{pb} \rangle_{H} \left(\lambda_j\int_{-CV}^{CV}\psi_j(t)dt-\int_{-CV/c}^{CV/c}\phi_j(t)dt\right)
    \end{equation}

  \noindent  Once again, changing $c$ does not affect $\widehat{\pi}_n^{pb}$ or the rate-optimal $J_n$! So a single $\widehat{\pi}_n^{pb}$ can be estimated and plugged into Equation (\ref{eq:delta_pihat_pb}) for each $c$ of interest.
\end{remark}

\section{Inference}\label{sec:inference}

\subsection{Intuition and Notation}

In this section I show the asymptotic normality of $\widehat{\Delta}_{c,n}^{pb}$ and derive a consistent variance estimator. Inference proceeds in three steps. First, I show that the estimator admits an asymptotically linear representation around a deterministic centering sequence. Second, I establish a CLT for the leading sample-average term. Third, I construct a feasible plug-in variance estimator by replacing the oracle influence function with a sample analogue.

These results require the introduction of several new pieces of notation.  Let $F,\widehat{F}_{n}$ denote the true CDF and empirical CDF of $|T|$ conditional on publication $R=1$. $\widetilde{\Delta}_{c,n}$ will denote the deterministic sequence of real numbers converging to $\Delta_c$ about which the estimator is centered.  $m_n^0(t)$ denotes the unknown oracle influence function and $\widehat{m}_n(t)$ is its feasible sample analogue. Exact definitions for all of these quantities are collected in Appendix \ref{app:inference_definitions} for convenience. These objects will be used to construct the oracle variance estimator  ${V}_n^0$ and its feasible analogue $\widehat{V}_n$.

\subsection{Asymptotic Normality}

 I show that $\widehat{\Delta}_{c,n}^{pb}$ is asymptotically normal in the usual way by expressing the estimation error as a weakly dependent sample average using Taylor approximations. Regularization bias will affect the centering of the estimator. The centering term is non-negligible because $\widehat{\Delta}_{c,n}^{pb}$ contains smoothing bias from two different sources. First, since only the first $J_n$ terms of the singular value decomposition are used, the rest of the terms are set to zero and this incurs regularization bias. Second, since the meta-analyst only observes a sample of $t$-scores, they estimate the discontinuities in $f_{T\mid R=1}$ by looking in a window of width $\epsilon_n$ on either side of each possible point of discontinuity. Since the density can change over this interval, smoothing bias is incurred here as well. 

After centering  properly, it is possible to linearize the sampling error of $\widehat{\Delta}_{c,n}^{pb}$. Lemma \ref{lem:linearization} below uses Taylor's Theorem several times to rewrite the estimator as a triangular array of sample means plus a dominated term. The function $m_{n}^0 (t)$ is deterministic. 
\begin{lemma}\label{lem:linearization}
Assume that publication bias follows Equations (\ref{eq:w_definition}) and (\ref{eq:pb_caliper_form}) and that Assumptions \ref{assum:normality_of_T}-\ref{assum:wnonsmooth} hold and that the meta-analyst chooses  sequences  $\epsilon_n \propto n^{-\frac{1}{3}}$ and  $\{J_n\}_{n\geq 1}\subseteq \mathbb{N}$. Then, for the deterministic sequence $\{\widetilde{\Delta}_{c,n}\}_{n\geq 1}\subset \mathbb{R}$ and the deterministic functions $m_n^0 \: : \: \mathbb{R}\to \mathbb{R}$ defined in Appendix \ref{app:inference_definitions}, all of the following hold: 

\begin{enumerate}
 \item $\widehat{\Delta}_{c,n}^{pb}-\widetilde{\Delta}_{c,n}
    = \frac{1}{n}\sum_{i=1}^n m_n^0 (t_i) -\mathbb{E}\left[m_n^0 (T)\mid R=1\right] +\mathcal{O}_p\left(\lambda_{J_n}^{-1}\left(n^{-1}\epsilon_n^{-1}+\epsilon_n^2+n^{-1/2}\right)\right)$
\item  $\mathbb{E}\left[m_n^0 (T)^2\mid R=1\right] = \mathcal{O}\left(\lambda_{J_n}^{-2}\epsilon_n^{-1}\right)$
\item $\sup_{n}\sup_{t\in\mathbb{R}}\epsilon_n\lambda_{J_n}|m_n^0(t)|<\infty$
    \item $\widetilde{\Delta}_{c,n}-\Delta_c = \mathcal{O}\left(\lambda_{J_n}^{-1}\epsilon_n+\eta_{J_n}\right)$
\end{enumerate}

\end{lemma}

The proof is in Appendix \ref{proof:linearization}. Next I show that the triangular array of sample means $ \frac{1}{n}\sum_{i=1}^n m_n^0(t_i)$ is asymptotically normal.  In order for this to guarantee that the estimator $ \widehat{\Delta}_{c,n}^{pb}$ is itself also normal, the sample means must dominate the Taylor residuals. To guarantee this I add  Assumption \ref{assum:varnotsmall} which says that the estimator does not converge too fast. This variance implicitly conditions on the publication of all $t_i$.

%This is now significantly weakened
\begin{assumption}\label{assum:varnotsmall}
    $$ \liminf_{n\to \infty} n\epsilon_n\lambda_{J_n}^2\mathbb{V}\left[\frac{1}{n}\sum_{i=1}^n m_{n}^0(t_i)\right] >0 $$
\end{assumption}

To show asymptotic normality of the triangular array of sample sums I invoke a Lindeberg-Feller type Central Limit Theorem. The key step is to check the Lyapunov Condition. While verifying this condition is a common tactic in ill-posed problems, my argument is quite different than \cite{CarrascoPaper}. Using the bounds on the magnitude and variance of $m_n^0(T)$ from Lemma \ref{lem:linearization}, we can verify that Assumption \ref{assum:varnotsmall} is sufficient to guarantee the Lyapunov Condition that uses the fourth moments. Theorem \ref{thm:clt} shows that the assumptions so far are enough to satisfy all of the hypotheses of the CLT for $m$-dependent triangular arrays proven by \cite{JansonPratelliRigo2024}. The formal argument is in Appendix \ref{proof:clt}.
\begin{theorem}\label{thm:clt}
Assume that publication bias follows Equations (\ref{eq:w_definition}) and (\ref{eq:pb_caliper_form}) and that Assumptions \ref{assum:normality_of_T}-\ref{assum:varnotsmall} hold and that the meta-analyst chooses  sequences  $\epsilon_n \propto n^{-\frac{1}{3}}$ and  $\{J_n\}_{n\geq 1}\subseteq \mathbb{N}$. Then:
   $$ \frac{\widehat{\Delta}_{c,n}^{pb}-\widetilde{\Delta}_{c,n}}{\sqrt{\mathbb{V}\left[\frac{1}{n}\sum_{i=1}^n m_{n}^0(t_i)\right]}} \to_d N(0,1)$$
\end{theorem}

\subsection{Variance Estimation}

Next I derive a consistent variance estimator. Equation (\ref{eq:var_covar_sum}) below expresses the variance of the sum as the sum of the covariances. Since Assumption \ref{assum:cross_study_independence} stipulates that two $t$-scores drawn from different studies are independent, the covariances are zero across studies. Recall that $\Lambda$ is the $n\times n$ block-diagonal matrix  where the element $\Lambda_{ij}$ indicates whether $t_i$ and $t_j$ were reported in the same study. This gives us the following expression for the variance:
\begin{equation}\label{eq:var_covar_sum}
   \mathbb{V}\left[\frac{1}{n}\sum_{i=1}^n m_n^0(t_i)\right] =  \frac{1}{n^2}\sum_{i=1}^n\sum_{k=1}^n\Lambda_{ik} \text{Cov}\left(m_n^0 (t_i),m_n^0(t_k)\right)
\end{equation}

This suggests the oracle variance estimator $V_n^0$ where $\overline{{m}^0_n}$ denotes the sample mean:
\begin{equation}
    {V}_n^0 \equiv \frac{1}{n^2}\sum_{i=1}^n\sum_{k=1}^n\Lambda_{ik} \left({m}_n^0 (t_i)-\overline{{m}^0_n}\right)\left({m}_n^0(t_k) -\overline{{m}^0_n}\right)
\end{equation}

\noindent Since the functions $m^0_n(t) $ depend on the unknown parameters $\theta_0$ and $\Pi_0$, the meta-analyst cannot compute $V_n^0$. Fortunately, there is a sample version $\widehat{m}_n(t) $ defined in Appendix \ref{app:inference_definitions} that the meta-analyst does observe and can be used for variance estimation.  Define $\overline{\widehat{m}}$ to be the sample mean of the $\widehat{m}_n(t_i)$. Let the feasible variance estimator $\widehat{V}_n$ be the analogue of the oracle estimator $V^0_n$:
\begin{equation}
        \widehat{V}_n \equiv \frac{1}{n^2}\sum_{i=1}^n\sum_{k=1}^n\Lambda_{ik} (\widehat{m}_n (t_i)-\overline{\widehat{m}_n})(\widehat{m}_n(t_k) -\overline{\widehat{m}_n})
\end{equation}
Theorem \ref{thm:variance_estimation} shows that $\widehat{V}_n$ is consistent and converges fast enough for inference.  
\begin{theorem}\label{thm:variance_estimation}
 Assume that publication bias follows Equations (\ref{eq:w_definition}) and (\ref{eq:pb_caliper_form}) and that Assumptions \ref{assum:normality_of_T}-\ref{assum:varnotsmall} hold. If the meta-analyst chooses  sequences  $\epsilon_n \propto n^{-\frac{1}{3}}$ and  $\{J_n\}_{n\geq 1}\subseteq \mathbb{N}$, then:
    \begin{align*}
     \frac{    \widehat{V}_n }{\mathbb{V}\left[\frac{1}{n}\sum_{i=1}^n m_n^0(t_i)\right]} \to_p 1 \quad \text{ and }\quad \frac{\widehat{\Delta}_{c,n}^{pb}-\widetilde{\Delta}_{c,n}}{\sqrt{\widehat{V}_n}}\to_d N(0,1)
\end{align*}
\end{theorem}

The proof is in Appendix \ref{proof:variance_estimation}. Theorem \ref{thm:variance_estimation} says that the meta-analyst can construct valid confidence intervals covering the centering sequence $\widetilde{\Delta}_{c,n}$ that includes both spectral cutoff and histogram smoothing bias. Lemma \ref{lem:linearization} showed that when $J_n$ and $\epsilon_n$ scale at the rates in Theorem \ref{thm:consistency_withpb}, the centering sequence converges to $\Delta_c$ at the same rate as the standard error decays. In theory, this could affect the coverage of the confidence intervals for $\Delta_c$ itself.  A natural solution is to ``undersmooth" or to change $J_n,\epsilon_n$ more quickly than the optimal rate in order to force $\widetilde{\Delta}_{c,n}$ to converge to $\Delta_c$ faster than the standard error decays. The simulations and empirical applications in this paper choose not to undersmooth but achieve nearly perfect coverage for the 95\%  confidence intervals in simulation nevertheless.

\section{Simulations}\label{sec:simulations}

This section presents simulations showing near-perfect coverage of $\Delta_c$ by the 95\% confidence intervals under a variety of circumstances. This section also illustrates how to set tuning parameters $\epsilon_n,$ $J_n$, and $\sigma_T$ to yield this good coverage. To guarantee the optimized rate of convergence in Theorem \ref{thm:consistency_withpb} the meta-analyst sets $\epsilon_n = Cn^{-1/3}$ and $J_n = \log\left(Dn^{-1/3}\right)/\log\left(\sqrt{\sigma_T^2/\left(1+\sigma_T^2\right)}\right)$ where $C,D>0$. This means that the meta-analyst's choice is actually over the triple of constants $\left\{C,D,\sigma_T\right\}$. The preceding theorems provide no specific guidance on how these constants are to be chosen and we must turn to simulations. I recommend that the meta-analyst should always at least disclose results using the following tuning parameters: $C=2$, $D=0.05$, and $\sigma_T=1$. These choices of tuning parameters yield good confidence interval coverage of $\Delta_c$ in simulation for a very wide variety of data generating processes. 

I generate simulated data in the following way. I use the simple model of publication bias from Section \ref{sec:pb} where $t$-scores are reported with probability $\theta_0$ if they do not clear $1.96$ and are reported with certainty otherwise. This matches the illustrative example from \cite{Andrews}. In this section I set $\theta_0=0.9$ but the results are not sensitive to this. Table \ref{tab:sims} reports simulations for several choices of $\Pi_0$ where we expect coverage to be poor for one reason or another. Theory predicts that coverage could be low when the distribution $\Pi_0$ is not very smooth or if the density $f_T$ is close to zero or has a steep slope at the critical threshold $1.96$. I report simulations against several different distributions $\Pi_0$, some of which are non-smooth.
\begin{enumerate}
    \item {\bf True Nulls:}  $H=0$ almost surely. As a simple default, consider a distribution where every null is true. Discrete distributions are not smooth and induce large regularization bias.  
     \item {\bf Cauchy:} $H\sim \text{Cauchy}$. Outlier treatment effects are relevant to the theory of experimental design \citep{ABfattails}. 
    \item {\bf Bimodal:} $H$ is distributed as a 50\% mixture of two normals with variance 1 with one centered at zero and the other at $2.8$. This is a mixture between studies with small effects and those with nearly 80\% true power. 
    \item  {\bf Large:}  $H \sim \text{N}\left(1.96,0.2^2\right)$. This makes $\Delta_c$ large. 
    \item {\bf Slope:} $H \sim \text{N}\left(0.96,0.2^2\right)$. This makes the slope of the $t$-curve steep near $1.96$ which makes the smoothing bias of $\widehat{\theta}_n$ large.
        \item {\bf Uniform:} $H \sim \text{Unif}(-3,3)$. The uniform distribution is not smooth so we expect high regularization bias. 
        \item {\bf Fitted:} This DGP was estimated from the full sample of \cite{bb} and is meant to generate a realistic $t$-curve with non-smooth $\Pi_0$ that will stress-test our confidence intervals. The model is that $\Pi_0$ is discrete with seven support points at $\{0,1,\cdots, 6\}$. Recall from Remark \ref{rem:signs} that the signs do not matter.  We estimate the model with the \verb|zcurve| R command from \cite{zcurve2}. The fitted model places mass $(0.02, 0.47, 0.01, 0.27, 0.14, 0.00, 0.09)$ at the seven support points. While this need not necessarily approximate the true $\Pi_0$ well, it does produce a $t$-curve that resembles a real one.
\end{enumerate}

The simulation results are reported in Table \ref{tab:sims}. Here I compare performance for these $\Pi_0$ under two different meta-sample sizes: a modest meta-sample of $50$ $t$-scores versus a large meta-sample of $500$ $t$-scores. The main message of Table \ref{tab:sims} is that coverage of $\Delta_c$ by the 95\% confidence intervals is close to the nominal level under a broad variety of potentially problematic $\Pi_0$ for both large and small $n$.

 \begin{table}[h!]
    \centering
      \caption{\label{tab:sims} Simulations $T-H\sim N(0,1)$} 
    \begin{threeparttable}
   % \vskip 0.05in
    
\begin{tabular}{ r l r r r r r r r }
\hline \hline 
\multicolumn{4}{c}{Parameters }& & \multicolumn{4}{c}{Results}\\
\cline{1-4}  \cline{6-9}\\
  $n$ & $\Pi_0$ & Unc. Pwr. & $\Delta_c$ & & Mean $\widehat{\Delta}_{c,n}^{pb}$ & SD $\widehat{\Delta}_{c,n}^{pb}$ & Mean St. Err. & Cover of 95\% CI\\
\hline 
  50   &   True Null & 0.05 & 0.00 & & 0.00 & 0.10 & 0.10 & 0.94\\
  50   &   Cauchy & 0.37 & 0.09 & & 0.07 & 0.08 & 0.08 & 0.94\\
  50   &   Bimodal & 0.44 & 0.12 & & 0.10 & 0.08 & 0.08 & 0.94\\
  50   &   Large & 0.50 & 0.28 & & 0.26 & 0.09 & 0.09 & 0.94\\
  50   &   Slope & 0.17 & 0.12 & & 0.13 & 0.10 & 0.10 & 0.94\\
  50   &   Uniform & 0.37 & 0.16 & & 0.15 & 0.09 & 0.09 & 0.95\\
  50   &   Fitted & 0.54 & 0.10 & & 0.09 & 0.08 & 0.08 & 0.94\\
\hline  
  500   &   True Null & 0.05 & 0.00 & & 0.00 & 0.06 & 0.06 & 0.95\\
  500   &   Cauchy & 0.37 & 0.09 & & 0.08 & 0.05 & 0.05 & 0.95\\
  500   &   Bimodal & 0.44 & 0.12 & & 0.12 & 0.05 & 0.05 & 0.95\\
  500   &   Large & 0.50 & 0.28 & & 0.28 & 0.05 & 0.05 & 0.95\\
  500   &   Slope & 0.17 & 0.12 & & 0.13 & 0.06 & 0.06 & 0.94\\
  500   &   Uniform & 0.37 & 0.16 & & 0.16 & 0.05 & 0.05 & 0.95\\
  500   &   Fitted & 0.54 & 0.10 & & 0.10 & 0.04 & 0.05 & 0.96\\
\hline
\end{tabular}

\begin{tablenotes}
\footnotesize
    \item {\it Notes:} Table showing simulation results. Each row is a parameterization. Each parameterization was run for 10000 repetitions. All simulations used the tuning parameters: $C=2$, $D=0.05$, and $\sigma_T=1$. Column 1 in the table is the number of $t$-scores. Column 2 is the distribution of true effects and each of these is described in Section \ref{sec:simulations}. Column 3 is unconditional (mean) statistical power at the status quo (before publication bias). Column 4 is the increase in unconditional power resulting from doubling the sample size of every experiment, so $c=\sqrt{2}$  (before publication bias). Columns 5 and 6 are the mean and standard deviation of the point estimates. The seventh column is the mean of the estimated standard errors and the final column is the fraction of confidence intervals that contained the true $\Delta_c$. Transparency R package: \url{https://github.com/sfaridan/Testing_for_Underpowered_Literatures_Transparency}
\end{tablenotes}
\end{threeparttable}
\end{table}

\subsection{Deviations from Normality}

In practice, $T-H$ is only approximately normally distributed, meaning that Assumption \ref{assum:normality_of_T} does not hold exactly. There are two reasons for this. First, the researcher must estimate the denominator of the $t$-score, which results in $T-H$ having a student-$t$ distribution. Table \ref{tab:sims_t} below shows that this is unlikely to be a concern when estimating $\Delta_c$. Here we repeat the Monte Carlo analysis in Table \ref{tab:sims}, but simulate data with $T - H\sim t(30)$. Despite the small number of degrees of freedom, the coverage rates remain largely the same as before.

A more serious  deviation from normality comes from the numerator of the $t$-score. When the effective sample size of the experiment is small, the Central Limit Theorem might not have fully ``kicked in" yet. The Edgeworth Series suggests that this is primarily a concern when the outcome variable studied by the experiment is skewed or has excess kurtosis. To address this concern, we run the simulation in Table \ref{tab:sims_lognorm}. Here $T-H$ is the scaled and centered mean of $185$ i.i.d. log-normally distributed random variables. The number $185$ was chosen to match the median number of treatment clusters in the RCTs from the \cite{bb} data.\footnote{ This is the median of a random sample of 100 of the RCTs in the dataset. A research assistant checked each paper individually to find the number of clusters. There was occasional ambiguity about the number of treatment clusters. In these cases we always used the smaller number. The median number of observations per study reported directly by \cite{bb} is 5202.} The log-normal distribution was chosen because it has substantial skew and excess kurtosis---making it an unfavorable distribution for the CLT. Nevertheless, coverage remains relatively high.  These simulations suggest that the empirical estimates in the next section are unlikely to be significantly distorted by the inexactness of the normal approximation.

 \begin{table}[h!]
    \centering
      \caption{\label{tab:sims_t} Simulations:  $T-H\sim t(30)$} 
    \begin{threeparttable}
   % \vskip 0.05in
    
  \begin{tabular}{ r l r r r r r r r }
\hline \hline 
\multicolumn{4}{c}{Parameters }& & \multicolumn{4}{c}{Results}\\
\cline{1-4}  \cline{6-9}\\
  $n$ & $\Pi_0$ & Unc. Pwr. & $\Delta_c$ & & Mean $\widehat{\Delta}_{c,n}^{pb}$ & SD $\widehat{\Delta}_{c,n}^{pb}$ & Mean St. Err. & Cover of 95\% CI\\
\hline  
  500   &   True Null & 0.06 & 0.00 & & 0.00 & 0.06 & 0.06 & 0.95\\
  500   &   Cauchy & 0.37 & 0.09 & & 0.08 & 0.05 & 0.05 & 0.96\\
  500   &   Bimodal & 0.45 & 0.12 & & 0.11 & 0.05 & 0.05 & 0.95\\
  500   &   Large & 0.50 & 0.28 & & 0.27 & 0.05 & 0.05 & 0.95\\
  500   &   Slope & 0.17 & 0.11 & & 0.12 & 0.06 & 0.06 & 0.94\\
  500   &   Uniform & 0.37 & 0.17 & & 0.16 & 0.05 & 0.05 & 0.95\\
  500   &   Fitted & 0.54 & 0.10 & & 0.10 & 0.04 & 0.05 & 0.95\\
\hline
\end{tabular}

\begin{tablenotes}
\footnotesize
    \item {\it Notes:} This table matches the lower panel of Table \ref{tab:sims} except that the conditional $t$-score has the Student-$t$ distribution with 30 degrees of freedom: $T=H+t(30)$. 
\end{tablenotes}
\end{threeparttable}
\end{table}

 \begin{table}[h!]
    \centering
      \caption{\label{tab:sims_lognorm} Simulations: $T- H$ distributed as scaled mean of log-normals} 
    \begin{threeparttable}
   % \vskip 0.05in
    
   \begin{tabular}{ r l r r r r r r r }
\hline \hline 
\multicolumn{4}{c}{Parameters }& & \multicolumn{4}{c}{Results}\\
\cline{1-4}  \cline{6-9}\\
  $n$ & $\Pi_0$ & Unc. Pwr. & $\Delta_c$ & & Mean $\widehat{\Delta}_{c,n}^{pb}$ & SD $\widehat{\Delta}_{c,n}^{pb}$ & Mean St. Err. & Cover of 95\% CI\\
\hline  
  500   &   True Null & 0.05 & 0.00 & & -0.02 & 0.06 & 0.06 & 0.94\\
  500   &   Cauchy & 0.36 & 0.09 & & 0.08 & 0.05 & 0.05 & 0.95\\
  500   &   Bimodal & 0.44 & 0.13 & & 0.12 & 0.05 & 0.05 & 0.95\\
  500   &   Large & 0.47 & 0.31 & & 0.28 & 0.05 & 0.05 & 0.92\\
  500   &   Slope & 0.16 & 0.11 & & 0.09 & 0.06 & 0.06 & 0.94\\
  500   &   Uniform & 0.37 & 0.17 & & 0.16 & 0.05 & 0.05 & 0.95\\
  500   &   Fitted & 0.54 & 0.09 & & 0.09 & 0.05 & 0.05 & 0.95\\
\hline
\end{tabular}

\begin{tablenotes}
\footnotesize
    \item {\it Notes:} This table matches the lower panel of Table \ref{tab:sims} except that the conditional $t$-score is the standardized sample mean of $185$ i.i.d. log-normal(0,1) random variables. The number $185$ matches a conservative estimate of the median number of treatment clusters for a random sample of 100 RCTs from the \cite{bb} data. 
\end{tablenotes}
\end{threeparttable}
\end{table}

\section{Empirical Application}\label{sec:empirical_applications}

This section applies the methods proposed above to address an empirical question with important policy implications: Are randomized controlled trials (RCTs) published in top economics journals too small? RCTs are lauded as the ``gold standard" of empirical evidence in social science partly because their design allows them to credibly control the rate of type-I errors. But what about type-II errors? If the conclusions reported by influential RCTs are sensitive to reasonable increases in sample size, then funders and researchers should run experiments with fewer treatment arms and allocate more resources toward data collection.

This is an empirical question and the answer will depend on the population of experiments under study. Here I study the experiments that have the most influence in academic economics: those published in top journals. The data source is \cite{bb}. This meta-study examined the universe of 684 articles published in top economics journals during 2015-2018 of which 145 were RCTs. The data contain 21,740 test statistics. All test statistics corresponded to main hypotheses of interest and excluded covariates, placebo tests, etc. Every $t$-score was produced using one of four empirical methods: Randomized Controlled Trials (RCTs),  Difference-in-Differences (DID), Discontinuity Designs (DD), and Instrumental Variables (IV). The $t$-scores are derounded.

I estimate $\Delta_c$ among RCTs and non-RCTs in Table \ref{tab:brodeur}. The estimate in column 1 says that counterfactually doubling the sample sizes of every RCT published in top economics journals would only increase the expected fraction of $t$-scores clearing the critical value of 1.96 by 7.2 percentage points with standard error $2.5$ percentage points.

\begin{table}[h!]
\caption{\label{tab:brodeur} Empirical Application: Randomized Trials in Economics} \vspace{0.05in}
\begin{center}
\begin{threeparttable}

\begin{tabular}{l c c c c c  }
\hline \hline 
 & \multicolumn{2}{c}{\textit{By Number of $t$-scores}} &  &\multicolumn{2}{c}{\textit{By Number of Studies}} \\
\cline{2-3} \cline{5-6} 
\\ & RCTs & Non-RCTs &  & RCTs & Non-RCTs \\
\hline
 $\hat{\Delta}^{pb}_{c}$ & .072 & .173  & & .107 & .167\\
  & (.025)  & (.023) & & (.010) & (.013) \\
  & [.02, .12] & [.13, .22] & & [.09, .13] & [.14, .19]\\ 
  $\hat{\theta}$ & 1.04 & .93 & & 1.17 & .93\\
    & (.12) & (.08) & & (.08) & (.05) \\ 
 \hline  
  J  & 17& 17 &  & 13 & 14\\
  $\epsilon$ & $.10$ & $.08$ &  & $.38$ & $.24$ \\ \hline
     Equality with RCTs (p-value)&  & .00 &  &  &.00 \\ \hline 
      $\hat{\Delta}$ \verb|zcurve| & .100 &  .122 & &  & \\
  & (.029)  & (.020) & &  &  \\ \hline 
        Status Quo Power & .38 & .54 && .38 & .54 \\
  $t$-scores & 7569 & 14171 &  & 7569 & 14171  \\
  Studies & 145& 559 &  & 145& 559 \\
\hline
\end{tabular}

\begin{tablenotes}
\footnotesize  
\item {\it Notes:}  Reports estimates $\widehat{\Delta}$ of the gain in unconditional statistical power of a two-sided size 5\% $t$-test resulting from counterfactually doubling every study's sample size from the status quo (c=$\sqrt{2}$). Compares $t$-scores testing main hypotheses from randomized trials published in top economics journals versus those reported by studies using difference-in-differences, regression discontinuity or instrumental variables. Tuning parameters are $D=.05$, $C=2$, and $\sigma_T=1$. Columns 1 and 2 report results when $J,\epsilon$ are chosen based on the number  of $t$-scores (more conservative). Columns 3 and 4 report results when tuning parameters are chosen based on the number of studies (less conservative). Standard errors are reported in round brackets and 95\% confidence intervals are in square brackets. Standard errors clustered by article. The rows labeled \verb|zcurve| show estimates of $\widehat{\Delta}_c$ using the mixture model method from \cite{zcurve2} implemented by the \verb|zcurve| R package. Average statistical power at status quo sample sizes can be estimated nonparametrically by simply computing the fraction of $t$-scores with absolute value larger than 1.96 in each sample. The bottom rows report the outcome of a test for equality of the RCT vs non-RCT values of $\Delta$, the status quo average power, the number of $t$-scores in each sample and the number of distinct research articles in each sample. To avoid causing problems for inference, we do not truncate $\widehat{\theta}_n$ at unity. Data: \cite{bb}. Replication: \url{https://github.com/sfaridan/Testing_for_Underpowered_Literatures_Transparency}
\end{tablenotes}
\footnotesize 
\end{threeparttable}
\end{center}
\end{table} I argue that 7.2 percentage points should be viewed as a small power gain and therefore that RCTs published in top economics journals are not very sensitive to sample size increases on average. This interpretation suggests that funders should sponsor more RCTs rather than fewer, larger ones. I defend the notion that 7.2 percentage points is small by comparing it to three benchmarks. 

As an initial benchmark, consider a literature where every experiment is run at exactly 80\% power. This is the traditional threshold for an RCT to be considered ``sufficiently powered" \citep{jpal_power_calcs}. We can calculate that doubling every sample size would increase power by 17.8 percentage points. The confidence interval in Column 1 of Table \ref{tab:brodeur} does not come close to this value. So RCTs are less than half as sensitive to sample size increases as a set of hypothetical ‘well-powered’ experiments. I therefore argue that they leave little easy power on the table.

\begin{figure} 
    \centering
    \includegraphics[width=5in]{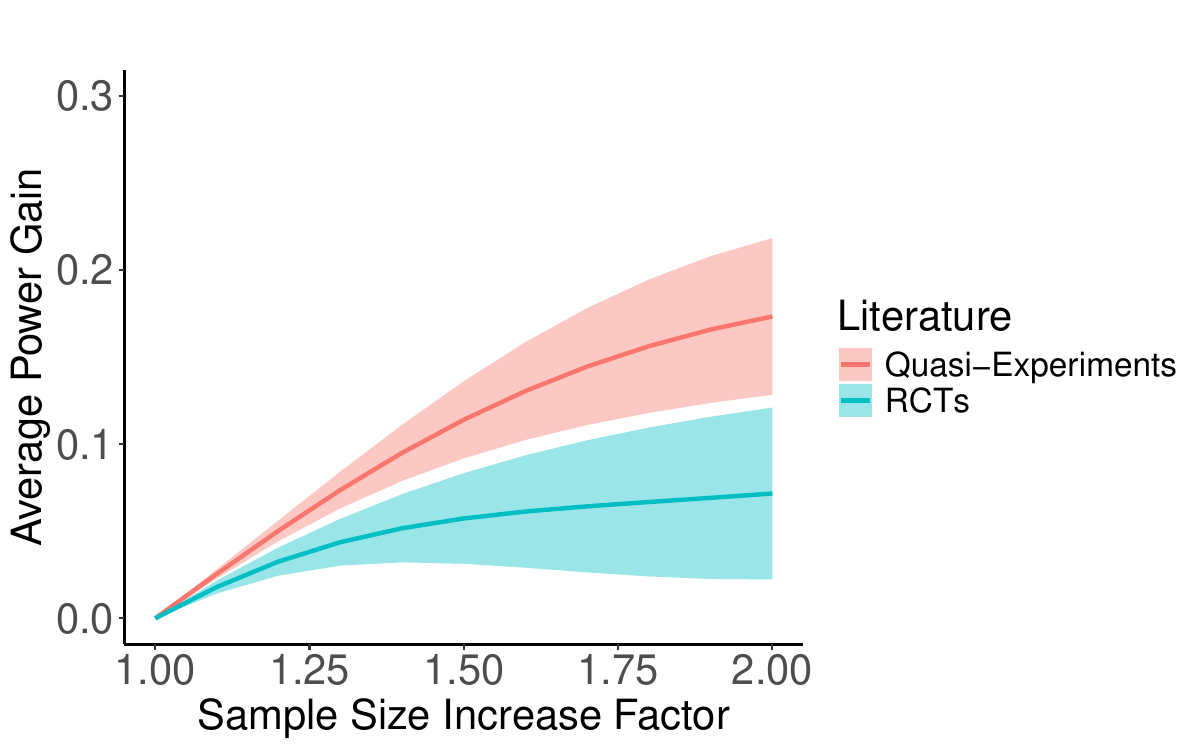}
    \caption{\footnotesize Compares power gain (y-axis) of RCTs vs non-RCTs over $c^2$ (x-axis). Shaded areas are pointwise 95\% confidence intervals. Data: \cite{bb}.}
    \label{fig:Two_plots_MM}
\end{figure}

As a second benchmark, I compare RCTs to natural experiments in the \cite{bb} dataset. The dataset consists of 14,171 $t$-scores reported in 559 articles that used observational methods to identify causal effects.\footnote{Twenty articles contained $t$-tests from both an RCT and an observational method.} Column 2 of Table \ref{tab:brodeur} reports that among tests conducted by non-RCTs, doubling sample sizes would increase power by $17.3$ percentage points, which is significantly larger than for RCTs $(p=0.00)$. Figure \ref{fig:Two_plots_MM} visualizes this contrast for a variety of sample size increases. It shows how power climbs faster for natural experiments than RCTs in the \cite{bb} sample across the board. 

\begin{remark}\normalfont
    These results are not very sensitive to the choice of tuning parameters. Columns 3 and 4 of Table \ref{tab:brodeur} show a robustness check where  $J,\epsilon$ are scaled by the number of studies instead of the number of $t$-scores. These confidence intervals are smaller but at greater risk of bias. The results are largely unchanged.
\end{remark}

\begin{remark}\normalfont
    Estimating $\widehat{\Delta}_c$ nonparametrically  substantively affects conclusions in practice. To show this, Table \ref{tab:brodeur} also includes an alternative deconvolution estimator from \cite{zcurve2} and \cite{Sotola} in rows 9 and 10. Under its default parameterization, the \verb|zcurve()| R function estimates a seven-mean mixture model of $\Pi_0$. \verb|zcurve()| finds a  gap between RCTs and non-RCTs that is only a fifth as large. This is possibly because \verb|zcurve| will only be consistent for $\Delta_c$ if $\Pi_0$ is discrete with probability mass at the numbers $\{0,1,\cdots,6\}$---which is probably not the case in the \cite{bb} setting. The fully nonparametric method proposed by this paper detects a large difference between RCTs and non-RCTs without making any assumptions about $\Pi_0$ at all. 
\end{remark}

The difference in $\Delta_c$ between RCTs and natural experiments is likely because experimenters who run RCTs choose their sample sizes. In contrast,  observational researchers typically cannot choose the sample size of a natural experiment. Even if power calculations involve guesswork, these results suggest that such calculations still provide useful information. Since randomized trials are already optimized in a way that quasi-experiments are not, it is not surprising that they have less to gain from sample size changes. The next subsection constructs a third benchmark that is perhaps more surprising. 

\subsection{Replication Studies in Laboratory Psychology}

This subsection presents a third benchmark to compare against the power gain of 7.2 percentage points estimated in the previous subsection. I construct the benchmark using a second set of experiments that has two special properties: (i) we have good reason to expect that all of the experiments in this set are unusually well powered and (ii) it is possible to construct a credible alternative estimate of the power gain as a robustness check. 

The Many Labs systematic replication project of \cite{ManyLabs} recruited 36 independent research teams who each attempted to replicate 13 published effects from laboratory psychology. Each team collected their own independent data and ran some or all of the thirteen experiments on their respective samples. Each sample contained at least 79 participants and the project included 6,344 participants in total. Eleven of the experiments were analyzed using $t$-tests of equality in means between a treatment group and a control group and I will limit my analysis to these. This yields a meta-sample of 385 $t$-scores. The aim of all of the experiments run by Many Labs was to replicate effects that had been published in top psychology journals and to investigate how consistently effects could be replicated across study sites. 

The key special property of replication projects is that the sample size of every experiment was chosen on the basis of data from a previous published experiment that studied the same effect. This means that we can expect the Many Labs experiments to be on average very well powered and to have a small $\Delta_c$ relative to what can reasonably be achieved in experiments run for the first time. In Table \ref{tab:manylabs} the 95\% confidence interval contains the power gain of 7.2 percentage points from economics RCTs. This means that I cannot reject the null hypothesis that $\Delta_c$ for Many Labs is the same as $\Delta_c$ for RCTs published in top economics journals $(p=0.178)$. That is to say, economics RCTs are not significantly more sensitive to sample-size increases than laboratory replications that we ex ante expect to be very well powered.  This is not simply a consequence of wide confidence intervals or high uncertainty because $\Delta_c$ for Many Labs is significantly different from $\Delta_c$ for non-RCTs in the \cite{bb} data. 

To show that the results are not sensitive to reasonable choices of the tuning parameters, Table \ref{tab:manylabs} presents four specifications. In columns 1 and 2 the tuning parameters  $J_n,\epsilon_n$ are scaled  by the total number of $t$-scores. In columns 3 and 4 the tuning parameters are instead scaled  by the total number of experimental sites, which is less conservative. In columns 1 and 3 I use the recommended values of the constants $C,D$ from Table \ref{tab:sims}. In columns 2 and 4 I vary these choices to show that the confidence sets do not change much even when the changes to $C,D$ are large.

 The Online Appendix presents another key robustness check. There I exploit a second key feature of the Many Labs setting: each laboratory was plausibly testing the same set of hypotheses. This special circumstance makes it possible to construct an alternative estimator that is much more precise than $\widehat{\Delta}_c$ in the spirit of \cite{PowerOfBias} and \cite{bundock}. I estimate that doubling the sample size of every Many Labs replication (this time conditional on the hypotheses themselves) would increase the fraction of statistically significant $t$-scores by 7.8 percentage points with a standard error of only 0.18 percentage points.

%\clearpage 

\begin{table}[h!]
\caption{\label{tab:manylabs} Empirical Application: Many Labs Replication Project} 
\centering
\begin{threeparttable}
%\vskip{0.05in}
\begin{tabular}{l c c c c c c }
\hline \hline 
 & \multicolumn{2}{c}{\textit{By Number of $t$-scores}} &  &\multicolumn{2}{c}{\textit{By Number of Sites}} & \\
\cline{2-3} \cline{5-6} 
\\ & Main Specification & Rob. Check &  & Main Specification & Rob. Check & \\
\hline
 $\hat{\Delta}^{pb}_{c}$ & .005 & .048 & & .033 & .058 & \\
& (.042) & (.025) & & (.026) & (.035) & \\
 &  [0, .088] & [0, .097] & & [0, .083] & [0, .126] & \\
    \hline 
  $\hat{\theta}$ & .92 & 1.20 & & 1.03 & .86 & \\
 & (.60) & (1.05) & & (.32) & (.49) & \\
 \hline  
 D  &  .05 &  .15 & & .05 & .15 & \\
  C  &  $2$ & $1$ & & $2$ & $1$ & \\
  J    & 14 & 11 & & 12& 8 & \\
  $\epsilon$  & $.27$ & $.14$ & &  $.61$ & $.30$ & \\
 \hline
     Unconditional Power & .61 & .61 &  & .61 &.61 & \\
  $t$-scores & 385 & 385 & & 385 & 385 & \\
  Sites & 36& 36 & & 36 & 36 & \\
  Treatments & 11 & 11& & 11 & 11 & \\
\hline
\end{tabular}

\begin{tablenotes}
\footnotesize 
\item {\it Notes:} Table reports estimates $\widehat{\Delta}$ of gain in unconditional statistical power  of a two-sided size 5\% $t$-test resulting from counterfactually doubling every study's sample size from the status quo. The left two columns report results where $J,\epsilon$ are scaled based on the number of $t$-scores (more conservative) while in the two columns on the right these tuning parameters are scaled based on the number of study sites (less conservative). Columns 1 and 3 use the preferred choice of $C,D$ while columns 2 and 4 present a further robustness check where we have significantly altered $C,D$. Standard errors are reported in round brackets and 95\% confidence intervals are in square brackets. Standard errors clustered by study site and treatment type. The bottom four rows present the fraction of $t$-scores larger than 1.96 in absolute value, the number of $t$-scores in total, the number of unique experimental research sites, and the number of unique experimental treatments. Confidence intervals are truncated at zero because $\Delta_c \geq 0$ by construction. To avoid causing problems for inference, we do not truncate $\widehat{\theta}_n$ at unity.  Data: \cite{ManyLabs}. Transparency package: \url{https://github.com/sfaridan/Testing_for_Underpowered_Literatures_Transparency}
\end{tablenotes}
\end{threeparttable}
\end{table}
%\clearpage

I conclude that RCTs published in top economics journals are on average relatively insensitive to counterfactual sample size increases. This can only happen if most trials are either studying very large effects relative to their status quo sample sizes (and are fully powered) or are investigating effects so small that even an experiment twice as large would not easily detect them. Power calculations---however imperfect---appear to be giving researchers enough information to know when power is cheap to increase and when it is not. Since there is no clear way to check whether a single sample size was ``well chosen" ex post, this new rigorous test of the aggregate downstream adequacy of power calculations was needed. The implication is that funders and researchers should consider alternative ways to raise power in addition to sample size increases---e.g. improving measurement quality---as suggested by \cite{McKenzie}.

\section{Conclusion}\label{sec:conclusion}

This paper proposes an estimator that is consistent for the  increase in the fraction of $t$-scores that would have been statistically significant had every experiment in a given population had its sample size counterfactually increased by a chosen factor $c^2>1$. Unlike existing work, no assumptions were imposed on the distribution of true intervention treatment effects---point masses and arbitrary densities are both allowed. The lack of such assumptions is important in theory and affects conclusions in practice. The proposed estimator is asymptotically normal and robust to simple forms of publication bias. A key technical contribution was to prevent uncertainty about publication bias from magnifying the smoothing error of the deconvolution step. 

The method is useful and can inform funding and design decisions. For example, an empirical application finds that the power of randomized trials in economics would only increase by 7.2 percentage points on average if every sample size had been doubled. I argue that this number is small by comparing it to three benchmarks. I conclude that---despite requiring guesswork about unknown parameters---power calculations appear to leave little easy power on the table  in practice. Improvements to measurement quality, compliance, and outcome choice may therefore offer lower-hanging fruit \citep{McKenzie}.  This suggests that funders looking to improve power should sponsor higher-quality experiments and not rely only on larger samples.  Future empirical work could apply the proposed method to specific subsets of experiments, e.g. those funded by different initiatives, in order to determine exactly which sorts of experiments should be made larger.

\clearpage

{\bf \noindent Declaration of generative AI and AI-assisted technologies in the manuscript preparation process.}

During the preparation of this work the author used Microsoft Copilot, Google Gemini and Refine in order to proofread, audit, and find typos. After using this tool/service, the author reviewed and edited the content as needed and takes full responsibility for the content of the published article.

\bibliographystyle{chicago}

\bibliography{biblio}
\appendix

\clearpage
\section{Technical Definitions of Objects used in the proofs}\label{app:inference_definitions}

This appendix collects the definitions of the technical objects used for inference. 

Recall that $F,\widehat{F}_n$ denote the true and empirical CDFs of $|T|$ conditional on publication $R=1$. Define ${B}_{\epsilon_n}^+,{B}_{\epsilon_n}^-$ as the expected fraction of $t$-scores just above and just below $1.96$ in absolute value and let $\widehat{B}_{\epsilon_n}^+,\widehat{B}_{\epsilon_n}^-$ be their sample analogues:
\begin{align}
    {B}_{\epsilon_n}^+ &\equiv {F}(1.96+\epsilon_n)-{F}(1.96)\\
    {B}_{\epsilon_n}^- &\equiv {F}(1.96)-{F}(1.96-\epsilon_n)\\
    \widehat{B}_{\epsilon_n}^+ &\equiv \widehat{F}_n(1.96+\epsilon_n)-\widehat{F}_n(1.96)\\
    \widehat{B}_{\epsilon_n}^- &\equiv \widehat{F}_n(1.96)-\widehat{F}_n(1.96-\epsilon_n)
\end{align}

Define the deterministic functions $S_n(t) \: : \: \mathbb{R}\to \mathbb{R} $ and $W(t;\theta,p) \: : \: \mathbb{R}^3\to \mathbb{R}$ as the following:
\begin{align}
    S_n(t) &\equiv \sum_{j=0}^{J_n} a_j\psi_j(t)\varphi_{\sigma_T^2}(t)\\
W(t;\theta,p) &\equiv \frac{1+\left({\theta}^{-1}-1\right)\mathbf{1}\left\{|t|< 1.96\right\}}{1+({\theta}^{-1}-1)p}
\end{align}
We can rewrite the estimator in terms of $S_n(t)$ and $W(t;\theta,p)$:
\begin{equation}
     \widehat{\Delta}_{c,n}^{pb} = \frac{1}{n}\sum_{i=1}^n S_n(t_i)W(t_i;\widehat{\theta}_n,\widehat{F}_n(1.96))
\end{equation}

Define the deterministic sequence of real numbers $Q_n$:
\begin{align}
    Q_n \equiv \mathbb{E}\left[ S_n(T) \frac{\mathbf{1}\left\{|T|< 1.96\right\}-{F}(1.96)}{\left(1+{F}(1.96)({\theta}_0^{-1}-1)\right)^2}\mid R=1\right]
\end{align}

Define its sample analogue $\widehat{Q}_n$ as:
\begin{align}
     \widehat{Q}_n &\equiv   \frac{1}{n}\sum_{i=1}^n S_n(t_i)  \left(\frac{\mathbf{1}\left\{|t_i|< 1.96\right\}-\widehat{F}_n(1.96)}{\left(1+\widehat{F}_n(1.96)(\widehat{\theta}_n^{-1}-1)\right)^2}\right)
\end{align}

Define the intermediate version $\widetilde{Q}_n$ with the true conditional CDF $F$ and true $\theta_0$ as:
\begin{align}
     \widetilde{Q}_n &\equiv  \frac{1}{n}\sum_{i=1}^n S_n(t_i)\left(\frac{\mathbf{1}\left\{|t_i|< 1.96\right\}-{F}(1.96)}{\left(1+{F}(1.96)({\theta}_0^{-1}-1)\right)^2}\right)
\end{align}

Define the function ${X}_{n} \: :\:\mathbb{R}^3\to \mathbb{R}$ (the influence function for $\widehat{\theta}_n^{-1}$) as:
\begin{align}
     {X}_{n}(t;b^+,b^-) &\equiv \frac{1}{b^-}\mathbf{1}\left\{|t|\in (1.96,1.96+\epsilon_n]\right\} -\frac{b^+}{\left(b^-\right)^2}\mathbf{1}\left\{|t|\in (1.96-\epsilon_n,1.96]\right\}
\end{align}

Define the oracle influence function ${m}_n^0(t)\: :\: \mathbb{R}\to \mathbb{R} $ as:
\begin{align*}
      {m}_n^0(t) &\equiv  S_n(t)W(t;{\theta}_0,{F}(1.96)) +{Q}_n{X}_{n}(t;{B}_{\epsilon_n}^+,{B}_{\epsilon_n} ^-)
\end{align*}

Define the feasible influence function $\widehat{m}_n(t)\: :\: \mathbb{R}\to \mathbb{R} $ as:
\begin{align*}
      \widehat{m}_n(t) &\equiv  S_n(t)W(t;\widehat{\theta}_n,\widehat{F}_n(1.96)) +\widehat{Q}_n{X}_{n}(t;\widehat{B}_{\epsilon_n}^+,\widehat{B}_{\epsilon_n} ^-)
\end{align*}

Define the deterministic sequence $\widetilde{\theta}_n$ as:
\begin{align}
    \widetilde{\theta}_n \equiv \frac{F(1.96)-F(1.96-\epsilon_n)}{F(1.96+\epsilon_n)-F(1.96)}
\end{align}

Define the deterministic sequence $\widetilde{\Delta}_{c,n}$ as:\begin{align}
 \widetilde{\Delta}_{c,n} &\equiv \sum_{j=0}^{J_n}\mathbb{E}\left[\psi_j(T)\varphi_{\sigma_T^2}(T)\right] a_j   +Q_n\left(\widetilde{\theta}_n^{-1}-{\theta}_0^{-1}\right) 
\end{align}

\clearpage
\section{Proofs}

\subsection{Proof of Lemma \ref{lem:conv_op}}\label{proof:conv_op} %audited 5/20/26

It is easier to decompose $\Delta_c$ into the difference of the non-rejection probabilities: $ \Delta_c = \beta_1-\beta_c$. Here $\beta_1$ is the non-rejection probability under the factual sample sizes: $\beta_1\equiv \int_{-CV}^{CV} f_{T}(t)dt$ and $\beta_c$ is the non-rejection probability under counterfactual sample sizes: $\beta_c \equiv  \int_{-CV}^{CV} f_{T_c}(u)du$. All integrands are non-negative, so we can use Tonelli's Theorem twice and a change of variables of integration to rewrite $\beta_c$. Then we substitute in the definition of $K_{c^{-2}}$. These steps yield:
\begin{align}
    \beta_c &= \int_{-CV}^{CV} f_{T_c}(u)du= \int_{-CV}^{CV} \int_{-\infty}^\infty \varphi(u-ch)d\Pi_0(h)  du &\text{Definitions}\\
    &=  \int_{-\infty}^\infty \int_{-CV}^{CV}\varphi(u-ch)  du \:d\Pi_0(h) &\text{Tonelli}\\
    &=  \int_{-\infty}^\infty\int_{-CV/c}^{CV/c} c\varphi(ct-ch)dt \:d\Pi_0(h) &\text{Change of variable: $u=ct$}\\
    &= \int_{-\infty}^\infty\int_{-CV/c}^{CV/c}  \varphi_{c^{-2}}(t-h) dt\: d\Pi_0(h)  &\text{Identity: $\varphi_{c^{-2}}(x)=c\varphi(cx)$}\\
    &= \int_{-CV/c}^{CV/c} \int_{-\infty}^\infty \varphi_{c^{-2}}(t-h)  d\Pi_0(h)\: dt &\text{Tonelli}\\
    &= \int_{-CV/c}^{CV/c} (K_{c^{-2}}\Pi_0)[t]  \: dt &\text{Definition of $K_{c^{-2}}$}
\end{align}

Applying the same steps with $c=1$ yields: $\beta_1 = \int_{-CV}^{CV} (K_{1}\Pi_0)[t] dt$. By Lemma \ref{lem:sumnoise}, $\beta_1 = \int_{-CV}^{CV}(K_{1-c^{-2}}K_{c^{-2}}\Pi_0)[t]dt$.  Taking the difference yields the claim of the lemma: $$\Delta_c = \int_{-CV}^{CV}(K_{1-c^{-2}}K_{c^{-2}}\Pi_0)[t]dt-\int_{-CV/c}^{CV/c}(K_{c^{-2}}\Pi_0)[t]dt$$

\subsection{Proof of Lemma \ref{lem:betac_singularvalues}}\label{proof:betac_singularvalues} %audited 5/20/26

As in Lemma \ref{lem:conv_op}, it is easier to first decompose $\Delta_c$ into the difference of the non-rejection probabilities: $ \Delta_c = \beta_1-\beta_c$. Here $\beta_1$ is the non-rejection probability under the factual sample sizes: $\beta_1\equiv \int_{-CV}^{CV} f_{T}(t)dt$ and $\beta_c$ is the non-rejection probability under counterfactual sample sizes: $\beta_c \equiv  \int_{-CV}^{CV} f_{T_c}(t)dt$. Using the same argument as in Lemma \ref{lem:conv_op}, we do a change of variables of integration to obtain:
\begin{align*}
   \beta_c&= \int_{-CV/c}^{CV/c} (K_{c^{-2}}\Pi_0)[t]  \: dt 
\end{align*}

Then we substitute in the singular value decomposition from Equation (\ref{eq:svdc2}):
\begin{align}\label{eq:betac_beforefubini}
    \beta_c= \int_{-CV/c}^{CV/c}\sum_{j=0}^\infty \eta_j \langle \chi_j,\pi_0 \rangle_{H} \phi_j(t)dt
\end{align}

We would like to exchange the sum and the integral with Fubini's Theorem. To do this it is sufficient to show that: $\sum_{j=0}^\infty \eta_j \left|\langle \chi_j,\pi_0 \rangle_{H}\right| \int_{-CV/c}^{CV/c}\left|\phi_j(t)\right| dt < \infty$. First,  use the fact that since $\pi_0$ is a PDF that integrates to one, H\"{o}lder's Inequality guarantees that $\left|\langle \chi_j,\pi_0 \rangle_{H}\right|\leq\sup_h|\chi_j(h)\varphi_{\sigma_T^2+1}(h)|$. Then, we use Lemma \ref{lem:bound_coeffs}, which says that $\sup_h|\chi_j(h)\varphi_{\sigma_T^2+1}(h)|\leq \frac{1}{\sqrt{2(\sigma_T^2+1)\pi }}$. So $\left|\langle \chi_j,\pi_0 \rangle_{H}\right| \leq \frac{1}{\sqrt{2(\sigma_T^2+1)\pi }}$. Second, since $\{\eta_j\}$ are a geometrically decaying sequence, we have $\sum_{j=0}^\infty |\eta_j|<\infty$. Finally, by Lemma \ref{lem:integral_hermite}, $\sup_j\int_{-CV/c}^{CV/c}|\phi_j(t)|dt$ is upper-bounded by a constant depending only on $CV,\sigma_T^2$ and $c$. Given these bounds: $\sum_{j=0}^\infty \eta_j \left|\langle \chi_j,\pi_0 \rangle_{H}\right| \int_{-CV/c}^{CV/c}\left|\phi_j(t)\right| dt < \infty$. Therefore we can use Fubini's Theorem to switch the integral and sum in Equation (\ref{eq:betac_beforefubini}):
\begin{align*}
     \beta_c = \sum_{j=0}^\infty \eta_j \langle \chi_j,\pi_0 \rangle_{H} \int_{-CV/c}^{CV/c}\phi_j(t)dt
\end{align*}

A symmetric argument that applies Fubini's Theorem to $f_T(t) = \sum_{j=0}^\infty {\lambda_j\eta_j}  {\langle \chi_j,\pi_0 \rangle_{H}} {\psi_j}(t) $ (Equation \ref{eq:fT}) lets us compute $\beta_1$. Taking the difference $\Delta_c=\beta_1-\beta_c$ lets us conclude the Lemma:
\begin{align*}
     \Delta_c = \sum_{j=0}^\infty \eta_j \langle \chi_j,\pi_0 \rangle_{H} \left(\lambda_j\int_{-CV}^{CV}\psi_j(t)dt-\int_{-CV/c}^{CV/c}\phi_j(t)dt\right)
\end{align*}

\subsection{Proof of Theorem \ref{thm:id}}\label{proof:id} %audited 5/20/26

For any probability distribution $\Pi_0$ define the sequence of continuous distributions $\Pi_n\equiv \Pi_0 * N\left(0,\tau_n^2\right)$  where $\tau_n\to 0$ and $*$ denotes convolution. So $\Pi_n$ has PDF $\pi_n\in \mathcal{L}_H$  and $\Pi_n\to_w \Pi_0$ (where $\to_w$ denotes weak convergence). Define $\Delta_{c,n}$ as the sequence of $\Delta_c$ under the distributions $\Pi_n$: 
$$\Delta_{c,n}  \equiv \mathbb{E}_{\Pi_n}\left[\text{Pr}\left(|T_c|>CV\mid H\right)-\text{Pr}\left(|T|>CV\mid H\right)\right]$$

The function $\text{Pr}\left(|T_c|>CV\mid H\right)-\text{Pr}\left(|T|>CV\mid H\right)$ is bounded and continuous in $H$. So by the Portmanteau Theorem:
$$\Delta_{c,n}\to \Delta_c $$

By Lemma \ref{lem:betac_singularvalues}:
\begin{align}
    \Delta_{c,n} = \sum_{j=0}^\infty \eta_j \langle \chi_j,\pi_n \rangle_{H} \left(\lambda_j\int_{-CV}^{CV}\psi_j(t)dt-\int_{-CV/c}^{CV/c}\phi_j(t)dt\right)
\end{align}

Since the inner product is an integral over a density, this is an expectation:
\begin{align}
    \Delta_{c,n} = \sum_{j=0}^\infty \eta_j \mathbb{E}_{\Pi_n}\left[\chi_j(H)\varphi_{\sigma_T^2+1}(H)\right] \left(\lambda_j\int_{-CV}^{CV}\psi_j(t)dt-\int_{-CV/c}^{CV/c}\phi_j(t)dt\right)
\end{align}

Our goal now is to push the limit inside the infinite sum using the Dominated Convergence Theorem.   Since the function $\chi_j(h)\varphi_{\sigma_T^2+1}(h)$ is uniformly bounded over all $j,h$ by Lemma \ref{lem:bound_coeffs}, the coefficients $\mathbb{E}_{\Pi_n}[\chi_j(H)\varphi_{\sigma_T^2+1}(H)]$  are uniformly bounded over all $\Pi_n$. This means that $\eta_j\mathbb{E}_{\Pi_n}[\chi_j(H)\varphi_{\sigma_T^2+1}(H)]$ are  bounded by a constant times $\eta_j$. By Lemma \ref{lem:integral_hermite}: $$\sup_j\left|\lambda_j\int_{-CV}^{CV}\psi_j(t)dt-\int_{-CV/c}^{CV/c}\phi_j(t)dt\right|\leq \sqrt{\frac{2 CV}{\varphi_{\sigma_T^2}(CV)}} +\sqrt{\frac{2 CV}{c\varphi_{\sigma_T^2+1-c^{-2}}(CV/c)}}<\infty $$ Therefore, $\eta_j\mathbb{E}_{\Pi_n}[\chi_j(H)\varphi_{\sigma_T^2+1}(H)]\left(\lambda_j\int_{-CV}^{CV}\psi_j(t)dt-\int_{-CV/c}^{CV/c}\phi_j(t)dt\right)$ are uniformly bounded for all $j$ by a constant times $\eta_j$. Since $\sum_{j=0}^\infty \eta_j$ is a geometric series, it converges. So by the Dominated Convergence Theorem we can push the limit inside the sum:
\begin{align*}
     \Delta_c=\lim_{n\to \infty}\Delta_{c,n} &= \lim_{n\to \infty}\sum_{j=0}^\infty \eta_j \mathbb{E}_{\Pi_n}\left[\chi_j(H)\varphi_{\sigma_T^2+1}(H)\right] \left(\lambda_j\int_{-CV}^{CV}\psi_j(t)dt-\int_{-CV/c}^{CV/c}\phi_j(t)dt\right)\\
     &= \sum_{j=0}^\infty \eta_j \left(\lim_{n\to \infty}\mathbb{E}_{\Pi_n}\left[\chi_j(H)\varphi_{\sigma_T^2+1}(H)\right]\right) \left(\lambda_j\int_{-CV}^{CV}\psi_j(t)dt-\int_{-CV/c}^{CV/c}\phi_j(t)dt\right)
\end{align*}

Since $\Pi_n\to_w \Pi_0$ and $\chi_j(h)\varphi_{\sigma_T^2+1}(h)$ is continuous and bounded, we can use the Portmanteau Theorem a second time:
\begin{align}\label{eq:limit_pin}
    \lim_{n\to \infty}\mathbb{E}_{\Pi_n}\left[\chi_j(H)\varphi_{\sigma_T^2+1}(H)\right]= \mathbb{E}_{\Pi_0}\left[\chi_j(H)\varphi_{\sigma_T^2+1}(H)\right]
\end{align}

Substituting (\ref{eq:limit_pin}) into the expression above:
\begin{align}\label{eq:Delta_identity}
     \Delta_c &= \sum_{j=0}^\infty \eta_j \mathbb{E}_{\Pi_0}\left[\chi_j(H)\varphi_{\sigma_T^2+1}(H)\right]\left(\lambda_j\int_{-CV}^{CV}\psi_j(t)dt-\int_{-CV/c}^{CV/c}\phi_j(t)dt\right)
\end{align}

By Lemma \ref{lem:moment_conversion}:
\begin{align}
   \mathbb{E}_{\Pi_0}\left[\chi_j(H)\varphi_{\sigma_T^2+1}(H)\right]&=    \frac{1}{\eta_j\lambda_j}\mathbb{E}_{\Pi_0}[\psi_j(T)\varphi_{\sigma_T^2}(T)] \label{eq:chi_h_identity}
\end{align}

Substituting (\ref{eq:chi_h_identity}) into (\ref{eq:Delta_identity}):
 $$\Delta_{c} = \sum_{j=0}^\infty \mathbb{E}_{\Pi_0}[\psi_j(T)\varphi_{\sigma_T^2}(T)]\left(\int_{-CV}^{CV}\psi_j(t)dt-\frac{1}{\lambda_j}\int_{-CV/c}^{CV/c}\phi_j(t)dt\right)=\sum_{j=0}^\infty \mathbb{E}_{\Pi_0}[\psi_j(T)\varphi_{\sigma_T^2}(T)]a_j $$

\subsection{Proof of Theorem \ref{thm:consistency_nopb}}\label{proof:consistency_nopb} %audited 5/20/26

The case $c=1$ is trivial: then $\lambda_j=1$, $\phi_j=\psi_j$, the two integration intervals coincide, so $a_j=0$ for all $j$. Hence $\widehat\Delta_{1,n}=0=\Delta_1$. We therefore from now on assume $c>1$.

First we decompose the estimation error. By Theorem \ref{thm:id}, the estimation error is:
\begin{align*}
  \Delta_c- \widehat{\Delta}_{c,n}  &= \sum_{j=0}^{J_n} \left(\mathbb{E}\left[\psi_j(T)\varphi_{\sigma_T^2}(T)\right]-\frac{1}{n}\sum_{i=1}^n \psi_j(t_{i})\varphi_{\sigma_T^2}(t_i) \right)a_j +\sum_{j=J_n+1}^\infty\mathbb{E}\left[\psi_j(T)\varphi_{\sigma_T^2}(T)\right]a_j
\end{align*}

\noindent The first sum has expectation zero and therefore contributes only to variance. The second sum has variance zero and therefore contributes only to bias. Since $\Delta_c$ is fixed:
\begin{align*}
    \mathbb{V}\left[\widehat{\Delta}_{c,n}\right] &= \mathbb{V}\left[\sum_{j=0}^{J_n}\frac{1}{n}\sum_{i=1}^n  \psi_j(t_{i})\varphi_{\sigma_T^2}(t_i) a_j\right] \\
     \Delta_c- \mathbb{E}\left[\widehat{\Delta}_{c,n}\right] &= \sum_{j=J_n+1}^\infty\mathbb{E}\left[\psi_j(T)\varphi_{\sigma_T^2}(T)\right]a_j 
\end{align*}

{\bf Variance Rate:} Now we bound the variance. Since all sums are finite, we can exchange the summations.
\begin{align*}
   \mathbb{V}\left[\sum_{j=0}^{J_n}\frac{1}{n}\sum_{i=1}^n  \psi_j(t_{i})\varphi_{\sigma_T^2}(t_i) a_j\right] &= \mathbb{V}\left[\frac{1}{n}\sum_{i=1}^n \sum_{j=0}^{J_n} \psi_j(t_{i})\varphi_{\sigma_T^2}(t_i) a_j\right] 
\end{align*}

By Assumption \ref{assum:cross_study_independence}, all observations are identically distributed and independent of all but at most $C_B$ other observations. Moreover, recall the fact that by Cauchy-Schwarz: $|\text{COV}(A,B)|\leq \mathbb{V}[A]$ for identically distributed $A,B$. Therefore we can bound the variance of the sum in terms of the individual variances:
\begin{align}\label{eq:varboundstart}
     \mathbb{V}\left[\frac{1}{n}\sum_{i=1}^n \sum_{j=0}^{J_n} \psi_j(t_{i})\varphi_{\sigma_T^2}(t_i) a_j\right]  &=  \frac{1}{n^2}\sum_{i=1}^n\sum_{k=1}^n \text{COV}\left(\sum_{j=0}^{J_n} \psi_j(t_{i})\varphi_{\sigma_T^2}(t_i) a_j,\sum_{j=0}^{J_n}\psi_j(t_{k})\varphi_{\sigma_T^2}(t_k) a_j\right)\\
     &= \frac{1}{n^2}\sum_{i=1}^n\sum_{k=1}^n \Lambda_{ik} \text{COV}\left(\sum_{j=0}^{J_n} \psi_j(t_{i})\varphi_{\sigma_T^2}(t_i) a_j,\sum_{j=0}^{J_n}\psi_j(t_{k})\varphi_{\sigma_T^2}(t_k) a_j\right)\\
     &\leq \frac{C_B}{n^2} \sum_{i=1}^n \mathbb{V}\left[\sum_{j=0}^{J_n} \psi_j(t_{i})\varphi_{\sigma_T^2}(t_i) a_j\right]\\
     &\leq \frac{C_B}{n}\mathbb{E}\left[\left(\sum_{j=0}^{J_n} \psi_j(T)\varphi_{\sigma_T^2}(T) a_j\right)^2\right]\\
     &\leq \frac{C_B}{n} \sup_{t\in\mathbb{R}}\left(\sum_{j=0}^{J_n} |\psi_j(t)\varphi_{\sigma_T^2}(t) a_j|\right)^2 \label{eq:varboundend}
\end{align}

By Lemma \ref{lem:bound_a_jpsi_j}, \begin{align*}
\sup_{t\in \mathbb{R}} \sum_{j=0}^{J_n} \left|\psi_j(t)\varphi_{\sigma_T^2}(t)a_j\right| &\leq \lambda_{J_n}^{-1} \frac{A_{c,\sigma_T,CV}}{(1-\lambda_1)\sqrt{2\sigma^2_T\pi}}
\end{align*}
where $A_{c,\sigma_T,CV}\equiv  \sqrt{\frac{2 CV}{\varphi_{\sigma_T^2}(CV)}} +\sqrt{\frac{2 CV}{c\varphi_{\sigma_T^2+1-c^{-2}}(CV/c)}}  $ is a finite constant.

Substituting, we obtain the variance rate:
\begin{equation}
   \mathbb{V}\left[\widehat{\Delta}_{c,n}\right] =   \mathbb{V}\left[\frac{1}{n}\sum_{i=1}^n \sum_{j=0}^{J_n} \psi_j(t_{i})\varphi_{\sigma_T^2}(t_i) a_j\right] \leq \frac{C_B}{n} \left(\lambda_{J_n}^{-1} \frac{A_{c,\sigma_T,CV}}{(1-\lambda_1)\sqrt{2\sigma^2_T\pi}}\right)^2  = \mathcal{O}\left(\lambda_{J_n}^{-2}n^{-1}\right)
\end{equation}

{\bf Bias Rate:} Next we bound the regularization bias:
\begin{align}\label{eq:basic_bias_expression}
    \Delta_c- \mathbb{E}\left[\widehat{\Delta}_{c,n}\right]  &= \sum_{j=J_n+1}^\infty\mathbb{E}\left[\psi_j(T)\varphi_{\sigma_T^2}(T)\right]a_j
\end{align}

By Lemma \ref{lem:moment_conversion}: 
\begin{align}
    \mathbb{E}\left[\psi_j(T)\varphi_{\sigma_T^2}(T)\right] = \eta_j\lambda_j\mathbb{E}\left[\chi_j(H)\varphi_{\sigma_T^2+1}(H)\right]
\end{align}

By Lemma \ref{lem:bound_coeffs}, $\left|\mathbb{E}\left[\chi_j(H)\varphi_{\sigma_T^2+1}(H)\right]\right| \leq \frac{1}{\sqrt{2(\sigma_T^2+1)\pi}}  $. By Lemma \ref{lem:integral_hermite}, $\sup_j |a_j|\lambda_j \leq A_{c,\sigma_T,CV} $. Applying these bounds:
\begin{align}\label{eq:bound1_bias}
    \left| \sum_{j=J_n+1}^\infty\eta_j\lambda_j\mathbb{E}\left[\chi_j(H)\varphi_{\sigma_T^2+1}(H)\right]a_j\right| &\leq \frac{A_{c,\sigma_T,CV}}{\sqrt{2(\sigma_T^2+1)\pi}} \sum_{j=J_n+1}^\infty \eta_j
\end{align}

Since this is a geometric series:
\begin{align}\label{eq:eta_geometric}
    \sum_{j=J_n+1}^\infty \eta_{j} = \eta_{J_n+1}\sum_{j=0}^\infty \eta_1^j =   \frac{\eta_{J_n+1}}{1-\eta_1} =\mathcal{O}\left(\eta_{J_n}\right)
\end{align}

Substituting Equation (\ref{eq:eta_geometric})
 into Equation (\ref{eq:bound1_bias}):
\begin{align}\label{eq:bias_bound_core}
     \left| \sum_{j=J_n+1}^\infty\eta_j\lambda_j\mathbb{E}\left[\chi_j(H)\varphi_{\sigma_T^2+1}(H)\right]a_j\right| &\leq \frac{A_{c,\sigma_T,CV}}{\sqrt{2(\sigma_T^2+1)\pi}}\frac{\eta_{J_n+1}}{1-\eta_1}  
\end{align}

Combining Equation (\ref{eq:bias_bound_core}) with Equation (\ref{eq:basic_bias_expression}) yields the rate for the bias:
\begin{align}
 \left|  \Delta_c- \mathbb{E}\left[\widehat{\Delta}_{c,n}\right] \right|  &\leq \frac{A_{c,\sigma_T,CV}}{\sqrt{2(\sigma_T^2+1)\pi}}\frac{\eta_{J_n+1}}{1-\eta_1}   = \mathcal{O}\left(\eta_{J_n}\right)
\end{align}

{\bf Optimizing the Rate of Convergence}
Finally we optimize the rate of convergence. Combining the bounds on bias and variance using Chebyshev's Inequality:
\begin{align*}
    \widehat{\Delta}_{c,n}-\Delta_{c} = \mathcal{O}_p\left(\eta_{J_n}+\lambda_{J_n}^{-1}n^{-1/2}\right)
\end{align*}
 The next task is to choose a growth rate for $J_n$ given $\sigma_T^2>0$ such that the order $\lambda_{J_n}^{-1}n^{-1/2}+\eta_{J_n}$ is balanced. To do this it is sufficient to set the orders of the two summands equal to each other. A sufficient condition for this is that for some $d>0$, $\frac{\eta_{J_n}}{\lambda_{J_n}^{-1}n^{-1/2}}\to d$. Recall the definitions of the singular values $\lambda_j$ and $\eta_j$ from Section \ref{sec:id}: $\eta_j = \left(\frac{1+\sigma_T^2-c^{-2}}{1+\sigma_T^2}\right)^{j/2}$ and $\lambda_j = \left(\frac{\sigma_T^2}{\sigma_T^2+1-c^{-2}}\right)^{j/2}$. Notice that we can rewrite $\eta_j$ as: $\eta_j = \lambda_j^{-1}\left(\frac{\sigma_T^2}{1+\sigma_T^2}\right)^{j/2}$. This makes their ratio: $\frac{\eta_{j}}{\lambda_{j}^{-1}n^{-1/2}} = \left(\frac{\sigma_T^2}{1+\sigma_T^2}\right)^{j/2}n^{1/2}$. So if the meta-analyst chooses $J_n$ such that for some $d>0$, $ n^{1/2}\left(\frac{\sigma_T^2}{\sigma_T^2+1}\right)^{J_n/2} \to d$, then: $\frac{\eta_{J_n}}{\lambda_{J_n}^{-1}n^{-1/2}}  \to d$. Thus, this choice makes the two terms $\eta_{J_n}$ and $n^{-1/2}\lambda_{J_n}^{-1}$ have the same order, which minimizes the order of the upper bound on their sum. Some algebra reveals the order:
\begin{align*}
     \left(\frac{1+\sigma_T^2-c^{-2}}{1+\sigma_T^2}\right)^{\log_{\frac{\sigma_T^2}{\sigma_T^2+1}}\left(n^{-1/2}\right)} 
     = n^{-q/2}
\end{align*}

\subsection{Proof of Lemma \ref{lem:pb}}\label{proof:pb}

By applying some algebra to Bayes' Rule we get: $f_T(t) = \frac{f_{T\mid R=1}(t)\text{Pr}(R=1) }{\text{Pr}(R=1|T=t)}=\frac{f_{T\mid R=1}(t)\mathbb{E}[ w_{\theta_0}(T)] }{w_{\theta_0}(t)}$.  Assumption \ref{assum:pblowerbound} ensures that all divisions by \(w_{\theta_0}\) are well-defined and that the denominator below is finite and strictly positive. For any bounded and measurable function $\gamma \: : \: \mathbb{R}\to\mathbb{R}$, 
\begin{align*}
\mathbb{E}[\gamma(T)] &= \int_{-\infty}^{\infty} \gamma(t)f_T(t)dt= \int_{-\infty}^\infty \gamma(t) \frac{f_{T\mid R=1}(t) }{w_{\theta_0}(t)}dt \mathbb{E}[ w_{\theta_0}(T)]= \mathbb{E}\left[\frac{\gamma(T) }{w_{\theta_0}(T)}\:\mid\: R=1\right] \mathbb{E}[ w_{\theta_0}(T)]
\end{align*}

We compute $\mathbb{E}[ w_{\theta_0}(T)]$ by taking a conditional expectation using the following trick:
\begin{align*}
\mathbb{E}\left[\frac{1}{w_{\theta_0}(T)}\mid R=1\right] &=  \int_{-\infty}^\infty \frac{1}{w_{\theta_0}(t)} f_{T\mid R=1}(t)dt=  \int_{-\infty}^\infty \frac{1}{w_{\theta_0}(t)}\frac{f_T(t) w_{\theta_0}(t)}{\mathbb{E}[ w_{\theta_0}(T)]}dt = \frac{1}{\mathbb{E}[ w_{\theta_0}(T)]}
\end{align*}

Set $\gamma(t)=\psi_j(t)\varphi_{\sigma_T^2}(t)$. This function is bounded by Lemma \ref{lem:bound_coeffs} so we can use the equations above to conclude: $
\mathbb{E}[\psi_j(T)\varphi_{\sigma_T^2}(T)]  = \frac{\mathbb{E}\left[\frac{\psi_j(T)\varphi_{\sigma_T^2}(T) }{w_{\theta_0}(T)}\:\mid\: R=1\right]  }{{\mathbb{E}\left[ \frac{1}{w_{\theta_0}(T)}\:\mid\: R=1\right]}}$

\subsection{Proof of Theorem \ref{thm:id_pb_general}}\label{proof:thm:id_pb_general}
%Audited 2/23/26 with meaningful cleanup.%audted 5/21/26

Consider the function $r_\theta(t) \equiv \frac{f_{T\mid R=1}(t)}{w_{\theta}(t)}$. By Assumption \ref{assum:pblowerbound}, $r_\theta(t)$ is always defined. Holding $\theta$ fixed, by Bayes' Rule: $$r_\theta(t) \equiv\frac{f_{T\mid R=1}(t)}{w_{\theta}(t)}=\frac{f_{T}(t)w_{\theta_0}(t)}{w_{\theta}(t)\int_{-\infty}^\infty f_{T}(s)w_{\theta_0}(s)ds }\propto f_T(t) \left(\frac{w_{\theta_0}(t)}{w_{\theta}(t)}\right) $$ Since the density $f_T(t)$ is the outcome of a convolution of $\Pi_0$ with a Gaussian, $f_T\in C^{\infty}$ and $f_T(t)>0$ for all $t\in\mathbb{R}$. Therefore, by the Product Rule,  $r_\theta(t)\in C^{\infty}$ if and only if $\frac{w_{\theta_0}(t)}{w_{\theta}(t)}\in  C^{\infty}$. If $\theta=\theta_0$, then $\frac{w_{\theta_0}(t)}{w_{\theta}(t)}=1$ and  $r_\theta(t) \in C^{\infty}$. If $\theta\neq \theta_0$, by Assumption \ref{assum:wnonsmooth}, $\frac{w_{\theta_0}(t)}{w_{\theta}(t)}\notin C^{\infty}$ because it has a jump discontinuity in itself or one of its derivatives. So $r_{\theta}(t)\in C^{\infty}$  if and only if $\theta=\theta_0$. Since $r_{\theta}(t)$ depends only on the distribution of the observed variable $T\mid R=1$, $\theta_0$ is identified. Corollary \ref{cor:id_pb} expresses $\Delta_c$ in terms of the distribution of published $t$-scores and $\theta_0$. Since $\theta_0$ is identified, so is $\Delta_c$.

\subsection{Proof of Theorem \ref{thm:consistency_withpb}}\label{proof:consistency_withpb}
%audited 4/6/26 % Audited 5/22/26

 Throughout the proof, expectations involving sample observations $t_i$ are understood to be conditional on publication, i.e. on $R_i=1$. Lemma \ref{lem:linearization}  states that:
    \begin{align}\label{eq:pb_esterror_withtilde}
    \widehat{\Delta}_{c,n}^{pb}-\widetilde{\Delta}_{c,n}
    &= \frac{1}{n}\sum_{i=1}^n m_n^0 (t_i) -\mathbb{E}\left[m_n^0 (T)\mid R=1\right] +\mathcal{O}_p\left(\lambda_{J_n}^{-1}\left(n^{-1}\epsilon_n^{-1}+\epsilon_n^2+n^{-1/2}\right)\right)
\end{align}
\noindent where the influence function $m_n^0$ is defined in Appendix \ref{app:inference_definitions}.

\vskip 0.1in
{\bf\noindent Step 1: Variance}

By Assumption \ref{assum:cross_study_independence}, the $m_n^0 (t_i)$ are identically distributed and each depends on at most $C_B$ others. Moreover, Lemma \ref{lem:linearization} states that $\mathbb{V}\left[m_n^0 (T)\mid R=1\right]=\mathcal{O}\left(\epsilon_n^{-1}\lambda_{J_n}^{-2}\right)$. Combining these:
\begin{align*}
    \mathbb{V}\left[ \frac{1}{n}\sum_{i=1}^n m_n^0 (t_i) \right] \leq \frac{C_B}{n}\mathbb{V}\left[m_n^0 (T)\mid R=1\right] =\mathcal{O}\left(n^{-1}\epsilon_n^{-1}\lambda_{J_n}^{-2}\right)
\end{align*}

Applying Chebyshev's Inequality to Equation (\ref{eq:pb_esterror_withtilde}) and ignoring dominated terms:
\begin{align}\label{eq:pb_minus_tilde}
     \widehat{\Delta}_{c,n}^{pb}-\widetilde{\Delta}_{c,n} &= \mathcal{O}_p\left(n^{-1/2}\epsilon_n^{-1/2}\lambda_{J_n}^{-1}+\epsilon_n^2\lambda_{J_n}^{-1}\right)
\end{align}

\vskip 0.1in
{\bf\noindent Step 2: Rate}

Lemma \ref{lem:linearization} already states that: 
\begin{align}\label{eq:rate_bias_pb}
 \widetilde{\Delta}_{c,n}-\Delta_c &= \mathcal{O}\left(\lambda_{J_n}^{-1}\epsilon_n+\eta_{J_n}\right)
\end{align}

Next we combine Equations (\ref{eq:rate_bias_pb}) and (\ref{eq:pb_minus_tilde}) with the triangle inequality. Since $\epsilon_n\to 0$ the $\epsilon_n^2\lambda_{J_n}^{-1}$ term is dominated by $\epsilon_n\lambda_{J_n}^{-1}$.
\begin{align}
  \widehat{\Delta}_{c,n}^{pb} - {\Delta}_{c}= \mathcal{O}_p\left(\lambda_{J_n}^{-1}\epsilon_n+\eta_{J_n} +n^{-1/2}\epsilon_n^{-1/2}\lambda_{J_n}^{-1}\right)
\end{align}

When we optimize by setting $\epsilon_n \sim n^{-1/3}$ we obtain the desired rate:
\begin{align}
    \widehat{\Delta}_{c,n}^{pb}-{\Delta}_{c} = \mathcal{O}_p\left(\lambda_{J_n}^{-1}n^{-1/3}+\eta_{J_n} \right)
\end{align}

 If the meta-analyst chooses $J_n$ such that for some $d>0$, $ n^{1/3}\left(\frac{\sigma_T^2}{\sigma_T^2+1}\right)^{J_n/2} \to d$ then again by an identical argument to the one in the proof of Theorem \ref{thm:consistency_nopb}, with $\frac{1}{2}$ exchanged for $\frac{1}{3}$ we obtain
$  \widehat{\Delta}_{c,n}^{pb}-\Delta_c = \mathcal{O}_p\left(n^{-q/3}\right)$ where $q\equiv \left. {\log\left(\frac{1+\sigma_T^2}{1+\sigma_T^2-c^{-2}}\right)}\middle/{\log\left(\frac{\sigma_T^2+1}{\sigma_T^2}\right)}\right. $

\subsection{Proof of Lemma \ref{lem:linearization}}\label{proof:linearization}
%audited 4/6/26 %audited 5/22/26

The proof proceeds in several steps. First, rewrite the estimator in terms of the shorthand functions $S_n$ and $W$ defined below (and also in Appendix \ref{app:inference_definitions}). Second, linearize the publication-bias weight $W$ around $(\theta_0,F(1.96))$ and the estimator $\widehat{\theta}_n^{-1}$ around $\widetilde{\theta}_n^{-1}$. Third, collect the leading sample-average term and define the oracle triangular-array influence function $m_n^0$. Fourth, bound the remainder, the size and variance of $m_n^0$, and the centering bias $\widetilde{\Delta}_{c,n}-\Delta_c$. Finally, we bound the squared moments and supremum of the influence function and the rate of the bias.

\vskip 0.1in
{\noindent\bf Step 1: Decompose Estimator}

 Start with the definition of the estimator and rearrange the sums:
\begin{align}\label{eq:deltahat_pb_sums_switched}
   \widehat{\Delta}_{c,n}^{pb} &= \frac{1}{n}\sum_{i=1}^n \left(\sum_{j=0}^{J_n}a_j \psi_j(t_i)\varphi_{\sigma_T^2}(t_i)\right) \left(\frac{ \frac{1}{w_{{\widehat{\theta}_n}}(t_i)}}{ \frac{1}{n}\sum_{k=1}^n \frac{1}{w_{{\widehat{\theta}_n}}(t_k)}}\right)
\end{align}

Let $F(t),\widehat{F}_n(t)$ denote the CDF and empirical CDF of $|T|$ conditional on publication $R=1$.  By Equation (\ref{eq:pb_caliper_form}): 
\begin{align}\label{eq:caliper_restated}
w_{\widehat{\theta}_n}(t_i) &= \mathbf{1}\left\{|t_i|\geq 1.96\right\}+\widehat{\theta}_n\mathbf{1}\left\{|t_i|< 1.96\right\}
\end{align}

Substituting (\ref{eq:caliper_restated}) into the ratio of weights on the right-hand side of (\ref{eq:deltahat_pb_sums_switched}) and rearranging with algebra:
\begin{align}
       \frac{ \frac{1}{w_{{\widehat{\theta}_n}}(t_i)}}{ \frac{1}{n}\sum_{k=1}^n \frac{1}{w_{{\widehat{\theta}_n}}(t_k)}} &=  \frac{1+\left(\widehat{\theta}_n^{-1}-1\right)\mathbf{1}\left\{|t_i|< 1.96\right\}}{1+(\widehat{\theta}_n^{-1}-1)\widehat{F}_n(1.96)} \label{eq:rearrange_pb_weights}
\end{align}

Substituting Equation (\ref{eq:rearrange_pb_weights}) into Equation (\ref{eq:deltahat_pb_sums_switched}):
\begin{align}\label{eq:deltahat_phin}
     \widehat{\Delta}_{c,n}^{pb} &= \frac{1}{n}\sum_{i=1}^n \left(\underbrace{\sum_{j=0}^{J_n}a_j \psi_j(t_i)\varphi_{\sigma_T^2}(t_i)}_{S_n(t_i)}\right) \left(\underbrace{\frac{1+\left(\widehat{\theta}_n^{-1}-1\right)\mathbf{1}\left\{|t_i|< 1.96\right\}}{1+(\widehat{\theta}_n^{-1}-1)\widehat{F}_n(1.96)}}_{W(t_i;\widehat{\theta}_n,\widehat{F}_n(1.96))} \right) 
\end{align}

It will be convenient to restate Equation (\ref{eq:deltahat_phin}) using the following shorthand. Let $W(t_i;\widehat{\theta}_n,\widehat{F}_n(1.96))$ be the weighting function that removes publication bias and let $S_n(t_i)$ be the deconvolution part.
\begin{align}\label{eq:deltahat_shorthand}
    \widehat{\Delta}_{c,n}^{pb} &= \frac{1}{n}\sum_{i=1}^n S_n(t_i)W(t_i;\widehat{\theta}_n,\widehat{F}_n(1.96))\\
      S_n(t) &\equiv \sum_{j=0}^{J_n} a_j\psi_j(t)\varphi_{\sigma_T^2}(t)\\
W(t;\theta,p) &\equiv \frac{1+\left({\theta}^{-1}-1\right)\mathbf{1}\left\{|t|< 1.96\right\}}{1+({\theta}^{-1}-1)p}
\end{align}

\vskip 0.1in
{\noindent\bf Step 2: Linearization}

Next we linearize the weight function $W$ about the true parameters. Lemma \ref{lem:linearize_w} states that:
   \begin{align*}
     & \sup_{t\in\mathbb{R}}  \left| W(t;\widehat{\theta}_n,\widehat{F}_n(1.96))-W(t;{\theta}_0,{F}(1.96))-\left(\widehat{\theta}_n^{-1}-{\theta}_0^{-1}\right) \frac{\mathbf{1}\left\{|t|< 1.96\right\}-{F}(1.96)}{\left(1+{F}(1.96)({\theta}_0^{-1}-1)\right)^2}  \right|\\
      &=  \mathcal{O}_p\left(\left(\widehat{\theta}_n^{-1}-\theta_0^{-1}\right)^2+n^{-1/2}\right) 
    \end{align*}
 
The first-order expansion above retains only the perturbation in $\widehat{\theta}_n^{-1}$; the effect of $\widehat{F}_n(1.96)-F(1.96)$ is absorbed into the $\mathcal{O}_p(n^{-1/2})$ remainder by Lemma \ref{lem:linearize_w}. Substituting this into Equation (\ref{eq:deltahat_shorthand}):
\begin{align*}
      \widehat{\Delta}_{c,n}^{pb} &= \frac{1}{n}\sum_{i=1}^n S_n(t_i)W(t_i;{\theta}_0,{F}(1.96))+\left(\widehat{\theta}_n^{-1}-\theta_0^{-1}\right)\frac{1}{n}\sum_{i=1}^n S_n(t_i)\left(\frac{\mathbf{1}\left\{|t_i|< 1.96\right\}-{F}(1.96)}{\left(1+{F}(1.96)({\theta}_0^{-1}-1)\right)^2}\right)\\
      &+\quad \mathcal{O}_p\left(\left(\left(\widehat{\theta}_n^{-1}-\theta_0^{-1}\right)^2+n^{-1/2}\right)\frac{1}{n}\sum_{i=1}^n |S_n(t_i)|\right)
\end{align*}

 Lemma \ref{lem:bound_a_jpsi_j} guarantees that $\sup_t|S_n(t)|=\mathcal{O}\left(\lambda_{J_n}^{-1}\right)$ and therefore $\frac{1}{n}\sum_{i=1}^n |S_n(t_i)| = \mathcal{O}_p\left(\lambda_{J_n}^{-1}\right)$. Moreover,  Assumption \ref{assum:pblowerbound} says that $\theta_0 \in [\overline{M}^{-1},1]$ and therefore: $\sup_t|W(t;\theta_0,F(1.96))| \leq  \overline{M}^2$. By Lemma \ref{lem:theta_estimation}, $\left(\widehat{\theta}_n^{-1}-\theta_0^{-1}\right)^2 = \mathcal{O}_p\left(\epsilon_n^2+\epsilon_n^{-1}n^{-1}\right)$. So the remainder rates simplify to:
\begin{align}
      \widehat{\Delta}_{c,n}^{pb} &= \frac{1}{n}\sum_{i=1}^n S_n(t_i)W(t_i;{\theta}_0,{F}(1.96))+\left(\widehat{\theta}_n^{-1}-\theta_0^{-1}\right)\underbrace{\frac{1}{n}\sum_{i=1}^n S_n(t_i)\left(\frac{\mathbf{1}\left\{|t_i|< 1.96\right\}-{F}(1.96)}{\left(1+{F}(1.96)({\theta}_0^{-1}-1)\right)^2}\right)}_{\widetilde{Q}_n}\label{eq:esterror_noQn}\\
      &+\quad \mathcal{O}_p\left(\lambda_{J_n}^{-1}\left(\widehat{\theta}_n^{-1}-{\theta}_0^{-1}\right)^2+\lambda_{J_n}^{-1}n^{-1/2}\right)
\end{align}

Next we will study the sum $\widetilde{Q}_n$ in line (\ref{eq:esterror_noQn}). The deterministic sequence $Q_n$ is defined below as the expectation of $\widetilde{Q}_n$. By Assumption \ref{assum:cross_study_independence}, the $t_i$ all share a marginal distribution, so $Q_n$ can be written as the expectation of a sample average.
\begin{align*}
 Q_n &\equiv \mathbb{E}\left[ S_n(T) \frac{\mathbf{1}\left\{|T|< 1.96\right\}-{F}(1.96)}{\left(1+{F}(1.96)({\theta}_0^{-1}-1)\right)^2} \mid R=1\right]\\
    &= \mathbb{E}\left[ \frac{1}{n}\sum_{i=1}^n S_n(t_i)\left(\frac{\mathbf{1}\left\{|t_i|< 1.96\right\}-{F}(1.96)}{\left(1+{F}(1.96)({\theta}_0^{-1}-1)\right)^2}\right) \mid R=1\right]\\
     &=\mathbb{E}\left[ \widetilde{Q}_n \mid R=1\right]
\end{align*}
\noindent Moreover, by Lemma \ref{lem:bound_a_jpsi_j},  $\sup_{t\in\mathbb{R}}|S_n(t)| = \mathcal{O}\left(\lambda_{J_n}^{-1}\right)$ and therefore:
\begin{align}\label{eq:Q_n_bound}
 |Q_n|\leq \frac{\sup_{t\in\mathbb{R}}|S_n(t)|}{(1+F(1.96)(\theta_0^{-1}-1))^2}=\mathcal{O}\left(\lambda_{J_n}^{-1}\right)   
\end{align}

Next we show that this sample average concentrates around its expectation $Q_n$ at rate $\mathcal{O}_p(\lambda_{J_n}^{-1}n^{-1/2})$. By Lemma \ref{lem:bound_a_jpsi_j}, the summands have variance upper-bounded by $\lambda_{J_n}^{-2}$ times a constant because they are almost surely upper bounded in absolute value. By Assumption \ref{assum:cross_study_independence}, all summands are independent of all but at most $C_B$ other summands. So by Chebyshev's Inequality:
\begin{align}\label{eq:Q_n_convergence}
  \widetilde{Q}_n = Q_n + \mathcal{O}_p\left(\lambda_{J_n}^{-1}n^{-1/2}\right)
\end{align}

Substituting Equation (\ref{eq:Q_n_convergence}) into line (\ref{eq:esterror_noQn}):
\begin{align}\label{eq:Deltahat_pb_finished}
     \widehat{\Delta}_{c,n}^{pb} &= \frac{1}{n}\sum_{i=1}^n S_n(t_i)W(t_i;{\theta}_0,{F}(1.96))
     +Q_n\left(\widehat{\theta}_n^{-1}-{\theta}_0^{-1}\right) + \mathcal{O}_p\left(\lambda_{J_n}^{-1}\left(\widehat{\theta}_n^{-1}-{\theta}_0^{-1}\right)^2+\lambda_{J_n}^{-1}n^{-1/2}\right)
\end{align}

We will center the estimation error around a deterministic sequence $\widetilde{\Delta}_{c,n}$ rather than $\Delta_c$ because $\widehat{\Delta}_{c,n}^{pb}$ contains both spectral cutoff and smoothing bias. The sequence $ \widetilde{\Delta}_{c,n}$ is defined as:\\ $\widetilde{\Delta}_{c,n} \equiv \sum_{j=0}^{J_n}\mathbb{E}\left[\psi_j(T)\varphi_{\sigma_T^2}(T)\right] a_j   +Q_n\left(\widetilde{\theta}_n^{-1}-{\theta}_0^{-1}\right) $. The first sum is $\Delta_c$ up to spectral cutoff bias and the second term is smoothing bias coming from the estimation of $\theta_0$. Here $  \widetilde{\theta}_n \equiv \frac{F(1.96)-F(1.96-\epsilon_n)}{F(1.96+\epsilon_n)-F(1.96)}$ is the deterministic centering sequence for $\widehat{\theta}_n$. Subtracting $ \widetilde{\Delta}_{c,n}$ from both sides of (\ref{eq:Deltahat_pb_finished}):
\begin{align}\label{eq:esterror_without_inf_functions}
    \widehat{\Delta}_{c,n}^{pb} - \widetilde{\Delta}_{c,n}  &=\frac{1}{n}\sum_{i=1}^n S_n(t_i) W(t_i;{\theta}_0,{F}(1.96)) - \sum_{j=0}^{J_n} \mathbb{E}\left[\psi_j(T)\varphi_{\sigma_T^2}(T)\right] a_j\\
     &\qquad +Q_n\left(\widehat{\theta}_n^{-1}-\widetilde{\theta}_n^{-1}\right)+ \mathcal{O}_p\left(\lambda_{J_n}^{-1}\left(\widehat{\theta}_n^{-1}-{\theta}_0^{-1}\right)^2+\lambda_{J_n}^{-1}n^{-1/2}\right)
\end{align}

\vskip 0.1in
{\noindent\bf Step 3: Influence-Function Representation}

Now we will rewrite (\ref{eq:esterror_without_inf_functions}) in terms of influence functions and their expectations. Lemma \ref{lem:theta_estimation} writes the estimation error of $ \widehat{\theta}_n^{-1}$ in terms of its influence function:
\begin{align}\label{eq:recall:lem:theta_estimation}
    \widehat{\theta}_n^{-1} -\widetilde{\theta}^{-1}_n = \frac{1}{n}\sum_{i=1}^n X_n(t_i;B_{\epsilon_n}^+,B_{\epsilon_n}^{-}) - \mathbb{E}\left[X_n(T;B_{\epsilon_n}^+,B_{\epsilon_n}^{-}) \mid R=1\right]+\mathcal{O}_p\left(n^{-1}\epsilon_n^{-1}\right)
\end{align}
\noindent where the influence function ${X}_{n} \: :\:\mathbb{R}^3\to \mathbb{R}$ (the influence function for $\widehat{\theta}_n^{-1}$) is defined as:
\begin{align}
     {X}_{n}(t;b^+,b^-) &\equiv \frac{1}{b^-}\mathbf{1}\left\{|t|\in (1.96,1.96+\epsilon_n]\right\} -\frac{b^+}{\left(b^-\right)^2}\mathbf{1}\left\{|t|\in (1.96-\epsilon_n,1.96]\right\}
\end{align}
\noindent and the sequences $B_{\epsilon_n}^+,B_{\epsilon_n}^{-}$ are defined as the probabilities of $|T|$ being just above or just below the critical value:
\begin{align}
     {B}_{\epsilon_n}^+ &\equiv {F}(1.96+\epsilon_n)-{F}(1.96)\quad \text{and} \quad {B}_{\epsilon_n}^- \equiv {F}(1.96)-{F}(1.96-\epsilon_n)
\end{align}

Define the oracle influence function ${m}_n^0(t)\: :\: \mathbb{R}\to \mathbb{R} $ as:
\begin{align}\label{eq:def_m0_innproof}
      {m}_n^0(t) &\equiv  S_n(t)W(t;{\theta}_0,{F}(1.96)) +{Q}_n{X}_{n}(t;{B}_{\epsilon_n}^+,{B}_{\epsilon_n} ^-)
\end{align}

\noindent Substituting  lines (\ref{eq:recall:lem:theta_estimation})-(\ref{eq:def_m0_innproof}) into line (\ref{eq:esterror_without_inf_functions}), we see that $\widehat{\Delta}_{c,n}^{pb} - \widetilde{\Delta}_{c,n}$ can be rewritten in terms of the influence functions $m_n^0$: 
\begin{align}\label{eq:Delta_minus_pb_nom0expectations}
    \widehat{\Delta}_{c,n}^{pb} - \widetilde{\Delta}_{c,n} &= \frac{1}{n}\sum_{i=1}^n m_n^0(t_i) - \sum_{j=0}^{J_n} \mathbb{E}\left[\psi_j(T)\varphi_{\sigma_T^2}(T)\right] a_j - Q_n \mathbb{E}\left[X_n(T;B_{\epsilon_n}^+,B_{\epsilon_n}^{-})\mid R=1\right]\\
    &+\mathcal{O}_p\left(Q_nn^{-1}\epsilon_n^{-1}+\lambda_{J_n}^{-1}\left(\widehat{\theta}_n^{-1}-{\theta}_0^{-1}\right)^2+\lambda_{J_n}^{-1}n^{-1/2}\right)
\end{align}

\noindent {\bf Centering } 

Next we will center the estimation error on the expectation of the influence functions. Taking the conditional expectation of the influence function: \begin{align}
    \mathbb{E}\left[m^0_n(T)\mid R=1\right] &= \sum_{j=0}^{J_n} a_j\mathbb{E}\left[\psi_j(T)\varphi_{\sigma_T^2}(T)  W(T;\theta_0,F(1.96))\mid R=1\right] \label{eq:m0_exp} \\
    &\qquad +Q_n \mathbb{E}\left[X_n(T;B_{\epsilon_n}^+,B_{\epsilon_n}^{-})\mid R=1\right]
\end{align}

By definition, $W(T;\theta_0,F(1.96)) =\frac{1+\left({\theta_0}^{-1}-1\right)\mathbf{1}\left\{|T|< 1.96\right\}}{1+({\theta_0}^{-1}-1)F(1.96)}$. By Lemma \ref{lem:pb}:
\begin{align*}
 \mathbb{E}\left[\psi_j(T)\varphi_{\sigma_T^2}(T)  W(T;\theta_0,F(1.96))\mid R=1\right]&= \mathbb{E}\left[  \psi_j(T)\varphi_{\sigma_T^2}(T)  \left(\frac{1+\left({\theta}_0^{-1}-1\right)\mathbf{1}\left\{|T|< 1.96\right\}}{1+({\theta}_0^{-1}-1){F}(1.96)} \right)\mid R=1\right]\\ 
 &=  \mathbb{E}\left[\psi_j(T)\varphi_{\sigma_T^2}(T)\right]
\end{align*}

Substituting this into (\ref{eq:m0_exp}), the expectation simplifies to
\begin{align}\label{eq:expectation_m0}
    \mathbb{E}\left[m^0_n(T)\mid R=1\right] &=\sum_{j=0}^{J_n}\mathbb{E}\left[\psi_j(T)\varphi_{\sigma_T^2}(T)\right] a_j + Q_n  \mathbb{E}\left[X_n(T;B_{\epsilon_n}^+,B_{\epsilon_n}^{-})\mid R=1\right]
\end{align}

Substituting Equation (\ref{eq:expectation_m0}) into line (\ref{eq:Delta_minus_pb_nom0expectations}), we have the desired linearization:
\begin{align}
    \widehat{\Delta}_{c,n}^{pb} - \widetilde{\Delta}_{c,n} &= \frac{1}{n}\sum_{i=1}^n m_n^0(t_i)-  \mathbb{E}\left[ m_n^0(T)\mid R=1\right]
    \\&\qquad +\mathcal{O}_p\left(Q_nn^{-1}\epsilon_n^{-1}+\lambda_{J_n}^{-1}\left(\widehat{\theta}_n^{-1}-{\theta}_0^{-1}\right)^2+\lambda_{J_n}^{-1}n^{-1/2}\right)\label{eq:remainder_line}
\end{align}

Finally we simplify the rate of the remainder in line (\ref{eq:remainder_line}). Lemma \ref{lem:theta_estimation} states that $(\widehat{\theta}_n^{-1}-{\theta}_0^{-1}) = \mathcal{O}_p\left(\epsilon_n+\epsilon_n^{-1/2}n^{-1/2}\right)$. Furthermore, $Q_n = \mathcal{O}\left(\lambda_{J_n}^{-1}\right)$ by Equation (\ref{eq:Q_n_bound}).  Therefore:
\begin{align*}
    \widehat{\Delta}_{c,n}^{pb} - \widetilde{\Delta}_{c,n} &= \frac{1}{n}\sum_{i=1}^n m_n^0(t_i)-   \mathbb{E}\left[ m_n^0(T)\mid R=1\right]  +\mathcal{O}_p\left(\lambda_{J_n}^{-1}\left(n^{-1}\epsilon_n^{-1}+\epsilon_n^2+n^{-1/2}\right)\right)
\end{align*}

\vskip 0.1in
{\noindent\bf Step 5: Second moment of influence function}

Recall the identity $\mathbb{E}[(A+B)^2]\leq 2\mathbb{E}[A^2]+2\mathbb{E}[B^2]$. We can use this identity to bound the second moment of the influence function:
\begin{align}\label{eq:variance_by_sum}
  &\mathbb{E}\left[m_n^0 (T)^2\mid R=1\right]  \\
  &\quad \leq 2\mathbb{E}\left[S_n(T)^2W(T;{\theta}_0,{F}(1.96))^2\mid R=1\right] + 2{Q}_n^2\mathbb{E}\left[{X}_{n}(T;{B}_{\epsilon_n}^+,{B}_{\epsilon_n} ^-)^2\mid R=1\right]
\end{align}

We bound the two second moments one at a time. By Lemma \ref{lem:bound_a_jpsi_j}, $\sup_{t\in\mathbb{R}}|S_n(t)|=\mathcal{O}\left(\lambda_{J_n}^{-1}\right)$. Similarly, $\sup_{t\in\mathbb{R}}\left|W(t;{\theta}_0,{F}(1.96)) \right|\leq \frac{1+\theta_0^{-1}}{1+({\theta_0}^{-1}-1)F(1.96)}\leq \frac{1+\overline{M}}{\overline{M}^{-1}}$ because $\theta_0 \in [\overline{M}^{-1},1]$ by Assumption \ref{assum:pblowerbound}. So $ \sup_{t\in\mathbb{R}}\left|W(t;{\theta}_0,{F}(1.96)) \right|< \infty$. Therefore,  $\mathbb{E}\left[ S_n(T)^2W(T;{\theta}_0,{F}(1.96))^2\mid R=1\right]=\mathcal{O}\left(\lambda_{J_n}^{-2}\right)$.

Next we bound the second squared expectation term on the right-hand side of (\ref{eq:variance_by_sum}). First, notice that $|Q_n|=\mathcal{O}\left(\lambda_{J_n}^{-1}\right)$ by Lemma \ref{lem:bound_a_jpsi_j}.  By Claim 4 of Lemma \ref{lem:theta_estimation},
\begin{align}
     \mathbb{E}\left[{X}_{n}(T;{B}_{\epsilon_n}^+,{B}_{\epsilon_n} ^-)^2\mid R=1\right] &= \mathcal{O}\left(\epsilon_n^{-1}\right)
\end{align}

Substituting these two individual variance bounds into Equation (\ref{eq:variance_by_sum}) yields:
\begin{align}
    \mathbb{E}\left[m_n^0 (T)^2\mid R=1\right] &= \mathcal{O}\left(\lambda_{J_n}^{-2}\epsilon_n^{-1}\right)
\end{align}

\vskip 0.1in
{\bf\noindent Step 6: Supremum of $m^0_n(t)$}

 To bound the influence function, recall the following bound on its components:
 \begin{align*}
    |m_n^0(t)| &\leq |S_n(t)W(t;{\theta}_0,{F}(1.96))| +|{Q}_n{X}_{n}(t;{B}_{\epsilon_n}^+,{B}_{\epsilon_n} ^-)| &\text{ Definition}\\
     \sup_{t\in\mathbb{R}}|S_n(t)| &=\mathcal{O}\left(\lambda_{J_n}^{-1}\right)  &\text{by Lemma \ref{lem:bound_a_jpsi_j}}\\
      Q_n &= \mathcal{O}\left(\lambda_{J_n}^{-1}\right) &\text{Equation (\ref{eq:Q_n_bound})}\\
      \sup_{t\in\mathbb{R}}|{X}_{n}(t;{B}_{\epsilon_n}^+,{B}_{\epsilon_n} ^-)|&\leq \frac{1}{{B}_{\epsilon_n} ^-}+\frac{{B}_{\epsilon_n}^+}{({B}_{\epsilon_n} ^-)^2}=\mathcal{O}\left(\epsilon_n^{-1}\right)&\text{Lemma \ref{lem:Bbounds}}\\
   \sup_{t\in\mathbb{R}}|W(t;{\theta}_0,{F}(1.96))| &\leq\frac{1+\left({\theta}_0^{-1}+1\right)}{1+({\theta}_0^{-1}-1)F(1.96)}=\mathcal{O}\left(1\right) 
 \end{align*}
Combining the above bounds yields:
$$\sup_n\sup_{t\in\mathbb{R}}\epsilon_n\lambda_{J_n}|m_n^0(t)|<\infty$$ 

\vskip 0.1in
{\bf\noindent Step 7: Bias of the Centering Sequence}

\noindent Finally, we show that $ \widetilde{\Delta}_{c,n}-\Delta_c$ converges.  Theorem \ref{thm:id} states that $\Delta_c = \sum_{j=0}^\infty \mathbb{E}\left[\psi_j(T)\varphi_{\sigma_T^2}(T)\right] a_j $. Subtracting this from the definition $\widetilde{\Delta}_{c,n} \equiv \sum_{j=0}^{J_n}\mathbb{E}\left[\psi_j(T)\varphi_{\sigma_T^2}(T)\right] a_j   +Q_n\left(\widetilde{\theta}_n^{-1}-{\theta}_0^{-1}\right) $ yields:
 \begin{align*}
 \widetilde{\Delta}_{c,n}-\Delta_c &= -\sum_{j=J_n+1}^{\infty} \mathbb{E}\left[\psi_j(T)\varphi_{\sigma_T^2}(T)\right] a_j +  \left(\widetilde{\theta}_n^{-1}-{\theta}_0^{-1}\right)Q_n
\end{align*}

By Equation (\ref{eq:bias_bound_core}) and Lemma \ref{lem:moment_conversion}, $\sum_{j=J_n+1}^{\infty} \mathbb{E}\left[\psi_j(T)\varphi_{\sigma_T^2}(T)\right] a_j =\mathcal{O}\left(\eta_{J_n}\right)$. By Lemma \ref{lem:theta_estimation},   $\widetilde{\theta}_n^{-1}-{\theta}_0^{-1} = \mathcal{O}\left(\epsilon_n\right)$. By Lemma \ref{lem:bound_a_jpsi_j}, $Q_n = \mathcal{O}\left(\lambda_{J_n}^{-1}\right)$.  Therefore:
\begin{align}
     \widetilde{\Delta}_{c,n}-\Delta_c &= \mathcal{O}\left(\eta_{J_n}+\lambda_{J_n}^{-1}\epsilon_n\right)
\end{align}

\subsection{Proof of Theorem \ref{thm:clt}}\label{proof:clt}
%audited 4/6/26 %audited 5/22/26

The proof strategy is the following: Lemma \ref{lem:linearization} shows that the estimator equals a centered sample average plus a negligible remainder; we then apply the Central Limit Theorem from \cite{JansonPratelliRigo2024} to the leading term. All expectations and variances below are with respect to the distribution of the reported
$t$-scores, i.e. conditional on $R_i=1$ for all $i$.

For convenience, define the centered triangular array $Y_{n} \equiv m_n^0(T)-\mathbb{E}\left[m_n^0(T)\mid R=1\right]$ and the sample version $Y_{ni} \equiv m_n^0(t_i)-\mathbb{E}\left[m_n^0(T)\mid R=1\right]$. Lemma \ref{lem:linearization} guarantees that: $$     \widehat{\Delta}_{c,n}^{pb}-\widetilde{\Delta}_{c,n}
    = \frac{1}{n}\sum_{i=1}^n Y_{ni} +\mathcal{O}_p\left(\lambda_{J_n}^{-1}\left(n^{-1}\epsilon_n^{-1}+\epsilon_n^2+n^{-1/2}\right)\right)$$
 Assumption \ref{assum:varnotsmall} guarantees that $\liminf_{n\to \infty}n\epsilon_n\lambda_{J_n}^2\mathbb{V}\left[\frac{1}{n}\sum_{i=1}^n m_{n}^0(t_i)\right] >0$. Notice that centering does not change the variance: $\mathbb{V}\left[\frac{1}{n}\sum_{i=1}^n m_{n}^0(t_i)\right]=\mathbb{V}\left[\frac{1}{n}\sum_{i=1}^n Y_{ni}\right]$. Thus, the remainder term is dominated and therefore:\begin{align}\label{eq:equivalency_Yn}
    \frac{    \widehat{\Delta}_{c,n}^{pb}-\widetilde{\Delta}_{c,n}}{\sqrt{\mathbb{V}\left[\frac{1}{n}\sum_{i=1}^n m_{n}^0(t_i)\right]}} = \frac{\frac{1}{n}\sum_{i=1}^n Y_{ni}}{\sqrt{\mathbb{V}\left[\frac{1}{n}\sum_{i=1}^n Y_{ni}\right]}}+o_p(1)
\end{align}
\noindent So it is sufficient to show that $\frac{\frac{1}{n}\sum_{i=1}^n Y_{ni}}{\sqrt{\mathbb{V}\left[\frac{1}{n}\sum_{i=1}^n Y_{ni}\right]}}$ is asymptotically normal.

We want to invoke the CLT from \cite{JansonPratelliRigo2024}. The theorem requires $m$-dependence. After relabeling the observations so that $t$-scores from the same study are contiguous,
Assumption \ref{assum:cross_study_independence} implies that the triangular array
$\{Y_{ni}\}_{i=1}^n$ is $m$-dependent for some fixed integer
$m \leq \lceil C_B\rceil$.

Let $\sigma_n^2 \equiv \mathbb{V}\left[\frac{1}{n}\sum_{i=1}^n Y_{ni}\right]$. Since $Y_{ni}$ are centered with finite moments, the only other requirement of their theorem that we need to check is their version of the Lindeberg Condition:
\begin{align}\label{eq:Lindeberg}
    \frac{1}{\sigma_n^2}\sum_{i=1}^n \mathbb{E}\left[(Y_{ni}/n)^2\mathbf{1}\left\{|Y_{ni}|/n>\nu \sigma_n\right\}\right]\to 0 \quad \forall \nu>0
\end{align}

Notice that $(Y_{ni}/n)^2\mathbf{1}\left\{|Y_{ni}|/n>\nu\sigma_n\right\}\leq (Y_{ni}/n)^4\nu^{-2}\sigma_n^{-2} $. Moreover, Assumption \ref{assum:cross_study_independence} guarantees that the $Y_{ni}$ are identically distributed. So the following Lyapunov condition  is sufficient:
\begin{align}
    \frac{\mathbb{E}\left[Y_{n}^4\right]}{n^3\sigma_n^4} \to 0 \implies (\ref{eq:Lindeberg})
\end{align}

By the triangle inequality, $|Y_n|\leq \sup_{t\in\mathbb{R}}2|m_n^0(t)|$. Lemma \ref{lem:linearization} states that:
$\sup_n\sup_{t\in\mathbb{R}}\epsilon_n\lambda_{J_n}|m_n^0(t)|<\infty$. So the following condition is sufficient. 
\begin{align}
    \frac{\epsilon_n^{-4}\lambda_{J_n}^{-4}}{n^3\sigma_n^4} \to 0 \implies (\ref{eq:Lindeberg})
\end{align}

By Assumption \ref{assum:varnotsmall}, $\liminf_{n\to \infty}n\epsilon_n\lambda_{J_n}^2\sigma_n^2 > 0$. So the following condition is sufficient:
\begin{align}\label{eq:final_lyp}
    \frac{\epsilon_n^{-4}\lambda_{J_n}^{-4}}{n\epsilon_n^{-2}\lambda_{J_n}^{-4}} = \frac{\epsilon_n^{-2}}{n}  \to 0 \implies (\ref{eq:Lindeberg})
\end{align}

Since $\epsilon_n\sim n^{-1/3}$ by assumption, the condition in (\ref{eq:final_lyp}) must hold. So (\ref{eq:Lindeberg}) holds and we can invoke the CLT from \cite{JansonPratelliRigo2024} to conclude that: $\frac{\frac{1}{n}\sum_{i=1}^n Y_{ni}}{\sqrt{\mathbb{V}[ \frac{1}{n}\sum_{i=1}^n Y_{ni}  ]}}\to_d N(0,1)$ Combining this with Equation (\ref{eq:equivalency_Yn}) via Slutsky, we have: $ \frac{    \widehat{\Delta}_{c,n}^{pb}-\widetilde{\Delta}_{c,n}}{\sqrt{\mathbb{V}\left[\frac{1}{n}\sum_{i=1}^n m_{n}^0(t_i)\right]}}\to_d N(0,1)$.

\subsection{Proof of Theorem \ref{thm:variance_estimation}}\label{proof:variance_estimation} %audited 5/23/26

The proof proceeds in three steps. In Step 1 we will show that the feasible variance estimator is asymptotically equivalent to the oracle variance estimator:
  $$\widehat{V}_n-V_n^0 = o_p\left(\frac{1}{n\lambda_{J_n}^2\epsilon_n}\right) $$
Then, in Step 2 we will show that the oracle variance estimator converges fast to the desired variance:
  $$  V_n^0 -\mathbb{V}\left[\frac{1}{n}\sum_{i=1}^n m_{n}^0(t_i)\right]= o_p\left(\frac{1}{n\lambda_{J_n}^2\epsilon_n}\right)$$
Finally, in Step 3 we will use the Continuous Mapping Theorem and the lower bound on the variance stipulated by  Assumption \ref{assum:varnotsmall} to conclude the proof.

  \vskip 0.1in
{\noindent\bf Step 1: Equivalence of Oracle and Feasible Variance Estimators } 

In this step we want to show that: $\widehat{V}_n-V^0_n= o_p\left(\frac{1}{n\lambda_{J_n}^2\epsilon_n}\right)$.

Define the scalars $\overline{\widehat{m}}_n \equiv \frac{1}{n}\sum_{i=1}^n  \widehat{m}_n(t_i)$ and $\overline{{m}}^0_n \equiv \frac{1}{n}\sum_{i=1}^n m_n^0(t_i)$. Define the $n \times 1$ vectors: $\widehat{\mathbf{m}}_n \equiv (\widehat{m}_n(t_1) -\overline{\widehat{m}}_n,\cdots \widehat{m}_n(t_n) -\overline{\widehat{m}}_n)$ and  $\mathbf{m}_n \equiv (m^0_n(t_1) -\overline{m}^0_n,\cdots m^0_n(t_n) -\overline{m}^0_n )$. Therefore in this vectorized notation we can rewrite the feasible and oracle variance estimators:\begin{align*}
  \widehat{V}_n&=  \frac{1}{n^2}\sum_{i=1}^n\sum_{k=1}^n\Lambda_{ik} (\widehat{m}_n (t_i)-\overline{\widehat{m}}_n)(\widehat{m}_n(t_k) -\overline{\widehat{m}}_n) =\frac{1}{n^2}\widehat{\mathbf{m}}_n'\Lambda \widehat{\mathbf{m}}_n\\
  {V}_n^0&=  \frac{1}{n^2}\sum_{i=1}^n\sum_{k=1}^n\Lambda_{ik} ({m}_n^0 (t_i)-\overline{{m}^0_n})({m}_n^0(t_k) -\overline{{m}^0_n}) =\frac{1}{n^2}{\mathbf{m}}_n'\Lambda {\mathbf{m}}_n
\end{align*}

 To bound the difference between the feasible and oracle variance estimators, use the operator norm of $\Lambda$:
\begin{align}
   \left| \frac{1}{n^2}\left(\widehat{\mathbf{m}}_n'\Lambda \widehat{\mathbf{m}}_n-\mathbf{m}_n'\Lambda \mathbf{m}_n  \right)\right| &=  \left|\frac{1}{n^2}(\mathbf{m}_n - \widehat{\mathbf{m}}_n)'\Lambda (\mathbf{m}_n + \widehat{\mathbf{m}}_n)\right|\\
    &\leq \frac{||\Lambda||_{op}}{n^2} ||\mathbf{m}_n - \widehat{\mathbf{m}}_n||_2||\mathbf{m}_n + \widehat{\mathbf{m}}_n||_2\\
    &\leq \frac{||\Lambda||_{op}}{n^2} ||\mathbf{m}_n - \widehat{\mathbf{m}}_n||_2(2||\mathbf{m}_n||_2+||\mathbf{m}_n - \widehat{\mathbf{m}}_n||_2) \label{eq:znzn_quad}
\end{align}

Assumption \ref{assum:cross_study_independence} says that $\Lambda$ is a block-diagonal matrix with entries zero and one with blocks of size at most $C_B$. So:
\begin{align}
    ||\Lambda||_{op} \leq C_B \label{eq:lambda_operator_norm}
\end{align}

Next we bound $   ||\mathbf{m}_n + \widehat{\mathbf{m}}_n||_2$. By Lemma \ref{lem:linearization}, $\mathbb{E}\left[m_n^0 (T)^2\mid R=1\right] = \mathcal{O}\left(\lambda_{J_n}^{-2}\epsilon_n^{-1}\right)$. By Chebyshev's Inequality:
\begin{align}
    ||\mathbf{m}_n||_2   =\mathcal{O}_p\left(\sqrt{n \mathbb{E}\left[m_n^0(T)^2|R\right] }\right) = \mathcal{O}_p\left(\frac{n^{1/2}}{\lambda_{J_n}\epsilon_n^{1/2}}\right) \label{eq:zn_2norm}
\end{align}

Lemma \ref{lem:oracle_feasible_variance} shows that the oracle and feasible influence functions are close in probability:
  $$\sup_{t\in \mathbb{R}}|m_n^0(t) - \widehat{m}_n(t)|= \mathcal{O}_p\left(\lambda_{J_n}^{-1}\right)$$
  
  Therefore: $||\mathbf{m}_n - \widehat{\mathbf{m}}_n||_{\infty}=\mathcal{O}_p\left(\lambda_{J_n}^{-1}\right)$. So by  Cauchy-Schwarz:
\begin{align}
   ||\mathbf{m}_n - \widehat{\mathbf{m}}_n||_2  \leq \sqrt{n \sup_{t\in\mathbb{R}}|{m}_{n}^0(t)-\widehat{m}_n(t)|^2 } = \mathcal{O}_p\left(\frac{n^{1/2}}{\lambda_{J_n}}\right)\label{eq:zhat_2norm}
\end{align}

Combining Equations (\ref{eq:znzn_quad}), (\ref{eq:lambda_operator_norm}), (\ref{eq:zn_2norm}), and (\ref{eq:zhat_2norm}) and since $\epsilon_n\to 0$:
\begin{align}
  \widehat{V}_n-V^0_n =    \frac{1}{n^2}\left(\widehat{\mathbf{m}}_n'\Lambda \widehat{\mathbf{m}}_n-\mathbf{m}_n'\Lambda \mathbf{m}_n \right)=o_p\left(\frac{1}{n\lambda_{J_n}^2\epsilon_n}\right)\label{eq:oracle_asgoodas}
\end{align}

\vskip 0.1in
{\noindent\bf Step 2: Consistency of the oracle variance estimator } 

In this step we will show that:
\begin{align*}
 V_n^0 -\mathbb{V}\left[\frac{1}{n}\sum_{i=1}^n m_{n}^0(t_i)\right]= o_p\left(\frac{1}{n\epsilon_n\lambda_{J_n}^2}\right)
\end{align*}

Define $Y_{ni} \equiv m_n^0(t_i)-\mathbb{E}\left[m_n^0(T)\mid R=1\right]$. Since each $ m_n^0(t_i)$ is independent of all but $C_B$ others by Assumption \ref{assum:cross_study_independence} and  $\mathbb{E}\left[m_n^0 (T)^2\mid R=1\right] = \mathcal{O}\left(\lambda_{J_n}^{-2}\epsilon_n^{-1}\right)$ by Lemma \ref{lem:linearization}, then $\overline{{m}^0_n} -\mathbb{E}[m_n^0(T)\mid R=1] = \mathcal{O}_p\left(\frac{1}{n^{1/2}\epsilon_n^{1/2}\lambda_{J_n}}\right)$. Thus, 
\begin{align}\label{eq:mn_and_Yn}
     \frac{1}{n^2}\mathbf{m}_n'\Lambda \mathbf{m}_n -\frac{1}{n^2}\sum_{i=1}^n\sum_{k=1}^n\Lambda_{ik}  Y_{ni}Y_{nk} &= o_p\left(\frac{1}{n\epsilon_n\lambda_{J_n}^2}\right)
\end{align}

Next we show that $\frac{1}{n^2}\sum_{i=1}^n\sum_{k=1}^n\Lambda_{ik}  Y_{ni}Y_{nk}$ is unbiased for the target variance. Using Assumption \ref{assum:cross_study_independence}:
\begin{align}\label{eq:sumY_unbiased}
  \mathbb{V}\left[\frac{1}{n}\sum_{i=1}^n m_n^0(t_i)\right]&= \mathbb{V}\left[\frac{1}{n}\sum_{i=1}^n Y_{ni}\right]
 = \mathbb{E}\left[\left(\frac{1}{n}\sum_{i=1}^n Y_{ni}\right)^2\right] \\&= \frac{1}{n^2}\sum_{i=1}^n\sum_{k=1}^n\Lambda_{ik} \mathbb{E}\left[Y_{ni}Y_{nk}\right]=\mathbb{E}\left[\frac{1}{n^2}\sum_{i=1}^n\sum_{k=1}^n\Lambda_{ik}  Y_{ni}Y_{nk}\right]
\end{align}

Next we show that the variance of $\frac{1}{n^2}\sum_{i=1}^n\sum_{k=1}^n\Lambda_{ik}  Y_{ni}Y_{nk}$ converges at the required rate. Expanding the sum:
\begin{align*}
    \mathbb{V}\left[\frac{1}{n^2}\sum_{i=1}^n\sum_{k=1}^n\Lambda_{ik}  Y_{ni}Y_{nk}\right]&= \frac{1}{n^4}\sum_{i=1}^n\sum_{k=1}^n\Lambda_{ik}\sum_{\ell=1}^n\sum_{m=1}^n\Lambda_{\ell m}  \text{COV}\left(Y_{ni}Y_{nk},Y_{n\ell}Y_{nm}\right)\\
     &= \frac{1}{n^4}\sum_{i=1}^n\sum_{k=1}^n\sum_{\ell=1}^n\sum_{m=1}^n\Lambda_{ik}\Lambda_{i\ell}\Lambda_{\ell m}  \text{COV}\left(Y_{ni}Y_{nk},Y_{n \ell}Y_{nm}\right)
\end{align*}

By Assumption \ref{assum:cross_study_independence}, in order for $ \text{COV}\left(Y_{ni}Y_{nk},Y_{n\ell}Y_{nm}\right)\neq 0$, observations $k,\ell,m$ must all come from the same article as observation $i$ and therefore there can be no more than $C_B^3$ such triples. Moreover, Lemma \ref{lem:linearization} proved that $\limsup_{n\to \infty}\sup_{t\in\mathbb{R}} \epsilon_n\lambda_{J_n}|m^0_n(t)|<\infty $.  Using these:
\begin{align*}
\frac{1}{n^4}\sum_{i=1}^n\sum_{k=1}^n\sum_{\ell=1}^n\sum_{m=1}^n\Lambda_{ik}\Lambda_{i\ell}\Lambda_{\ell m}  \text{COV}\left(Y_{ni}Y_{nk},Y_{n\ell}Y_{nm}\right)
    &\lesssim \frac{C_B^3}{n^4}\sum_{i=1}^n   \sup_{t\in \mathbb{R}}2^4|m^0_n(t)|^4\\
    &\leq \frac{C_B^3}{n^3} \sup_{t\in \mathbb{R}}2^4|m^0_n(t)|^4 =\mathcal{O}\left(  \frac{1}{n^3\epsilon_n^4\lambda_{J_n}^4}\right)
\end{align*}

So we have shown that $ \mathbb{V}\left[\frac{1}{n^2}\sum_{i=1}^n\sum_{k=1}^n\Lambda_{ik}  Y_{ni}Y_{nk}\right]=\mathcal{O}\left(  \frac{1}{n^3\epsilon_n^4\lambda_{J_n}^4}\right)$. Recall that (\ref{eq:sumY_unbiased}) showed unbiasedness of this double sum over $Y_{ni}$ for the target variance. Applying Chebyshev's Inequality and then the assumption that $\epsilon_n\sim n^{-1/3}$ yields: 
\begin{align}\label{eq:Yn_converges}
    \frac{1}{n^2}\sum_{i=1}^n\sum_{k=1}^n\Lambda_{ik} Y_{ni}Y_{nk} -  \mathbb{V}\left[\frac{1}{n}\sum_{i=1}^n m_n^0(t_i)\right] = \mathcal{O}_p\left( \frac{1}{n^{3/2}\epsilon_n^2\lambda_{J_n}^2}\right) = o_p\left(\frac{1}{n\epsilon_n\lambda_{J_n}^2}\right)
\end{align}

Applying the triangle inequality to (\ref{eq:Yn_converges}) and (\ref{eq:mn_and_Yn}) lets us conclude that the oracle estimator converges at the required rate:
\begin{align}
  V_n^0-\mathbb{V}\left[\frac{1}{n}\sum_{i=1}^n m_n^0(t_i)\right]= \frac{1}{n^2}\mathbf{m}_n'\Lambda \mathbf{m}_n -\mathbb{V}\left[\frac{1}{n}\sum_{i=1}^n m_n^0(t_i)\right]= o_p\left(\frac{1}{n\epsilon_n\lambda_{J_n}^2}\right)  \label{eq:oracle_consistent}
\end{align}

\vskip 0.1in
{\noindent\bf Step 3: Conclusion}

Assumption \ref{assum:varnotsmall} states that $ \liminf_{n\to \infty} n\epsilon_n\lambda_{J_n}^2\mathbb{V}\left[\frac{1}{n}\sum_{i=1}^n m_n^0(t_i)\right] >0 $.  Therefore, the estimation error of the variance estimator is dominated by the variance itself:
\begin{align}\label{eq:variance_convergence}
    \frac{    \widehat{V}_n }{\mathbb{V}\left[\frac{1}{n}\sum_{i=1}^n m_n^0(t_i)\right]} = \frac{    \mathbb{V}\left[\frac{1}{n}\sum_{i=1}^n m_n^0(t_i)\right]+o_p\left(\frac{1}{n\epsilon_n\lambda_{J_n}^2}\right) }{\mathbb{V}\left[\frac{1}{n}\sum_{i=1}^n m_n^0(t_i)\right]}=1+\frac{o_p\left(\frac{1}{n\epsilon_n\lambda_{J_n}^2}\right)}{\mathbb{V}\left[\frac{1}{n}\sum_{i=1}^n m_n^0(t_i)\right]}= 1+o_p(1)
\end{align}

Using the Continuous Mapping Theorem we can combine Equation (\ref{eq:variance_convergence}) with Theorem \ref{thm:clt} to obtain the final result:
\begin{align*}
    \frac{\widehat{\Delta}_{c,n}^{pb}-\widetilde{\Delta}_{c,n}}{\sqrt{\widehat{V}_n}}\to_d N(0,1)
\end{align*}

\clearpage
\section{ Technical Lemmas}

\subsection{Lemma \ref{lem:integral_hermite} and proof}\label{proof:lem:integral_hermite} %Audited 3/16/2026 % audited 5/21/26

\begin{lemma}\label{lem:integral_hermite}
Define the constant $A_{c,\sigma_T,CV} \equiv \sqrt{\frac{2 CV}{\varphi_{\sigma_T^2}(CV)}} +\sqrt{\frac{2 CV}{c\varphi_{\sigma_T^2+1-c^{-2}}(CV/c)}}$. All three of the following statements are true for all  ${j\in \{0,1,2,\cdots\}}$ and $c\geq 1$ and $\sigma_T>0$:
    \begin{align*}
       \int_{-CV/c}^{CV/c} \left|\phi_j(t)\right|dt &\leq  \sqrt{\frac{2 CV/c}{\varphi_{\sigma_T^2+1-c^{-2}}(CV/c)}}\\
        \int_{-CV}^{CV} \left|\psi_j(t)\right|dt &\leq  \sqrt{\frac{2 CV}{\varphi_{\sigma_T^2}(CV)}}\\
       \left|\lambda_j\int_{-CV}^{CV}\psi_j(t)dt-\int_{-CV/c}^{CV/c}\phi_j(t)dt\right|&\leq A_{c,\sigma_T,CV}
    \end{align*}
\end{lemma}

\begin{proof}
    
To prove the first claim, by Cauchy-Schwarz:
\begin{align}\label{eq:cw_1}
     \int_{-CV/c}^{CV/c} \left|\phi_j(t)\right|dt  \leq \sqrt{2CV/c}\left( \int_{-CV/c}^{CV/c} \phi_j(t)^2dt \right)^{1/2}
\end{align}

Since $\varphi_{\sigma_T^2+1-c^{-2}}(t)$ is non-negative, even, and decreasing in $|t|$, $\frac{\varphi_{\sigma_T^2+1-c^{-2}}(t)}{\varphi_{\sigma_T^2+1-c^{-2}}(CV/c)}\geq 1$ for all $|t|\leq CV/c$. Therefore:
\begin{align*}
    \int_{-CV/c}^{CV/c} \phi_j(t)^2dt \leq \int_{-CV/c}^{CV/c} \phi_j(t)^2\frac{\varphi_{\sigma_T^2+1-c^{-2}}(t)}{\varphi_{\sigma_T^2+1-c^{-2}}(CV/c)}dt \leq  \frac{\int_{-\infty}^{\infty} \phi_j(t)^2\varphi_{\sigma_T^2+1-c^{-2}}(t)dt}{\varphi_{\sigma_T^2+1-c^{-2}}(CV/c)} = \frac{\langle \phi_j,\phi_j\rangle_{T_c}}{\varphi_{\sigma_T^2+1-c^{-2}}(CV/c)}
\end{align*}

By Example 1 of \cite{CarrascoPaper}, the basis polynomials $\{\phi_j\}$ are orthonormal in $\langle \cdot ,\cdot \rangle_{T_c}$ and therefore last integral in the line above equals one (to see this in their notation, let $\sigma^2 =1-c^{-2}$ and $\sigma_Y^2=\sigma_T^2$). Substituting this into Equation (\ref{eq:cw_1}):
\begin{align*}
     \int_{-CV/c}^{CV/c} \left|\phi_j(t)\right|dt \leq \sqrt{\frac{2CV/c}{\varphi_{\sigma_T^2+1-c^{-2}}(CV/c)}}
\end{align*}

When $c=1, \phi_j(t)=\psi_j(t)$, so the second claim of the lemma is a special case of the first when $c=1$. The third claim is an application of the triangle inequality to the first two claims:
\begin{align*}
     \left|\lambda_j\int_{-CV}^{CV}\psi_j(t)dt-\int_{-CV/c}^{CV/c}\phi_j(t)dt\right| &\leq |\lambda_j|\int_{-CV}^{CV}|\psi_j(t)|dt+\int_{-CV/c}^{CV/c}|\phi_j(t)|dt
\end{align*}
Since   $\lambda_j=\left(\frac{\sigma_T^2}{\sigma_T^2+1-c^{-2}}\right)^{j/2} \leq 1$ for all $j$, the stated constant $A_{c,\sigma_T,CV} $ bounds the expression uniformly in $j$.
\end{proof}

\subsection{Lemma \ref{lem:bound_coeffs} and proof}\label{proof:lem:bound_coeffs} %audited 3/16/2026 % audited 5/21/26

\begin{lemma}\label{lem:bound_coeffs}
For all ${j\in \{0,1,2,\cdots\}}$ and $c\geq 1$ and $\sigma_T>0$:
    \begin{align*}
\sup_{t\in\mathbb{R}}\left|He_j(t)\varphi(t)\right| &\leq \frac{1}{\sqrt{2\pi}}  \\
\sup_{t\in\mathbb{R}}\left|\psi_j(t)\varphi_{\sigma_T^2}(t)\right| &\leq \frac{1}{\sqrt{2\sigma_T^2\pi}}  \\
\sup_{t\in\mathbb{R}}\left|\phi_j(t)\varphi_{\sigma_T^2+1-c^{-2}}(t)\right| &\leq \frac{1}{\sqrt{2(\sigma_T^2+1-c^{-2})\pi}}  \\
\sup_{t\in\mathbb{R}}\left|\chi_j(t)\varphi_{\sigma_T^2+1}(t)\right| &\leq \frac{1}{\sqrt{2(\sigma_T^2+1)\pi}}  
    \end{align*}

   % Proof: Section \ref{proof:lem:bound_coeffs}
\end{lemma}
\begin{proof}

 We now prove the first claim, from which the others are immediate. Converting from the normalized Hermite polynomials $He_j$ to  physicist  version in the notation of \cite{HermiteBound} $H_j$: \begin{align*}
He_j(t)&= H_j(t/\sqrt{2})\frac{2^{-j/2}}{\sqrt{j!}} \\
 \sup_{t\in\mathbb{R}}\left|He_j(t)\varphi(t)\right| &=\sup_{t\in\mathbb{R}}\left|He_j(t)e^{-t^2/2}/(\sqrt{2\pi})\right|\\
     &=\sup_{t\in\mathbb{R}} \left|H_j(t/\sqrt{2})2^{-j/2}e^{-t^2/2}/\sqrt{2\pi j!}\right|
\end{align*}

\noindent From \cite{HermiteBound}, the physicist's Hermite polynomials are bounded by:
    $|H_j(t/\sqrt{2})| \leq (2^j j!)^{1/2}e^{t^2/4}$. 
\begin{align*}
    \left|H_j(t/\sqrt{2})\right|2^{-j/2}e^{-t^2/2}/\sqrt{2\pi j!} &\leq (2^j j!)^{1/2}e^{t^2/4}2^{-j/2}e^{-t^2/2}/\sqrt{2\pi j!}=e^{-t^2/4}/\sqrt{2\pi}\leq \frac{1}{\sqrt{2\pi}}
\end{align*}

To see the second claim, notice that: $\varphi_{\sigma^2}(t)=\varphi(t/\sigma)/\sigma$ and $\psi_j(t)\varphi_{\sigma_T^2}(t) = He_j(t/\sigma_T)\varphi(t/\sigma_T)/\sigma_T$ and apply the first claim to it. Similarly, for the third claim:
\begin{align*}
    \phi_j(t)\varphi_{\sigma_T^2+1-c^{-2}}(t) &= He_j\left(t/\sqrt{\sigma_T^2+1-c^{-2}}\right)\varphi\left(t/\sqrt{\sigma_T^2+1-c^{-2}}\right)/\sqrt{(\sigma_T^2+1-c^{-2})}
\end{align*}

 And again similarly for the fourth claim: $$\chi_j(t)\varphi_{\sigma_T^2+1}(t) = He_j\left(t/\sqrt{\sigma_T^2+1}\right)\varphi\left(t/\sqrt{\sigma_T^2+1}\right)/\sqrt{(\sigma_T^2+1)}$$
 \end{proof}
 
\subsection{Lemma \ref{lem:bound_a_jpsi_j} and proof}\label{proof:lem:bound_a_jpsi_j} %audited 3/16/26 %audited 5/21/26

\begin{lemma}\label{lem:bound_a_jpsi_j}
If $c>1$ and $\sigma_T>0$, then for all $J_n\in \{0,1,2,\cdots\}$:

    $$ \sup_{t\in \mathbb{R}} \sum_{j=0}^{J_n} \left|a_j\psi_j(t)\varphi_{\sigma_T^2}(t)\right| \leq \lambda_{J_n}^{-1} \frac{1}{(1-\lambda_1)\sqrt{2\sigma^2_T\pi}}\left(\sqrt{\frac{2 CV}{\varphi_{\sigma_T^2}(CV)}} +\sqrt{\frac{2 CV}{c\varphi_{\sigma_T^2+1-c^{-2}}(CV/c)}}\right) $$
\end{lemma}
\begin{proof}
     Notice that: $\sup_{t\in \mathbb{R}} \sum_{j=0}^{J_n} \left|a_j\psi_j(t)\varphi_{\sigma_T^2}(t)\right|\leq \sup_{t,\ell}|\psi_\ell(t)\varphi_{\sigma_T^2}(t)|\sum_{j=0}^{J_n}|a_j|$. By Lemma \ref{lem:bound_coeffs}, $\sup_{t,\ell}|\psi_\ell(t)\varphi_{\sigma_T^2}(t)| \leq \frac{1}{\sqrt{2\sigma_T^2\pi}}$. Therefore:
    \begin{equation}\label{eq:aj_bound_1}
        \sup_{t\in \mathbb{R}} \sum_{j=0}^{J_n} \left|a_j\psi_j(t)\varphi_{\sigma_T^2}(t)\right|\leq \frac{1}{\sqrt{2\sigma_T^2\pi}}\sum_{j=0}^{J_n}|a_j|
    \end{equation}
    
    Next we bound the sum over $a_j$. By Lemma \ref{lem:integral_hermite}, $\sup_j |\lambda_j a_j| \leq \left(\sqrt{\frac{2 CV}{\varphi_{\sigma_T^2}(CV)}} +\sqrt{\frac{2 CV}{c\varphi_{\sigma_T^2+1-c^{-2}}(CV/c)}}\right) $. 
    
    Thus:
\begin{align}\label{eq:bound_aj}
    \sum_{j=0}^{J_n}|a_j| &\leq  \left(\sqrt{\frac{2 CV}{\varphi_{\sigma_T^2}(CV)}} +\sqrt{\frac{2 CV}{c\varphi_{\sigma_T^2+1-c^{-2}}(CV/c)}}\right)   \sum_{j=0}^{J_n}\frac{1}{\lambda_j}
\end{align}

Since $c>1$, $0<\lambda_1<1$.  Use the fact that $\lambda_j$ is a power sequence, i.e. $\lambda_j = \lambda_1^{j}$. This gives us the identity:
\begin{align}\label{eq:bound_lambdasum}
\sum_{j=0}^{J_n}\lambda_j^{-1}
= \sum_{j=0}^{J_n}\lambda_1^{-j}
= \lambda_{J_n}^{-1}\sum_{k=0}^{J_n}\lambda_1^k
\le \frac{\lambda_{J_n}^{-1}}{1-\lambda_1}
\end{align}

Substituting this into Equation (\ref{eq:bound_aj}):
\begin{equation}
      \sum_{j=0}^{J_n}|a_j| \leq  \lambda_{J_n}^{-1}\left(\sqrt{\frac{2 CV}{\varphi_{\sigma_T^2}(CV)}} +\sqrt{\frac{2 CV}{c\varphi_{\sigma_T^2+1-c^{-2}}(CV/c)}}\right)   \frac{1}{1-\lambda_1} 
\end{equation}
And substituting this into Equation (\ref{eq:aj_bound_1}), we have the desired result:
 $$ \sup_{t\in \mathbb{R}} \sum_{j=0}^{J_n} \left|a_j\psi_j(t)\varphi_{\sigma_T^2}(t)\right| \leq \lambda_{J_n}^{-1} \frac{1}{(1-\lambda_1)\sqrt{2\sigma^2_T\pi}}\left(\sqrt{\frac{2 CV}{\varphi_{\sigma_T^2}(CV)}} +\sqrt{\frac{2 CV}{c\varphi_{\sigma_T^2+1-c^{-2}}(CV/c)}}\right) $$
\end{proof}

\subsection{Lemma \ref{lem:moment_conversion} and proof}
\begin{lemma}\label{lem:moment_conversion} %audited 3/17/26 % audited 5/21/26

If $c>1$, $\sigma_T>0$, and  Assumption \ref{assum:normality_of_T} holds,  then for any probability distribution $\Pi$ of $H$:
    $$\mathbb{E}_{\Pi}\left[\chi_j(H)\varphi_{\sigma_T^2+1}(H)\right]  = \frac{1}{\eta_j\lambda_j}\mathbb{E}_{\Pi}[\psi_j(T)\varphi_{\sigma_T^2}(T)]\qquad \forall j \in \{0,1,2,\cdots\}  $$
\end{lemma}
\begin{proof}

Define $K_{c^{-2}}^*$ as the adjoint of the operator $K_{c^{-2}}$ with respect to $\langle \cdot,\cdot\rangle_{H}$ and $\langle \cdot,\cdot\rangle_{T_c}$. We want to show that $ K_{c^{-2}}^* \phi_j= \eta_j\chi_j$. To see this, consider any function $g\in \mathcal{L}_H$. From the definition of the adjoint $\langle K^*_{c^{-2}}\phi_j, g\rangle_H=\langle \phi_j,K_{c^{-2}}g\rangle_{T_c}$. By the singular value decomposition from Example 1 of \cite{CarrascoPaper}: $K_{c^{-2}}g = \sum_{\ell=0}^\infty \eta_\ell\langle \chi_\ell,g \rangle_H \phi_\ell $.   By orthonormality of the Hermite polynomials: $\langle \phi_j,K_{c^{-2}}g\rangle_{T_c} = \eta_j \langle \chi_j ,g\rangle_{H}$. So $ \langle K^*_{c^{-2}}\phi_j, g\rangle_H=\langle \eta_j \chi_j, g\rangle_H$ for any $j\in \{0,1,\cdots\}$ and any $g\in \mathcal{L}_H$. Since this holds for all $g\in \mathcal{L}_H$, then by the uniqueness of the Riesz representation, we have the desired equation:
\begin{align}\label{eq:adj1}
    K_{c^{-2}}^* \phi_j= \eta_j\chi_j
\end{align}

 Let $K_{1-c^{-2}}^*$ be the adjoint of the operator $K_{1-c^{-2}}$ with respect to $\langle \cdot,\cdot\rangle_{T_c}$ and $\langle \cdot,\cdot\rangle_{T}$. By an identical argument we have the analogous useful equation:
\begin{align}\label{eq:adj2}
    K_{1-c^{-2}}^* \psi_j= \lambda_j\phi_j
\end{align}

 We first prove the identity for probability distributions with densities in
$\mathcal L_H$, and then extend it to arbitrary probability distributions by smoothing and weak convergence. For any probability distribution $\Pi$ we can define a sequence of continuous distributions $H_n\sim\Pi_n\equiv \Pi * N\left(0,\tau_n^2\right)$  where $\tau_n\to 0$ and $*$ denotes convolution. So $\Pi_n$ has PDF $\pi_n$ and $\Pi_n\to_w \Pi$ (where $\to_w$ denotes weak convergence). Since each $\pi_n$ has finite height, we have $\pi_n\in\mathcal L_H$. Define $f_{T,n}\equiv K_1\pi_n$ as the density of $T$ under $\Pi_n$. Using Equations (\ref{eq:adj1}) and (\ref{eq:adj2}), the definition of the adjoint,  and Lemma \ref{lem:sumnoise}, for each $\Pi_n$ we have:
\begin{align}\label{eq:start_adjoint}
  \mathbb{E}_{\Pi_n}\left[\chi_j(H)\varphi_{\sigma_T^2+1}(H)\right] &= \langle \chi_j,\pi_n \rangle_{H}\\
  &=  \frac{1}{\eta_j}\langle K^*_{c^{-2}}\phi_j,\pi_n \rangle_{H}\\
    &=  \frac{1}{\eta_j}\langle \phi_j,K_{c^{-2}}\pi_n \rangle_{T_c}\\
      &=  \frac{1}{\eta_j\lambda_j}\langle K^*_{1-c^{-2}}\psi_j,K_{c^{-2}}\pi_n \rangle_{T_c}\\
      &=  \frac{1}{\eta_j\lambda_j}\langle \psi_j,K_{1-c^{-2}}K_{c^{-2}}\pi_n \rangle_{T}\\
    &=\frac{1}{\eta_j\lambda_j}\langle \psi_j,f_{T,n} \rangle_{T}\\
    &=   \frac{1}{\eta_j\lambda_j}\mathbb{E}_{\Pi_n}[\psi_j(T)\varphi_{\sigma_T^2}(T)] \label{eq:end_adjoint}
\end{align}

By Lemma \ref{lem:bound_coeffs}, $\chi_j(h)\varphi_{\sigma_T^2+1}(h)$ and $\psi_j(t)\varphi_{\sigma_T^2}(t)$ are bounded  and by their definition as Hermite polynomials times Gaussian densities they are continuous. Since $\Pi_n \to_w \Pi$, by the Portmanteau Theorem:
\begin{align}
\mathbb{E}_{\Pi_n}\left[\chi_j(H)\varphi_{\sigma_T^2+1}(H)\right] &\to \mathbb{E}_{\Pi}\left[\chi_j(H)\varphi_{\sigma_T^2+1}(H)\right]
\end{align}

Since $H_n\to_w H$ and $Z$ is independent of both $H_n$ and $H$,  we have $T_n=H_n+Z\to_w H+Z=T$. Therefore, Portmanteau applies to bounded and continuous functions of $T_n$.  Since we already showed that $\psi_j(t)\varphi_{\sigma_T^2}(t)$ is a bounded and continuous function we can apply Portmanteau again to show:
 \begin{align}
     \mathbb{E}_{\Pi_n}[\psi_j(T)\varphi_{\sigma_T^2}(T)] &\to \mathbb{E}_{\Pi}[\psi_j(T)\varphi_{\sigma_T^2}(T)]
 \end{align}

Thus:
$$\mathbb{E}_{\Pi}\left[\chi_j(H)\varphi_{\sigma_T^2+1}(H)\right]  = \frac{1}{\eta_j\lambda_j}\mathbb{E}_{\Pi}[\psi_j(T)\varphi_{\sigma_T^2}(T)]  $$

\end{proof}

\subsection{Lemma \ref{lem:Bbounds} and proof}\label{proof:lem:Bbounds}
%audited 4/3/26 % audited 5/23/26

Recall that $F,\widehat{F}_n$ denote the true and empirical cdfs of $|T|$ conditional on $R=1$. Recall also that $B_{\epsilon_n}^-\equiv F(1.96)-F(1.96-\epsilon_n)$ and $B_{\epsilon_n}^+\equiv F(1.96+\epsilon_n)-F(1.96)$. Their definitions are collected in Appendix \ref{app:inference_definitions} for convenience. 

\begin{lemma}\label{lem:Bbounds}
    Under the assumptions of Theorem \ref{thm:consistency_withpb}, the distribution of $|T|$ conditional on $R=1$ is continuous and all of the following hold:
\begin{enumerate}
    \item $ F(t)>0 \quad \forall t>0$
    \item  $\liminf_{n\to \infty} \epsilon_n^{-1}B_{\epsilon_n}^{-} > 0$
      \item     $B_{\epsilon_n}^{+} =\mathcal{O}(\epsilon_n)$
       \item   $B_{\epsilon_n}^{-} =\mathcal{O}(\epsilon_n)$
\end{enumerate}

\end{lemma}
\begin{proof}
   
   \noindent {\bf Proof of Claim 1:}  
   
   By Bayes' Rule: \begin{align*}
f_{|T|\mid R=1}(s)
=
\frac{w_{\theta_0}(s)f_T(s)+w_{\theta_0}(-s)f_T(-s)}
{\Pr(R=1)}, \qquad s>0.
   \end{align*}
 By Equation (\ref{eq:pb_caliper_form}), $w_{\theta_0}$ is continuously differentiable except possibly at $\pm 1.96$. By Assumption \ref{assum:pblowerbound} $w_{\theta_0}(t)>\overline{M}^{-1}>0$ everywhere. Since $T$ is the outcome of a Gaussian convolution, $f_T$ is continuously differentiable everywhere and bounded away from zero on every compact interval. So $f_{|T|\mid R=1}$ is continuously differentiable except possibly at $1.96$, bounded away from zero on all compact intervals, and thus $F(t) = \int_{0}^tf_{|T|\mid R=1}(s)ds >0$ for all $t>0$.

\vskip 0.1in
 \noindent {\bf Proof of Claim 2:} 

Next we lower-bound $B_{\epsilon_n}^{-}$.  Note that 
$\mathbb P(R=1)\leq 1$. Moreover, Assumption 
\ref{assum:pblowerbound} implies $\text{Pr}\left(R=1\mid T=t\right)\geq  \overline M^{-1}$ for all $t$.  Using Bayes' Rule:
\begin{align*}
    B_{\epsilon_n}^{-} &= \text{Pr}\left(|T|\in (1.96-\epsilon_n,1.96] |R=1\right)
    = \frac{\theta_0\text{Pr}\left(|T|\in (1.96-\epsilon_n,1.96]  \right)}{\text{Pr}\left(R=1 \right)}\\
    &\geq  \overline{M}^{-1}\text{Pr}\left(|T|\in (1.96-\epsilon_n,1.96]  \right) \\
    &=  \overline{M}^{-1} \left(\int_{1.96-\epsilon_n}^{1.96}f_T(t)dt +\int_{-1.96}^{-1.96+\epsilon_n}f_T(t)dt\right)\\
    &\geq  \overline{M}^{-1} \epsilon_n \min_{t\in [-1.96-\epsilon_n,1.96+\epsilon_n]}f_T(t)
\end{align*}
\noindent (The endpoint $1.96$ is immaterial because $|T|$ is continuously distributed.)

Since $f_T(t) = \int_{-\infty}^\infty \varphi(t-h)d\Pi(h)$ is the PDF of a Gaussian convolution, it is continuous and bounded away from zero on every compact interval, so $\liminf_{n\to \infty}\min_{t\in [-1.96-\epsilon_n,1.96+\epsilon_n]}f_T(t)>0$. Therefore:
\begin{equation*}
    \liminf_{n\to \infty} \epsilon_n^{-1} B_{\epsilon_n}^{-} > 0
\end{equation*}

 \noindent {\bf Proof of Claim 3:} 

An upper bound can be derived in a similar way. Notice that Assumption \ref{assum:pblowerbound} guarantees that $\text{Pr}\left(R=1 \right) \geq \overline{M}^{-1}$.
\begin{align*}
    B_{\epsilon_n}^{+} &= \text{Pr}\left(|T|\in (1.96,1.96+\epsilon_n] |R=1\right)
    \leq \frac{\text{Pr}\left(|T|\in (1.96,1.96+\epsilon_n] \right)}{\text{Pr}\left(R=1 \right)}\\
    &\leq  \overline{M}\text{Pr}\left(|T|\in (1.96,1.96+\epsilon_n] \right)\\
    &=  \overline{M} \left(\int_{1.96}^{1.96+\epsilon_n}f_T(t)dt +\int_{-1.96-\epsilon_n}^{-1.96}f_T(t)dt\right)\\
    &\leq  \overline{M} 2\epsilon_n \sup_{t\in\mathbb{R}}f_T(t) \\
      &\leq  \overline{M} 2\epsilon_n \frac{1}{\sqrt{2\pi}} 
\end{align*}
 
 The final inequality holds because $f_T(t) = \int_{-\infty}^\infty \varphi(t-h)d\Pi(h) \leq \sup_{x\in\mathbb{R}}\varphi(x) = \frac{1}{\sqrt{2\pi}} $. 

 \vskip 0.1in
{\bf\noindent Proof of Claim 4:} 

Using Assumption \ref{assum:pblowerbound} twice yields $\frac{\theta_0}{\text{Pr}\left(R=1 \right)}\leq \overline{M}^2$. An analogous argument to the proof of Claim 3 above yields:
\begin{align*}
    B_{\epsilon_n}^{-} &= \text{Pr}\left(|T|\in (1.96-\epsilon_n,1.96] |R=1\right)
    \leq \frac{\theta_0\text{Pr}\left(|T|\in (1.96-\epsilon_n,1.96] \right)}{\text{Pr}\left(R=1 \right)}\\
    &\leq  \overline{M}^2\text{Pr}\left(|T|\in (1.96-\epsilon_n,1.96] \right)\leq  \overline{M}^2 2\epsilon_n \frac{1}{\sqrt{2\pi}} 
\end{align*}
\end{proof}

\subsection{Lemma \ref{lem:theta_estimation} and proof}\label{proof:lem:theta_estimation}
% audited 4/4/26 % audited 5/23/26

The definitions of $F(t),\widehat{F}_n(t),B_{\epsilon_n}^-,B_{\epsilon_n}^+,\widehat{B}_{\epsilon_n}^-,\widehat{B}_{\epsilon_n}^+$ and $X_n(t;b^+,b^{-})$ are given in Appendix \ref{app:inference_definitions}. Recall that $F,\widehat{F}_n$ denote the true and empirical CDFs of $|T|$ conditional on publication $R=1$ and that $  \widetilde{\theta}_n \equiv \frac{F(1.96)-F(1.96-\epsilon_n)}{F(1.96+\epsilon_n)-F(1.96)}$.

\begin{lemma}\label{lem:theta_estimation}
     Assume that publication bias follows Equations (\ref{eq:w_definition}) and (\ref{eq:pb_caliper_form}), that Assumptions \ref{assum:normality_of_T}-\ref{assum:wnonsmooth} hold, and that the meta-analyst chooses $\epsilon_n$ such that $\epsilon_n\to 0$ and $n\epsilon_n \to \infty$. Then all of the following hold: 

\begin{enumerate}
    \item $  {\theta}_0^{-1} -\widetilde{\theta}^{-1}_n = \mathcal{O}\left(\epsilon_n\right)$
    \item $  \widehat{\theta}_n^{-1} -\widetilde{\theta}^{-1}_n =  \mathcal{O}_p\left((n\epsilon_n)^{-1/2}\right)$
    \item $\widehat{\theta}_n^{-1} -\widetilde{\theta}^{-1}_n = \frac{1}{n}\sum_{i=1}^n X_n(t_i;B_{\epsilon_n}^+,B_{\epsilon_n}^{-}) - \mathbb{E}\left[X_n(T;B_{\epsilon_n}^+,B_{\epsilon_n}^{-})\mid R=1\right]+\mathcal{O}_p\left(n^{-1}\epsilon_n^{-1}\right)$

  \item $ \mathbb{E}[X_n(T;B_{\epsilon_n}^+,B_{\epsilon_n}^{-})^2\mid R=1]=\mathcal{O}\left(\epsilon_n^{-1}\right)$
\end{enumerate}

\end{lemma}

\begin{proof}
    
Throughout this proof recall that the data $t_i$ are all published $t$-scores, so by Assumption \ref{assum:cross_study_independence} they each have the marginal distribution of $T\mid R=1$.

\vskip 0.1in
{\bf Proof of Claim 1}

 To show that $\widetilde{\theta}_n^{-1}$ converges to $\theta_0^{-1}$, we use Bayes' Rule:
\begin{align*}
    \widetilde{\theta}_n^{-1} &\equiv  \frac{\text{Pr}\left(|T| \in (1.96,1.96+\epsilon_n]\:\mid\: R=1\right)}{\text{Pr}\left(|T| \in (1.96-\epsilon_n,1.96]\:\mid\: R=1\right)}\\
    &= \frac{\text{Pr}\left(|T| \in (1.96,1.96+\epsilon_n]\right)\text{Pr}\left(R=1\:|\:|T| \in (1.96,1.96+\epsilon_n]\right)}{\text{Pr}\left(|T| \in (1.96-\epsilon_n,1.96]\right)\text{Pr}\left(R=1\:|\:|T| \in (1.96-\epsilon_n,1.96]\right)}\\
    &= \theta_0^{-1}\frac{\text{Pr}\left(|T| \in (1.96,1.96+\epsilon_n]\right)}{\text{Pr}\left(|T| \in (1.96-\epsilon_n,1.96]\right)} 
\end{align*}

Since $T$ is the outcome of a Gaussian convolution, $T$ has a $C^\infty$ density $f_T$ on $\mathbb{R}$.
Hence $|T|$ has density $
f_{|T|}(x)=f_T(x)+f_T(-x), \: x>0,$
which is continuously differentiable in a neighborhood of $1.96$ and strictly positive at $1.96$. By the mean value theorem applied separately to the numerator and denominator, there are $\zeta_1,\zeta_2 \in [1.96-\epsilon_n,1.96+\epsilon_n]$ such that:
\begin{align*}
    \widetilde{\theta}_n^{-1}  &= \theta_0^{-1} \frac{f_{|T|}(\zeta_1)}{f_{|T|}(\zeta_2)} =\theta_0^{-1} \left(1+\mathcal{O}\left(\epsilon_n\right)\right) = \theta_0^{-1}+\mathcal{O}\left(\epsilon_n\right)
\end{align*}

\vskip 0.1in
{\noindent\bf Proof of Claim 2 }

{\it Linearization:} Next we linearize $\widehat{\theta}_n^{-1}$ about the centering sequence $\widetilde{\theta}_n^{-1}$ using a standard ratio argument. Recall from Appendix \ref{app:inference_definitions} that $F,\widehat{F}_n$ denote the true and empirical CDFs of $|T|$ conditional on $R=1$. We rewrite the reciprocal $\widehat{\theta}_n^{-1}$ in terms of $\widehat{F}_n$ and linearize it about $\widetilde{\theta}_n^{-1}$.
 \begin{align*}
    \widehat{\theta}_n^{-1} -\widetilde{\theta}^{-1}_n &= \frac{\widehat{F}_n(1.96+\epsilon_n)-\widehat{F}_n(1.96)}{\widehat{F}_n(1.96)-\widehat{F}_n(1.96-\epsilon_n)}-\frac{F(1.96+\epsilon_n)-F(1.96)}{F(1.96)-F(1.96-\epsilon_n)} = \frac{\widehat{B}_{\epsilon_n}^+}{\widehat{B}_{\epsilon_n}^-}-\frac{{B}_{\epsilon_n}^+}{{B}_{\epsilon_n}^-}
 \end{align*}

Now we will rewrite the difference in fractions so that the dominant terms have non-stochastic denominators. First use the identity $\frac{a}{b}-\frac{c}{d} =\frac{ad-cb}{bd}=\frac{(a-c)d-(b-d)c}{bd}$. Then use the identity $\frac{1}{b}=\frac{1}{d}-\frac{b-d}{bd}$.  Combining these identities yields: $\frac{a}{b}-\frac{c}{d} =\frac{a-c}{d}- \frac{c(b-d)}{d^2} -(a-c)\frac{b-d}{bd}+\frac{c(b-d)}{d}\frac{b-d}{bd}$. Substituting in the $B_{\epsilon_n}$ terms yields:
\begin{align*}
    \frac{\widehat{B}_{\epsilon_n}^+}{\widehat{B}_{\epsilon_n}^-}-\frac{{B}_{\epsilon_n}^+}{{B}_{\epsilon_n}^-} 
      &= \frac{\widehat{B}_{\epsilon_n}^+-{B}_{\epsilon_n}^+}{{B}_{\epsilon_n}^-} -\left(\frac{{B}_{\epsilon_n}^+}{{B}_{\epsilon_n}^-}\right)\frac{\widehat{B}_{\epsilon_n}^--{B}_{\epsilon_n}^-}{{B}_{\epsilon_n}^-}+\left(\frac{\widehat{B}_{\epsilon_n}^--{B}_{\epsilon_n}^-}{{B}_{\epsilon_n}^-\widehat{B}_{\epsilon_n}^-}\right)\left(\left(\frac{{B}_{\epsilon_n}^+}{{B}_{\epsilon_n}^-}\right)(\widehat{B}_{\epsilon_n}^--{B}_{\epsilon_n}^-) -(\widehat{B}_{\epsilon_n}^+-{B}_{\epsilon_n}^+)\right)
\end{align*}

We will think of the last term as the remainder term and call it $r_n$. So the estimation error can be expressed in linearized form:
\begin{align}\label{eq:theta_linearized}
       \widehat{\theta}_n^{-1} -\widetilde{\theta}^{-1}_n &=  \frac{\widehat{B}_{\epsilon_n}^+-{B}_{\epsilon_n}^+}{{B}_{\epsilon_n}^-} -\left(\frac{{B}_{\epsilon_n}^+}{{B}_{\epsilon_n}^-}\right)\frac{\widehat{B}_{\epsilon_n}^--{B}_{\epsilon_n}^-}{{B}_{\epsilon_n}^-}+r_n
\end{align}
\noindent where $r_n \equiv \left(\frac{\widehat{B}_{\epsilon_n}^--{B}_{\epsilon_n}^-}{{B}_{\epsilon_n}^-\widehat{B}_{\epsilon_n}^-}\right)\left(\left(\frac{{B}_{\epsilon_n}^+}{{B}_{\epsilon_n}^-}\right)(\widehat{B}_{\epsilon_n}^--{B}_{\epsilon_n}^-) -(\widehat{B}_{\epsilon_n}^+-{B}_{\epsilon_n}^+)\right)$.

\vskip 0.1in
{\it\noindent Linear Term:} By Lemma \ref{lem:Bhat},  $\widehat{B}_{\epsilon_n}^+-{B}_{\epsilon_n}^+ =\mathcal{O}_p\left(\epsilon_n^{1/2}n^{-1/2}\right)$ and $\widehat{B}_{\epsilon_n}^--{B}_{\epsilon_n}^- =\mathcal{O}_p\left(\epsilon_n^{1/2}n^{-1/2}\right)$. Since Lemma \ref{lem:Bbounds} guarantees that $\liminf_{n\to \infty}\epsilon_n^{-1}B_{\epsilon_n}^{-} >0$ and $\left(\frac{{B}_{\epsilon_n}^+}{{B}_{\epsilon_n}^-}\right) = \mathcal{O}(1)$, the linear part of the estimation error converges:
\begin{align}\label{eq:Bhat_convergence}
    \frac{\widehat{B}_{\epsilon_n}^+-{B}_{\epsilon_n}^+}{{B}_{\epsilon_n}^-} -\left(\frac{{B}_{\epsilon_n}^+}{{B}_{\epsilon_n}^-}\right)\frac{\widehat{B}_{\epsilon_n}^--{B}_{\epsilon_n}^-}{{B}_{\epsilon_n}^-} &= \mathcal{O}_p\left((n\epsilon_n)^{-1/2}\right)
\end{align}

\vskip 0.1in
{\it\noindent Remainder Term:} Next we bound the remainder term $r_n$. First consider the event: $\widehat{B}_{\epsilon_n}^- \geq B_{\epsilon_n}^{-}/2$. Conditional on this event:
\begin{align}\label{eq:rn_bound}
 \left|  r_n \right| &\leq 2\left|   \left(\frac{\widehat{B}_{\epsilon_n}^--{B}_{\epsilon_n}^-}{{B}_{\epsilon_n}^-{B}_{\epsilon_n}^-}\right)(|\widehat{B}_{\epsilon_n}^+-{B}_{\epsilon_n}^+|+\left(\frac{{B}_{\epsilon_n}^+}{{B}_{\epsilon_n}^-}\right)|\widehat{B}_{\epsilon_n}^--{B}_{\epsilon_n}^-| ) \right|= \mathcal{O}_p\left((n\epsilon_n)^{-1}\right)
\end{align}
\noindent By Lemma \ref{lem:Bhat}, $\text{Pr}\left( \widehat{B}_{\epsilon_n}^- \geq B_{\epsilon_n}^{-}/2\right) \to 1$. Since the probability limit in Equation (\ref{eq:rn_bound}) holds conditional on an event with probability going to one, the following holds unconditionally:
\begin{align}\label{eq:theta_remainder}
   r_n &= \mathcal{O}_p\left((n\epsilon_n)^{-1}\right)
\end{align}

\vskip 0.1in
{\it\noindent Conclusion:} Substituting Equations (\ref{eq:Bhat_convergence})  and (\ref{eq:theta_remainder}) into Equation (\ref{eq:theta_linearized}) yields Claim 2:
\begin{align}
     \widehat{\theta}_n^{-1} -\widetilde{\theta}^{-1}_n &= \mathcal{O}_p\left((n\epsilon_n)^{-1/2}\right)
\end{align}

\vskip 0.1in
{\bf\noindent Proof of Claim 3}

Since $\widehat{B}_{\epsilon_n}^+,\widehat{B}_{\epsilon_n}^-$ are sample averages, the estimation error in Equation (\ref{eq:theta_linearized}) can be rewritten in terms of an influence function $X_n(t;b^+,b^{-})$ defined below:
\begin{align}
     \widehat{\theta}_n^{-1} -\widetilde{\theta}^{-1}_n &= \frac{1}{n}\sum_{i=1}^n X_n(t_i;B_{\epsilon_n}^+,B_{\epsilon_n}^{-}) - \mathbb{E}\left[X_n(T;B_{\epsilon_n}^+,B_{\epsilon_n}^{-})\mid R=1\right]+\mathcal{O}_p\left((n\epsilon_n)^{-1}\right)\\
     X_n(t;b^+,b^{-}) &\equiv \frac{1}{b^{-}}\mathbf{1}\left\{|t|\in (1.96,1.96+\epsilon_n]\right\} - \frac{b^{+}}{(b^{-})^2}\mathbf{1}\left\{|t|\in (1.96-\epsilon_n,1.96]\right\}
\end{align}

\vskip 0.1in
{\bf\noindent Proof of Last Claim}

Since $(a+b)^2 \leq 2a^2+2b^2$ we have:
\begin{align*}
    \mathbb{E}[X_n(T;B_{\epsilon_n}^+,B_{\epsilon_n}^{-})^2\mid R=1] \leq 2\frac{B_{\epsilon_n}^+}{(B_{\epsilon_n}^{-})^2} +2\frac{(B_{\epsilon_n}^+)^2 B_{\epsilon_n}^{-}}{(B_{\epsilon_n}^{-})^4}
\end{align*}

 Lemma \ref{lem:Bbounds} guarantees that $B_{\epsilon_n}^+ = \mathcal{O}\left(\epsilon_n\right)$ and $\liminf_{n\to \infty}\epsilon_n^{-1}B_{\epsilon_n}^- >0$. Substituting these in:
\begin{align*}
 \mathbb{E}[X_n(T;B_{\epsilon_n}^+,B_{\epsilon_n}^{-})^2\mid R=1]
&=\mathcal{O}\left(\epsilon_n^{-1}\right)
\end{align*}

\end{proof}

\subsection{Lemma \ref{lem:linearize_w} and proof}\label{proof:lem:linearize_w}
%audited 4/3/26 % audited 5/23/26

Recall the definition of $W(t;\theta,p) \equiv \frac{1+\left({\theta}^{-1}-1\right)\mathbf{1}\left\{|t|< 1.96\right\}}{1+({\theta}^{-1}-1)p}$ from Appendix \ref{app:inference_definitions}.

\begin{lemma}\label{lem:linearize_w}
  Assume that publication bias follows Equations (\ref{eq:w_definition}) and (\ref{eq:pb_caliper_form}), that Assumptions \ref{assum:normality_of_T}-\ref{assum:wnonsmooth} hold,  that $\epsilon_n \to 0$, and that $n\epsilon_n\to \infty$. Then:
  \begin{align*}
     & \sup_{t\in\mathbb{R}}  \left| W(t;\widehat{\theta}_n,\widehat{F}_n(1.96))-W(t;{\theta}_0,{F}(1.96))-\left(\widehat{\theta}_n^{-1}-{\theta}_0^{-1}\right) \frac{\mathbf{1}\left\{|t|< 1.96\right\}-{F}(1.96)}{\left(1+{F}(1.96)({\theta}_0^{-1}-1)\right)^2}  \right|\\
      &=  \mathcal{O}_p\left(\left(\widehat{\theta}_n^{-1}-\theta_0^{-1}\right)^2+n^{-1/2}\right) 
    \end{align*}
\end{lemma}
\begin{proof}
    We will prove this in two steps. In step 1 we show that replacing $\widehat{F}_n(1.96)$ with $F(1.96)$ incurs an error of at most $\mathcal{O}_p\left(n^{-1/2}\right)$. In step 2 we linearize about $\widehat{\theta}_n^{-1}=\theta_0^{-1}$.

\vskip 0.1in
{\bf Step 1:  Replacing $\widehat{F}_n(1.96)$ with $F(1.96)$}

From the definition in Appendix \ref{app:inference_definitions}:
\begin{align*}
    \left|W(t;\widehat{\theta}_n,\widehat{F}_n(1.96))-W(t;\widehat{\theta}_n,{F}(1.96))\right| &= \left| \frac{1+\left(\widehat{\theta}_n^{-1}-1\right)\mathbf{1}\left\{|t|< 1.96\right\}}{1+(\widehat{\theta}_n^{-1}-1)\widehat{F}_n(1.96)}-\frac{1+\left(\widehat{\theta}_n^{-1}-1\right)\mathbf{1}\left\{|t|< 1.96\right\}}{1+(\widehat{\theta}_n^{-1}-1){F}(1.96)}\right|\\
    &\leq (1+ \widehat{\theta}_n^{-1})\left| \frac{1 }{1+(\widehat{\theta}_n^{-1}-1)\widehat{F}_n(1.96)}-\frac{1 }{1+(\widehat{\theta}_n^{-1}-1){F}(1.96)}\right|\\
   &=  (1+ \widehat{\theta}_n^{-1})\left|  \frac{(\widehat{\theta}_n^{-1}-1)({F}(1.96)-\widehat{F}_n(1.96))}{(1+(\widehat{\theta}_n^{-1}-1){F}(1.96))(1+(\widehat{\theta}_n^{-1}-1)\widehat{F}_n(1.96))}\right|
\end{align*} Since the only dependence on $t$ is through the indicator
$\mathbf{1}\{|t|<1.96\}$, all bounds are uniform in $t\in\mathbb R$.

To show the convergence of the right-hand side, we will lower-bound its denominator on an event that has high probability. First use the identity that $1+(a-1)p = 1-p +ap \geq \min\{1,a\}$ for $p\in[0,1]$ and $a\geq 0$. So $(1+(\widehat{\theta}_n^{-1}-1){F}(1.96))(1+(\widehat{\theta}_n^{-1}-1)\widehat{F}_n(1.96)) \geq \min\{1,\widehat{\theta}_n^{-2}\}$. Consider the event $\{\widehat{\theta}_n^{-1}\in [\overline M^{-1}/2,2\overline M]\}$. Conditional on this event, the denominator is lower-bounded by $\overline{M}^{-2}/4$. Thus, on the event $\{\widehat{\theta}_n^{-1}\in[\overline M^{-1}/2,2\overline M]\}$,
\begin{align*}
    \sup_{t\in\mathbb R}
\left|W(t;\widehat{\theta}_n,\widehat{F}_n(1.96))
      -W(t;\widehat{\theta}_n,F(1.96))\right|
\leq C\,|\widehat{F}_n(1.96)-F(1.96)|,
\end{align*}
for a deterministic constant $C>0$ depending only on $\overline M$. Since
$\Pr(\widehat{\theta}_n^{-1}\in[\overline M^{-1}/2,2 \overline{M}])\to 1$ by Lemma \ref{lem:theta_estimation},
the same bound holds with probability tending to one.

By Assumption \ref{assum:cross_study_independence},  the indicators $\mathbf{1}\left\{|t_i| \leq 1.96\right\}$ are identically distributed and independent across bounded-size blocks, so their sample mean has variance $\mathcal{O}(n^{-1})$. Hence,  ${F}(1.96)-\widehat{F}_n(1.96) = \mathcal{O}_p\left(n^{-1/2}\right)$. Therefore:
\begin{align}\label{eq:remove_Fhat}
    \sup_{t\in\mathbb{R}}\left|W(t;\widehat{\theta}_n,\widehat{F}_n(1.96))-W(t;\widehat{\theta}_n,{F}(1.96))\right| &= \mathcal{O}_p\left(n^{-1/2}\right)
\end{align}

\vskip 0.1in
{\bf Step 2: Linearizing about $\widehat{\theta}_n^{-1} = \theta_0^{-1}$}

In this step we linearize the $\theta^{-1}$ argument. Taking the first-order Taylor expansion about $\widehat{\theta}_n^{-1} = \theta_0^{-1}$:
\begin{align*}
     W(t;\widehat{\theta}_n,{F}(1.96)) &= W(t;{\theta}_0,{F}(1.96)) +\left(\widehat{\theta}_n^{-1}-{\theta}_0^{-1}\right)  \frac{\mathbf{1}\left\{|t|< 1.96\right\}-{F}(1.96)}{\left(1+{F}(1.96)({\theta}_0^{-1}-1)\right)^2}\\
     &- \left(\widehat{\theta}_n^{-1}-{\theta}_0^{-1}\right)^2 \frac{\mathbf{1}\left\{|t|< 1.96\right\}-{F}(1.96)}{\left(1+{F}(1.96)(\zeta-1)\right)^3}{F}(1.96)
\end{align*}
for some random number $\zeta$ between $\widehat{\theta}_n^{-1}$ and $\theta_0^{-1}$. Rearranging and using the bound $|\mathbf{1}\left\{|t|< 1.96\right\}-{F}(1.96)|{F}(1.96)\leq 1$ yields:
  \begin{align*}
     & \sup_{t\in\mathbb{R}}  \left| W(t;\widehat{\theta}_n,{F}(1.96))-W(t;{\theta}_0,{F}(1.96))-\left(\widehat{\theta}_n^{-1}-{\theta}_0^{-1}\right) \frac{\mathbf{1}\left\{|t|< 1.96\right\}-{F}(1.96)}{\left(1+{F}(1.96)({\theta}_0^{-1}-1)\right)^2}  \right|\\
      &\leq  \left(\widehat{\theta}_n^{-1}-{\theta}_0^{-1}\right)^2 \left|\frac{1}{\left(1+{F}(1.96)(\zeta-1)\right)^3}\right|
    \end{align*}

 Assumption \ref{assum:pblowerbound} says that  $\theta_0^{-1}\in [1, \overline{M}]$. We already showed that $\text{Pr}\left(\widehat{\theta}_n^{-1} \in [\overline{M}^{-1}/2,2 \overline{M}]\right)\to 1$ by Lemma \ref{lem:theta_estimation}. Since $\zeta$ is between $\theta_0^{-1}$ and $\widehat{\theta}_n^{-1}$, this implies that  $\text{Pr}\left(\zeta \in [\overline{M}^{-1}/2,2\overline{M}]\right)\to 1$ and $\text{Pr}\left(\left(1+{F}(1.96)(\zeta-1)\right)^3 \geq \frac{1}{8\overline{M}^3}\right)\to 1$. Therefore:
\begin{align*}
 \left(\widehat{\theta}_n^{-1}-{\theta}_0^{-1}\right)^2\frac{1}{\left(1+{F}(1.96)(\zeta-1)\right)^3}=\mathcal{O}_p\left(\left(\widehat{\theta}_n^{-1}-{\theta}_0^{-1}\right)^2\right)
\end{align*}

By substitution:
  \begin{align*}
     & \sup_{t\in\mathbb{R}}  \left| W(t;\widehat{\theta}_n,{F}(1.96))-W(t;{\theta}_0,{F}(1.96))-\left(\widehat{\theta}_n^{-1}-{\theta}_0^{-1}\right) \frac{\mathbf{1}\left\{|t|< 1.96\right\}-{F}(1.96)}{\left(1+{F}(1.96)({\theta}_0^{-1}-1)\right)^2}  \right|\\
      &=  \mathcal{O}_p\left(\left(\widehat{\theta}_n^{-1}-\theta_0^{-1}\right)^2\right) 
    \end{align*}

    Combining this with Equation (\ref{eq:remove_Fhat}) using the triangle inequality yields the final result.
\end{proof}

\subsection{Lemma \ref{lem:oracle_feasible_variance} and proof}\label{proof:lem:oracle_feasible_variance}
%audited 4/5/26 %auidted 5/24/26

Recall that $m^0_n$ is the true influence function and $\widehat{m}_n$ is its empirical counterpart. Recall that $F,\widehat{F}_n$ denote the true and empirical CDFs of $|T|$ conditional on $R=1$. Recall also that $B_{\epsilon_n}^-$ denotes $F(1.96)-F(1.96-\epsilon_n)$, $B_{\epsilon_n}^+$ denotes $F(1.96+\epsilon_n)-F(1.96)$, and $\widehat{B}_{\epsilon_n}^-,\widehat{B}_{\epsilon_n}^+$ denote their empirical counterparts.  These are defined formally in Appendix \ref{app:inference_definitions}. 

\begin{lemma}\label{lem:oracle_feasible_variance}
    Under the conditions of Theorem \ref{thm:variance_estimation},
  $$\sup_{t\in \mathbb{R}}|m_n^0(t) - \widehat{m}_n(t)|= \mathcal{O}_p(\lambda_{J_n}^{-1})$$

\end{lemma}
\begin{proof}
   We will use the mean value theorem to show that replacing the oracle influence function $m_n^0(t)$ with the feasible $\widehat{m}_n(t)$ leads to an error that grows at a sufficiently slow rate for consistent variance estimation. The proof proceeds in five steps. First we bound the difference with the mean value theorem. Second, we bound the rate of convergence for each nuisance parameter one by one. Third, we take derivatives and fourth we show with high probability that those derivatives are all bounded. Fifth, we combine the previous steps to conclude the proof.

\vskip 0.1in
{\noindent \bf Step 1:  Mean Value Theorem}

Define $\nu \in \mathbb{R}^5$ as the vector of nuisance parameters. Let $\nu_0$ be the vector of true values and $\widehat{\nu}_n$ be their estimates.
\begin{align*}
    \nu &\equiv ( \theta^{-1}, p,b^+,b^-,q  )\\
    \nu_0 &\equiv ( \theta_0^{-1}, F(1.96),B^+_{\epsilon_n},B^-_{\epsilon_n},Q_n  )\\
    \widehat{\nu}_n &\equiv ( \widehat{\theta}_n^{-1}, \widehat{F}_n(1.96),\widehat{B}^+_{\epsilon_n},\widehat{B}^-_{\epsilon_n},\widehat{Q}_n  )
\end{align*}

We can define $m_n(t;\nu) \equiv \left(\sum_{j=0}^{J_n}a_j \psi_j(t)\varphi_{\sigma_T^2}(t)\right) \left(\frac{1+\left({\theta}^{-1}-1\right)\mathbf{1}\left\{|t|< 1.96\right\}}{1+({\theta}^{-1}-1)p} \right)+qX_n(t;b^+,b^{-})$ and rewrite $m_n^0(t)= m_n(t;\nu_0)$ and $\widehat{m}_n(t)=m_n(t;\widehat{\nu}_n)$. Define $\Gamma_n\subseteq \mathbb{R}^5$ as the set of all possible vectors $\nu$ such that all components lie in between $\widehat{\nu}_n$ and $\nu_0$ (i.e. the coordinate-wise rectangle). 

Next we verify differentiability of the influence function. Consider the event:
\begin{align}\label{eq:def_Tn}
    \mathcal{T}_n \equiv \widehat{B}_{\epsilon_n}^- \geq {B}_{\epsilon_n}^-/2\text{ and }\widehat{\theta}_n^{-1}\in [ \overline{M}^{-1}/2, 2\overline{M}] \text{ and } \widehat{F}_n(1.96)\geq F(1.96)/2
\end{align}
First, we will show that  $\text{Pr}\left(\mathcal{T}_n\right)\to 1$. Since $\theta_0\geq \overline{M}^{-1}>0$ by Assumption \ref{assum:pblowerbound} and $\widehat{\theta}_n\to_p \theta_0$ by Lemma \ref{lem:theta_estimation}, we have $\text{Pr}\left(\widehat{\theta}_n^{-1}\leq 2\overline{M}\right)\to 1$. Similarly,  $\text{Pr}\left(\widehat{\theta}_n^{-1}\geq \overline{M}^{-1}/2\right)\to 1$. By Lemma \ref{lem:Bhat}, $\text{Pr}\left( \widehat{B}_{\epsilon_n}^-\geq  {B}_{\epsilon_n}^-/2\right)\to 1$. By Lemma \ref{lem:Bbounds}, $F(1.96)>0$. By Assumption \ref{assum:cross_study_independence} and Chebyshev's Inequality, $\widehat{F}_n(1.96)\to_p F(1.96)>0$. So $\text{Pr}\left(\widehat{F}_n(1.96)\geq F(1.96)/2\right)\to 1$. Since all three events occur with probability approaching one, their intersection also occurs with probability approaching one. So  $ \text{Pr}\left( \mathcal{T}_n\right) \to 1 $.

On the event $\mathcal{T}_n$, the function $m_n(t;\nu)$ is differentiable in all arguments of $\nu$ for all $\nu \in \Gamma_n$ for every fixed $t$, because it is a rational function of $\nu$ with denominators bounded away from zero on $\mathcal{T}_n$.  By the Mean Value Theorem:
\begin{align}
 \sup_{t\in\mathbb{R}} \left|m_n^0(t)-\widehat{m}_n(t)\right| &= \sup_{t\in\mathbb{R}} \left|m_n(t;{\nu}_0)-m_n(t;\widehat{\nu}_n)\right|\\
 &\leq  
 |B^-_{\epsilon_n}-\widehat{B}^-_{\epsilon_n}| \sup_{\nu \in \Gamma_n,t\in\mathbb{R}}\left|\frac{\partial}{\partial b^-}m_n(t;\nu)\right|\\
  &\quad +  |B^+_{\epsilon_n}-\widehat{B}^+_{\epsilon_n}|  \sup_{\nu \in \Gamma_n,t\in\mathbb{R}}\left|\frac{\partial}{\partial b^+}m_n(t;\nu)\right|  \\
   &\quad + |\theta_0^{-1}-\widehat{\theta}_n^{-1}| \sup_{\nu \in \Gamma_n,t\in\mathbb{R}}\left|\frac{\partial}{\partial \theta^{-1}}m_n(t;\nu)\right| \\
    &\quad +  |F(1.96)-\widehat{F}_n(1.96)| \sup_{\nu \in \Gamma_n,t\in\mathbb{R}}\left|\frac{\partial}{\partial p}m_n(t;\nu)\right|  \\
    &\quad + |Q_n-\widehat{Q}_n|\sup_{\nu \in \Gamma_n,t\in\mathbb{R}}\left|\frac{\partial}{\partial q}m_n(t;\nu)\right| 
\end{align}

\vskip 0.1in
{\noindent\bf Step 2:  Bounding Estimation errors}

 Now we bound all the estimation errors one by one. By Assumption \ref{assum:cross_study_independence} and Chebyshev's Inequality, we have $\widehat{F}_n(1.96)-F(1.96)=\mathcal{O}_p\left(n^{-1/2}\right)$. Lemma \ref{lem:theta_estimation} already proved that $\widehat{\theta}_n^{-1}-\theta_0^{-1} = \mathcal{O}_p\left(\epsilon_n+\epsilon_n^{-1/2}n^{-1/2}\right)$. By Lemma \ref{lem:Bhat}, $\widehat{B}_{\epsilon_n}^+-B_{\epsilon_n}^+ =\mathcal{O}_p\left(\epsilon_n^{1/2}n^{-1/2}\right)$ and $\widehat{B}_{\epsilon_n}^--B_{\epsilon_n}^- =\mathcal{O}_p\left(\epsilon_n^{1/2}n^{-1/2}\right)$. 

 Finally we bound $\widehat{Q}_n-Q_n$. We will do this in two steps. Define $\widehat{Q}^*_n$ and recall the definition of $\widehat{Q}_n$:
 \begin{align*}
     \widehat{Q}^*_n &= \frac{1}{n}\sum_{i=1}^n\sum_{j=0}^{J_n} a_j \psi_j(t_i)\varphi_{\sigma_T^2}(t_i) \left(\frac{\mathbf{1}\left\{|t_i|< 1.96\right\}-{F}(1.96)}{\left(1+{F}(1.96)({\theta}_0^{-1}-1)\right)^2}\right)\\
     \widehat{Q}_n &= \frac{1}{n}\sum_{i=1}^n\sum_{j=0}^{J_n} a_j \psi_j(t_i)\varphi_{\sigma_T^2}(t_i) \left(\frac{\mathbf{1}\left\{|t_i|< 1.96\right\}-\widehat{F}_n(1.96)}{\left(1+\widehat{F}_n(1.96)(\widehat{\theta}_n^{-1}-1)\right)^2}\right)
 \end{align*}

By the triangle inequality: $|\widehat{Q}_n-Q_n|\leq |\widehat{Q}_n-\widehat{Q}^*_n|+|\widehat{Q}^*_n-Q_n|$. We will bound the two terms on the right-hand side one at a time. First we take derivatives:
\begin{align*}
     \frac{ \partial\widehat{Q}_n}{\partial\widehat{F}_n(1.96)} &= \frac{1}{n}\sum_{i=1}^n\sum_{j=0}^{J_n} a_j \psi_j(t_i)\varphi_{\sigma_T^2}(t_i) \left(\frac{-1}{\left(1+\widehat{F}_n(1.96)(\widehat{\theta}_n^{-1}-1)\right)^2}\right)\\
     &\qquad -\frac{2}{n}\sum_{i=1}^n\sum_{j=0}^{J_n} a_j \psi_j(t_i)\varphi_{\sigma_T^2}(t_i)\left(\frac{(\mathbf{1}\left\{|t_i|< 1.96\right\}-\widehat{F}_n(1.96))(\widehat{\theta}_n^{-1}-1)}{\left(1+\widehat{F}_n(1.96)(\widehat{\theta}_n^{-1}-1)\right)^3}\right)\\
     \frac{ \partial\widehat{Q}_n}{\partial\widehat{\theta}_n^{-1}} &= -\frac{2}{n}\sum_{i=1}^n\sum_{j=0}^{J_n} a_j \psi_j(t_i)\varphi_{\sigma_T^2}(t_i) \left(\frac{\left(\mathbf{1}\left\{|t_i|< 1.96\right\}-\widehat{F}_n(1.96)\right)\widehat{F}_n(1.96)}{\left(1+\widehat{F}_n(1.96)(\widehat{\theta}_n^{-1}-1)\right)^3}\right)
\end{align*}

By Lemma \ref{lem:bound_a_jpsi_j}, $\sup_{t\in\mathbb{R}}\left|\sum_{j=0}^{J_n} a_j \psi_j(t)\varphi_{\sigma_T^2}(t)\right|=\mathcal{O}\left(\lambda_{J_n}^{-1}\right)$. Recall the event $\mathcal{T}_n$ defined in Equation (\ref{eq:def_Tn}). On   $\mathcal{T}_n$, both derivatives times $\lambda_{J_n}$ are uniformly bounded. So by the mean value theorem:
\begin{align}\label{eq:Q_conditional}
    \widehat{Q}_n- \widehat{Q}^*_n = \mathcal{O}_p\left(\lambda_{J_n}^{-1}|\widehat{\theta}_n^{-1}-\theta_0^{-1}|+\lambda_{J_n}^{-1}|\widehat{F}_n(1.96)-F(1.96)|\right) \qquad \text{conditional on $\mathcal{T}_n$}
\end{align}

Since  $\text{Pr}\left(\mathcal{T}_n\right)\to 1$, and the bound in Equation (\ref{eq:Q_conditional}) holds on $\mathcal{T}_n$, the same $\mathcal{O}_p(\cdot)$ bound holds unconditionally.  By Lemma \ref{lem:theta_estimation} and Chebyshev, $ |\widehat{\theta}_n^{-1}-\theta_0^{-1}|+|\widehat{F}_n(1.96)-F(1.96)| =\mathcal{O}_p\left(\epsilon_n+\epsilon_n^{-1/2}n^{-1/2}\right)$. So $\widehat{Q}_n- \widehat{Q}^*_n = \mathcal{O}_p\left(\lambda_{J_n}^{-1}(\epsilon_n+\epsilon_n^{-1/2}n^{-1/2})\right)$. Since the summands of $\lambda_{J_n}\widehat{Q}^*_n$ are uniformly bounded almost surely by Lemma \ref{lem:bound_a_jpsi_j}, and the $T$ are weakly dependent by Assumption \ref{assum:cross_study_independence}, $\mathbb{V}[\widehat{Q}^*_n] =\mathcal{O}\left(\lambda_{J_n}^{-2}n^{-1}\right)$. So by Chebyshev's inequality we have $\widehat{Q}^*_n-\mathbb{E}[\widehat{Q}^*_n] = \mathcal{O}_p\left(\lambda_{J_n}^{-1}n^{-1/2}\right)$. Notice that by Assumption \ref{assum:cross_study_independence} the $T$ are identically distributed, so $\mathbb{E}[\widehat{Q}^*_n]=Q_n$. So  $\widehat{Q}^*_n-Q_n = \mathcal{O}_p\left(\lambda_{J_n}^{-1}n^{-1/2}\right)$.  Finally, by the triangle inequality: 
\begin{align}
    \widehat{Q}_n- {Q}_n = \mathcal{O}_p\left(\lambda_{J_n}^{-1}(\epsilon_n+\epsilon_n^{-1/2}n^{-1/2})\right)
\end{align}

\vskip 0.1in
{\bf \noindent Step 3:  Bounding Derivatives}

 In this step we take derivatives with respect to each nuisance parameter in $\nu$. Upper-bounding the derivatives with respect to $b^+,b^-$:
\begin{align*}
    \left|\frac{\partial}{\partial b^+}m_n(t;\nu) \right| &=  \left|q\frac{\partial}{\partial b^+}X_{n}(t;b^+,b^-) \right| \leq |q| \frac{1}{(b^-)^2} \\
    \left|\frac{\partial}{\partial b^-}m_n(t;\nu) \right| &=  \left|q\frac{\partial}{\partial b^-}X_{n}(t;b^+,b^-) \right| \leq |q|  \left(\frac{1}{(b^-)^2}+\frac{2b^+}{(b^-)^3}  \right)
\end{align*}

Bounding the next derivative with respect to $\theta^{-1}$:
\begin{align*}
        \left|\frac{\partial}{\partial \theta^{-1}}m_n(t;\nu) \right|&=\left|\sum_{j=0}^{J_n}a_j \psi_j(t)\varphi_{\sigma_T^2}(t)\right| \left|\frac{\mathbf{1}\left\{|t|< 1.96\right\}-p}{(1+({\theta}^{-1}-1)p)^2}\right|\leq \left|\sum_{j=0}^{J_n}a_j \psi_j(t)\varphi_{\sigma_T^2}(t)\right| \left|\frac{2}{(1+({\theta}^{-1}-1)p)^2}\right|
\end{align*}

Bounding the next derivative with respect to $p$:
\begin{align*}
  \left|\frac{\partial}{\partial p}m_n(t;\nu) \right|&=\left|\sum_{j=0}^{J_n}a_j \psi_j(t)\varphi_{\sigma_T^2}(t)\right| \left|\frac{(\theta^{-1}-1)(1+\left({\theta}^{-1}-1\right)\mathbf{1}\left\{|t|< 1.96\right\})}{(1+({\theta}^{-1}-1)p)^2}\right|\\
  &\leq \left|\sum_{j=0}^{J_n}a_j \psi_j(t)\varphi_{\sigma_T^2}(t)\right| \frac{(1+\theta^{-1})^2}{(p\theta^{-1}+1-p)^2}
\end{align*}

Bounding the next derivative with respect to $q$:
\begin{align*}
    \left|\frac{\partial}{\partial q}m_n(t;\nu)\right| &= \left|X_n(t;b^+,b^{-})\right| 
    \leq \frac{1}{b^-}+\frac{b^+}{(b^-)^2} 
\end{align*}

\vskip 0.1in
{\bf \noindent Step 4: Suprema of the Derivatives}

To maximize the derivatives over the set of nuisance parameters $\Gamma_n$, we first show that with high probability the estimated vector of nuisance parameters $\widehat{\nu}_n$ is not too large or too close to zero. First, by Lemma \ref{lem:bound_a_jpsi_j}, $Q_n = \mathcal{O}\left(\lambda_{J_n}^{-1}\right)$. By Lemma \ref{lem:Bbounds}, $B^+_{\epsilon_n},B^{-}_{\epsilon_n} = \mathcal{O}\left(\epsilon_n\right)$ and $\liminf_{n\to\infty}\epsilon_n^{-1}B^{-}_{\epsilon_n} >0$. Combining these facts with the bounds on estimation errors from Step 2 above, we conclude that there are constants $L_p,L_Q,L^-_b,L^+_b$ such that with probability approaching one,  all $\nu \in \Gamma_n$ must have elements that satisfy: $ p > L_p, \: 
    |q| <  L_Q \lambda_{J_n}^{-1}, \: 
    \theta^{-1} \in \left[\overline{M}^{-1}/2, 2\overline{M}\right], \:
   b^- > L^-_{b}\epsilon_n, \text{ and }  
    b^+ < L^+_{b}\epsilon_n$. Since  these bounds hold with probability approaching one, we can substitute them into the derivative bounds from the previous step to obtain:
\begin{align}
  \sup_{\nu \in \Gamma_n,t\in\mathbb{R}} \left|\frac{\partial}{\partial b^+}m_n(t;\nu) \right| &= \mathcal{O}_p\left(\lambda_{J_n}^{-1}\epsilon_n^{-2}\right)\label{eq:begin_derivative_suprema} \\
   \sup_{\nu \in \Gamma_n,t\in\mathbb{R}} \left|\frac{\partial}{\partial b^-}m_n(t;\nu) \right| &= \mathcal{O}_p\left(\lambda_{J_n}^{-1}\epsilon_n^{-2}\right)\\
   \sup_{\nu \in \Gamma_n,t\in\mathbb{R}}   \left|\frac{\partial}{\partial \theta^{-1}}m_n(t;\nu) \right| &= \mathcal{O}_p\left(\lambda_{J_n}^{-1} \right)\\
     \sup_{\nu \in \Gamma_n,t\in\mathbb{R}}  \left|\frac{\partial}{\partial p}m_n(t;\nu) \right|   &=  \mathcal{O}_p\left(\lambda_{J_n}^{-1} \right)\\
           \sup_{\nu \in \Gamma_n,t\in\mathbb{R}}    \left|\frac{\partial}{\partial q}m_n(t;\nu)\right| &=  \mathcal{O}_p\left(\epsilon_n^{-1}\right) \label{eq:end_derivative_suprema}
\end{align}

{\noindent\bf Step 5: Combining with Mean Value Theorem}

Now we can substitute the bounds on the estimation errors and the bounds on the suprema of the derivatives in Equations (\ref{eq:begin_derivative_suprema})-(\ref{eq:end_derivative_suprema}) into the mean value theorem. 
\begin{align*}
  &\sup_{t\in\mathbb{R}}  \left|m_n^0(t)-\widehat{m}_n(t)\right| \\
  &= \mathcal{O}_p\left(\lambda_{J_n}^{-1}\epsilon_n^{-2}\epsilon_n^{1/2}n^{-1/2}+\lambda_{J_n}^{-1}(\epsilon_n+\epsilon_n^{-1/2}n^{-1/2})+\lambda_{J_n}^{-1}n^{-1/2}+\epsilon_n^{-1}\lambda_{J_n}^{-1}(\epsilon_n+\epsilon_n^{-1/2}n^{-1/2})\right)
\end{align*}

Since $\epsilon_n\sim n^{-1/3}$ by assumption,  the dominant rate is $\lambda_{J_n}^{-1}$.

The preceding bounds hold on the event $\mathcal{T}_n$. Since $ \text{Pr}\left( \mathcal{T}_n\right) \to 1 $, the same
$\mathcal{O}_p(\lambda_{J_n}^{-1})$ conclusion holds unconditionally.

\end{proof}

\subsection{Lemma \ref{lem:Bhat} and proof}\label{proof:lem:Bhat}
%audited 4/4/26 % Audited 5/23/26

\begin{lemma}\label{lem:Bhat}
      Assume that publication bias follows Equations (\ref{eq:w_definition}) and (\ref{eq:pb_caliper_form}), that Assumptions \ref{assum:normality_of_T}-\ref{assum:wnonsmooth} hold,  that $\epsilon_n \to 0$, and that $n\epsilon_n\to \infty$. Then the following hold:
      \begin{enumerate}
      \item   $\widehat{B}_{\epsilon_n}^+-{B}_{\epsilon_n}^+ =\mathcal{O}_p\left(\epsilon_n^{1/2}n^{-1/2}\right)$ and $\widehat{B}_{\epsilon_n}^--{B}_{\epsilon_n}^- =\mathcal{O}_p\left(\epsilon_n^{1/2}n^{-1/2}\right)$
          \item $\text{Pr}\left( \widehat{B}_{\epsilon_n}^-\geq  {B}_{\epsilon_n}^-/2\right)\to 1$
      \end{enumerate}
\end{lemma}

\begin{proof}

{\bf\noindent Proof of Claim 1 } Recall from Appendix \ref{app:inference_definitions} that $F,\widehat{F}_n$ denote the true and empirical CDFs of $|T|$ conditional on publication $R=1$. Recall also that $ \widehat{B}_{\epsilon_n}^+$ is defined as:
\begin{align}
{B}_{\epsilon_n}^+ &\equiv {F}(1.96+\epsilon_n)-{F}(1.96) \\
     \widehat{B}_{\epsilon_n}^+ &\equiv \widehat{F}_n(1.96+\epsilon_n)-\widehat{F}_n(1.96) = \frac{1}{n}\sum_{i=1}^n \mathbf{1}\left\{|t_i| \in (1.96,1.96+\epsilon_n]\right\}
\end{align}

Notice that $\mathbb{E}\left[\widehat{B}_{\epsilon_n}^+ \right] = {B}_{\epsilon_n}^+$. So we need only bound the variance. Since $\widehat{B}_{\epsilon_n}^+$ is a sample average, its variance can be expressed as:
$$\mathbb{V}\left[\widehat{B}_{\epsilon_n}^+\right] = \frac{1}{n^2}\sum_{i=1}^n\sum_{j=1}^n \text{COV}\left(\mathbf{1}\left\{|t_i| \in (1.96,1.96+\epsilon_n]\right\},\mathbf{1}\left\{|t_j| \in (1.96,1.96+\epsilon_n]\right\} \right)$$

 By Assumption \ref{assum:cross_study_independence}, each observation is identically distributed and independent of all but at most $C_B$ others. So, the variance can be upper-bounded
\begin{align}
    \mathbb{V}\left[\widehat{B}_{\epsilon_n}^+ \right] &\leq \frac{C_B}{n}\mathbb{V}\left[\mathbf{1}\left\{|T| \in (1.96,1.96+\epsilon_n]\right\}\mid R=1\right]\\
    &\leq \frac{C_B}{n}\text{Pr}\left(|T| \in (1.96,1.96+\epsilon_n]\mid R=1\right) 
    = \frac{C_B}{n}B_{\epsilon_n}^{+}
\end{align}
  Lemma \ref{lem:Bbounds} guarantees that $B_{\epsilon_n}^{+} = \mathcal{O}\left(\epsilon_n\right)$ and $B_{\epsilon_n}^{-} = \mathcal{O}\left(\epsilon_n\right)$. So by Chebyshev's Inequality:
\begin{equation}\label{eq:Bplus_conv_exp2}
     \widehat{B}_{\epsilon_n}^+-{B}_{\epsilon_n}^+ =\mathcal{O}_p\left(\epsilon_n^{1/2} n^{-1/2}\right)
\end{equation}

A similar argument shows that $\widehat{B}_{\epsilon_n}^--{B}_{\epsilon_n}^- =\mathcal{O}_p\left(\epsilon_n^{1/2}n^{-1/2}\right)$.

\vskip 0.1in
{\noindent \bf Proof of Claim 2}

 By claim 1, $\widehat{B}_{\epsilon_n}^-- {B}_{\epsilon_n}^- = \mathcal{O}_p\left(\epsilon_n^{1/2}n^{-1/2}\right)$. By Lemma \ref{lem:Bbounds}, $\liminf_{n\to \infty}\epsilon_n^{-1}B_{\epsilon_n}^->0$, so $\frac{\widehat{B}_{\epsilon_n}^-- {B}_{\epsilon_n}^-}{{B}_{\epsilon_n}^-} =\mathcal{O}_p\left(\epsilon_n^{-1/2}n^{-1/2}\right)$. By assumption, $n\epsilon_n\to \infty$, so  $\frac{\widehat{B}_{\epsilon_n}^-- {B}_{\epsilon_n}^-}{{B}_{\epsilon_n}^-} ={o}_p\left(1\right)$. So we have: 
    \begin{align}\label{eq:Bhatminus_not_small}
          \text{Pr}\left( \widehat{B}_{\epsilon_n}^- \geq B_{\epsilon_n}^{-}/2\right) =\text{Pr}\left( \frac{\widehat{B}_{\epsilon_n}^-- {B}_{\epsilon_n}^-}{{B}_{\epsilon_n}^-}\geq -\frac{1}{2}\right)\to 1
    \end{align}
    
\end{proof}

\clearpage

\section{Online Appendix} 
\subsection{Many Labs Robustness Check}

I exploit a second special property of the Many Labs project to construct a key robustness check that does not suffer from statistical imprecision or wide confidence intervals. A key feature of the Many Labs setting is that it is plausible to assume that all of the research teams were investigating the same (or very similar) true effects. \cite{ManyLabs} conclude that variation in the true effect size across study sites was found to be very small compared to the variation in the effect sizes across the experimental treatments. This finding is plausible because the experiments took place in controlled laboratory environments, the researchers were all using the same protocols, and the primary aim of every experiment was to replicate existing results consistently. Given that the research teams had no incentive to selectively report their results, it is also reasonable to assume that there was no publication bias.

Assuming that the true effects did not vary across study sites makes it possible to estimate $\Delta_c^{(n)}$, defined as the power gain conditional on sample draws of $H$. Notice that $\mathbb{E}\left[ \Delta_c^{(n)}\right]= \Delta_c$ and in large samples, $ \Delta_c^{(n)}\to\Delta_c$.
\begin{equation}
    \Delta_c^{(n)}\equiv \frac{1}{n}\sum_{i=1}^n\left[\text{Pr}\left(|T_c|>1.96\:|\:H=h_i\right)-\text{Pr}\left(|T|>1.96\:|\:H=h_i\right) \right]
\end{equation} 

To estimate $ \Delta_c^{(n)}$ proceed as follows. For each of the 11 experimental treatments that were studied with $t$-tests, I take the (sample-size-weighted) mean $\overline{b}$ of the reported effects across the 36 study sites to estimate the true effect $b$ for that treatment. Then I replace each of the 385 $t$-scores with the ratio $\frac{\overline{b}}{s_i}$ where $s_i$ is the standard error used to compute the $i$th $t$-score. To compute unconditional power, I plug each ratio $\frac{\overline{b}}{s_i}$ into the power function for the size 5\% $t$-test and take the mean over study sites. To compute power were the sample sizes all to have been counterfactually doubled, I plug $\sqrt{2} \frac{\overline{b}}{s_i}$ into the power function instead. Taking the average yields the point estimate: $ \widehat{\Delta}_c^{(n)} = 0.078$ with standard error $0.0007$. This estimation technique is similar in spirit to those used by \cite{PowerOfBias,bundock}. It assumes a common true effect across study sites for each treatment and that the $s_i$ are true standard errors rather than estimates.

\begin{figure}
    \centering
    \includegraphics[width=5in]{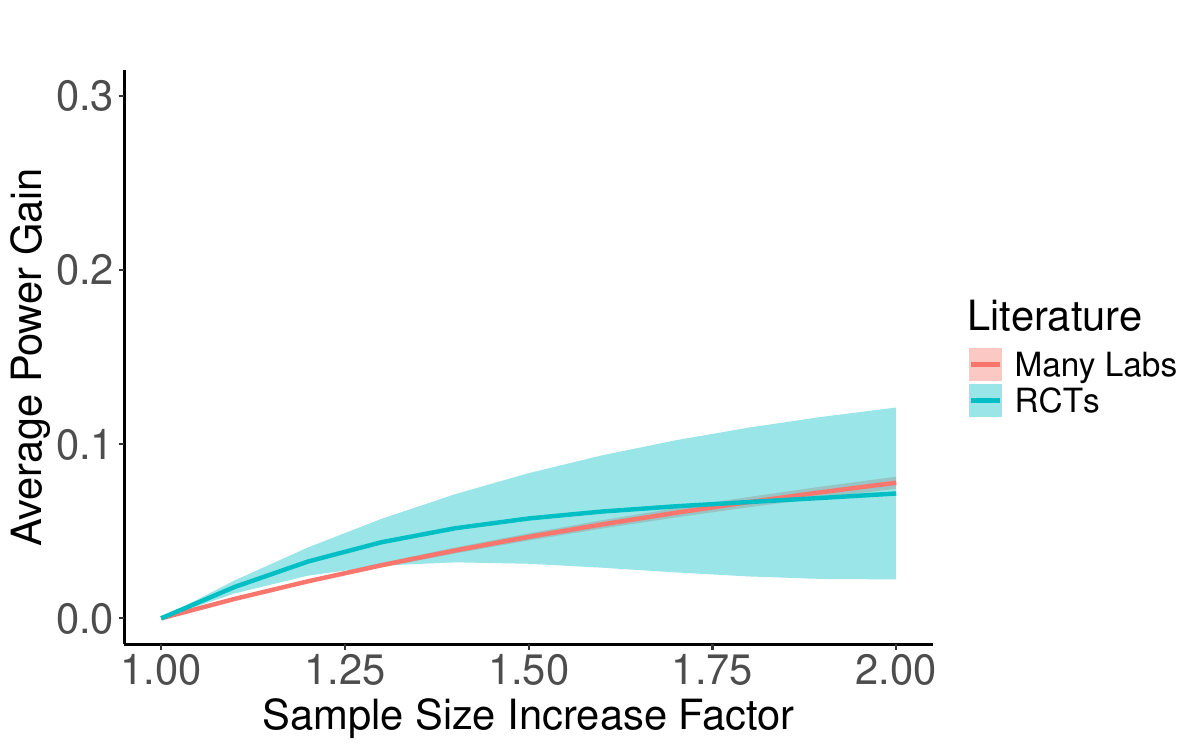}
    \caption{\footnotesize Compares power gain $\widehat{\Delta}_{c}^{(n)}$ for Many Labs  vs $\widehat{\Delta}_{c,n}$ Economics RCTs  (y-axis) over many $c^2$ (x-axis) . Data: \cite{bb} and \cite{ManyLabs}. }
    \label{fig:RCts_Alt}
\end{figure}

Notice that $\widehat{\Delta}_c^{(n)}$ is extremely close to the 7.2 percentage point power gain that we estimated for RCTs in economics and is estimated very precisely. Figure \ref{fig:RCts_Alt} visualizes how close the two power gain curves are. Computing the standard error under the conservative worst-case assumption that two experiments in the same lab are perfectly correlated yields standard error $0.0018$. This reinforces the earlier conclusion that RCTs in economics are about as well powered as laboratory replications. 

\end{document}